\documentclass[twocolumn, prb, reprint, amsfonts, amsmath, amssymb, superscriptaddress, longbibliography, floatfix]{revtex4-2}

\usepackage{graphicx}
\usepackage{bm}
\usepackage{physics}
\usepackage[colorlinks,
    linkcolor=blue,
    anchorcolor=red,
    citecolor=green
]{hyperref}
\usepackage{wasysym}
\usepackage{MnSymbol}
\usepackage{xcolor}
\usepackage[thinc]{esdiff}
\usepackage[normalem]{ulem}

\newcommand{\comment}[1]{}

\newcommand{\eqnref}[1]{Eq.~(\ref{#1})}
\newcommand{\eqnsref}[3]{Eqs.~(\ref{#1},\ref{#2},\ref{#3})}
\newcommand{\figref}[1]{Fig.~\ref{#1}}

\newcommand{\sfigref}[2]{Fig.~\hyperref[#1]{\ref{#1}#2}}

\newcommand{\secref}[1]{Sec.~\ref{#1}}
\newcommand{\appref}[1]{Appendix~\ref{#1}}

\newcommand{\DT}[1]{color{blue}}

\usepackage{wasysym}

\begin{document}
\title{
Design and Benchmarks for Emulating Kondo Dynamics on a Quantum Chip
}
\author{Soumyadeep Sarma}
\affiliation{Undergraduate Programme, Indian Institute of Science, CV Raman Road, Bengaluru 560012, India}

\author{Jukka I. V\"ayrynen}
\affiliation{%
Department of Physics and Astronomy, Purdue University, West Lafayette, Indiana 47907, USA
}%

\author{Elio J. K\"onig}
\affiliation{%
Department of Physics, University of Wisconsin-Madison, Madison, Wisconsin 53706, USA
}%
\affiliation{%
Max-Planck Institute for Solid State Research, 70569 Stuttgart, Germany
}%

\begin{abstract}

Motivated by recent advances in digital quantum simulation and the overall prospective of solving correlated many-electron problems using quantum algorithms, we design a gate-based quantum circuit that emulates the dynamics of the Kondo impurity model. We numerically determine the impurity magnetization, entanglement between impurity and fermionic sites and energy as a function of time (i.e.~circuit depth) for various initial states and find universal long-time dynamics. We complement the numerical simulations for moderate system size with an asymptotically exact analytical solution that is effective in the limit of large system sizes and for starting states corresponding to a filled Fermi sea. This work opens up the perspective of studying the dynamics of electronic quantum many-body states on quantum chips of the NISQ era.

\end{abstract}

\maketitle

\section{Introduction}

    The noisy-intermediate scale quantum (NISQ) era~\cite{Preskill2018} has awakened unique possibilities of realizing and studying exotic quantum many-body phases of matter and their dynamics. 
    While conventional analog quantum emulators, as they have long been realized, e.g., in cold-atomic gases, trapped ions, photonics and Josephson junction arrays, opened up unprecedented scientific fields, they {used to} require highly trained experts to operate them. {With the ongoing second quantum revolution the situation is about to change. In particular digital} quantum chips with predefined gate sets are about to enable almost anyone to run their own quantum simulations {with cloud access 
    for users who are not scientific experts in the given field or who might be geographically unable to access a 
    given quantum device}.  

    Within the subfield of simulations for many-body physics, the study of quantum dynamics has emerged as a particularly interesting application of NISQ devices. Contrary to classic solid state realizations, the high degree of control enables, firstly, to explore the entirety of the Hilbert space and, secondly, to design at will and locally read out very specific dynamic quantum many-body models. For example, this opens the road to experimental information about the build-up of quantum entanglement between various constituents of quantum many-body systems.  

    Needless to say, gate-based quantum simulations lead to discrete time dynamics, and, therefore, their study falls under the general umbrella of Floquet dynamics~\cite{Mori2023,BukovPolkovnikov2015}, which harbors dynamical many-body phases of matter that can be distinct from their equilibrium counterparts. Moreover, it is a nontrivial question to study when, if, and how heating effects destroy the features of equilibrium ground state phase diagrams relevant to the low-energy physics applicable 
    to solid state realizations. 

    Recent experimental studies of many-body dynamics in gate-based quantum simulators of spin chains include bound state formation \cite{MorvanRoushan2022} and Kardar-Parisi-Zhang growth of magnetization \cite{RosenbergRoushan2024}, as well as Floquet symmetry protected phases~\cite{MiRoushan2022,ZhangWang2022}.
    Theoretically, there is a particular focus towards integrable, or nearly integrable systems \cite{GritsevPolkovnikov2017, VanicatProsen2018,LjubotinaProsen2019,Dupont2020,Aleiner2021,VernierMitra2024,VernierPiroli2023}. Combining gate-based, digital quantum simulation of quantum dynamics with recent advances in state preparations also opens up the prospective for studying quench dynamics~\cite{Mitra2018} in NISQ devices.

    Beyond the previously discussed spin simulations, a major scientific application of quantum computing is the solution of the correlated many-electron problem pertinent to quantum chemistry~\cite{McArdleYuan2020}, materials science~\cite{OfteliedeJong2021} and solid state physics~\cite{AyralGuzman2023}. At the same time, the digital simulation of quantum dynamics for strongly correlated electron systems has not yet enjoyed the same amount of attention as for its spin system counterparts. 

    In this paper, we {theoretically} study a gate-based quantum simulation of the Kondo impurity problem, arguably the simplest strongly correlated electronic many-body problem~\cite{Bravyi2017}. 
    Using a previously experimentally realized gate set we design a circuit and study its quench dynamics upon sudden inclusion of Kondo interactions in the Floquet unitary. Both impurity magnetization and entanglement properties are studied numerically {running simulations on a classical computer} and using a quantum field theoretical, asymptotically exact solution.  

    The Kondo problem~\cite{HewsonBook}, i.e.~the interaction between a localized magnetic impurity and a sea of non-interacting conduction electrons in a metal, is effectively a boundary problem of a one-dimensional system. Numerous numerical and analytical methods have been applied to solve this problem at thermodynamic equilibrium, including numerical renormalization group~\cite{Wilson1975}, Bethe Ansatz~\cite{Andrei1980,Wiegman1980}, {parton constructions}~\cite{Coleman1983,ColemanBook}, bosonization~\cite{EmeryKivelson1992}, and conformal field theory~\cite{AffleckLudwig1991}. A notable feature is the formation of a many-body singlet state in which surrounding electrons screen the impurity spin. The singlet forms below the characteristic Kondo temperature $T_K$.

    Given its outstanding conceptual importance, the Kondo problem has a long tradition of being implemented in solid state analogue quantum emulators, in particular quantum dots coupled to metallic wires~\cite{PustilnikGlazman2004}. Generalizations to quantum critical emulators of standard (non-standard) symmetry have been studied experimentally~\cite{IftikharPierre2018} (theoretically~\cite{BeriCooper2012, MitchellAffleck2021, LiVayrynen2023,KoenigTsvelik2023, BollmannKoenig2024}). Theoretically, quench~\cite{NordlanderLangreth1999,LobaskinKehrein2005, HeylKehrein2010, VasseurSaleur2013} and Floquet dynamics~\cite{LobaskinKehrein2006, HeylKehrein2010b} have revealed the importance of the time scale $\hbar/(k_BT_K)$ for dynamics, a feature that, as we will see, only partially translates to the present gate-based simulator.
    
    In the context of the design of the circuit under consideration it is important to emphasize that several schemes optimizing quantum solvers for the Kondo model and related quantum impurity models~\cite{Bravyi2017} have been designed over the years~\cite{YaoOrth2021,AyralGuzman2023}, including the use of natural orbitals~\cite{He2014,Yang2017,BesserveAyral2022} and Gaussian circuits ~\cite{WuStoudenmire2022}, which essentially exploit non-local basis changes to degrees of freedom which are closer to the exact collective excitations on top of the quantum many-body ground state. This is an important trick if the goal is to design (quantum) algorithms targeted towards the low-energy sector of the quantum many-body phase of matter. Instead, this work intends to benchmark the design of quantum dynamics in strongly correlated electronic systems, their entanglement dynamics and local observables, starting from a toy model that is well understood with the prospective to extend to non-integrable models in the long run. As such we deliberately refrain from non-local basis changes in the circuit design and, instead, stay physically as close as possible to the actual Kondo problem. Needless to say, we do, of course, exploit various non-local transformations in the analytical solution of the problem. 

The remainder of the paper is structured as follows: In \secref{sec:sec2}, we explain the design of the quantum circuit involving the starting states, the free fermion and Kondo unitary gates. In \secref{sec:sec3} we present the observables of impurity magnetization, internal energy, 
local and non-local entanglement that we measure on the output state of the quantum circuit after a given number of Floquet steps. The corresponding numerical results in \secref{sec:sec4} highlight the main features of the graphs obtained via exact diagonalization and circuit simulations. In \secref{sec:Analytics} we present an analytical solution of our Floquet system close to the thermodynamic limit using bosonization and refermionization. We conclude with a summary and outlook in \secref{sec:Discussion}  and delegate important technical details to the appendices. 

Throughout the paper, we use the convention $\hbar = {k_B} = 1$.

\section{Design of the quantum circuit}
\label{sec:sec2}
In this section, we {review and introduce}
{
two different but related types of Kondo models: First, the \textit{Hamiltonian} Kondo model (the exponential of which defines continuum time evolution) and, second, the Kondo model \textit{(unitary) circuit} (corresponding to Floquet time evolution).}  


\subsection{{Hamiltonian Kondo model}}

The Kondo Hamiltonian is described as
\begin{subequations}
\label{eqn:main-ham}
\begin{align}
    &H = H_{\rm kin} + H_K,\\ 
    \label{eq:main-ham_b}
    &H_{\rm kin} = -t_0 \sum_{n \geq 0 , \sigma}({\Psi}^\dag_{n,\sigma}{\Psi}_{n+1,\sigma} + H.c.),\\ 
    &H_K = J_K({\Psi}^\dag_0 \sigma_+ {\Psi}_0 S^- + H.c.) +  J_z {\Psi}^\dag_0 \sigma_z {\Psi}_0 S_z,
\end{align}
\end{subequations}
where $\sigma_{x,y,z}, \sigma^\pm = [\sigma_x \pm i \sigma_y]/2$ are Pauli matrices acting in the space of electron spin and $S_\pm = (x_0 \pm i y_0)/2, S_z = z_0/2$
are spin operators acting on the {impurity site}, where $x_0,y_0,z_0$ are Pauli operators. The index $n=0,1,\dots,N-1$ labels fermionic sites ($n=0$ being coupled to the impurity),
and ${\Psi}_{n,\sigma}$ is the annihilation operator of electrons at site $n$ with spin $\sigma \in \{ \uparrow, \downarrow \}$ (index occasionally suppressed for simplified notation), while $H.c.$ means Hermitian conjugate. The hopping amplitude is denoted $t_0$ while $J_K \, (J_z)$ are the in-plane (out-of-plane) exchange coupling constants, {(we choose} {
the quantization axis for the spin 
along $z$ }
{when resorting to the notion of in-plane vs. out-of-plane).} 

\subsection{Unfolded model}
The first step is to convert the Hamiltonian Kondo model 
to a configuration as shown in~\figref{fig:main-img} a) (left of `` $\simeq$ " sign), with 
$\sigma = \uparrow$ on the top and $\sigma = \downarrow$ on the bottom half. We use the convention of setting the impurity site position as $x = 0$, and the spin-resolved fermionic site positions as $x = \pm \frac{2n+1}{2}$, $n$ is a non-negative integer.
An electron in site $x = - \frac{2n+1}{2}$, 
means that a spin-down is occupying site $n$ in the spin-degenerate chain, and so on. In the unfolded model, \eqnref{eqn:main-ham} becomes \cite{WuStoudenmire2022}
\begin{subequations}
\label{eqn:unfolded1}
\begin{align}
    &H_{\rm kin} = -t_0 \sum_{x;x \neq -\frac{1}{2}}({\Psi^\dag(x)\Psi(x+1)}  + H.c.),\\
    &H_K = J_K({\Psi^\dag\left(-\frac{1}{2}\right)\Psi\left(\frac{1}{2}\right)}S^+ + H.c.) + \frac{J_z}{2} z_0(\hat n_{\frac{1}{2}} - \hat n_{-\frac{1}{2}}),
\end{align}
\end{subequations}
where $\hat n_x = {\Psi^\dag(x)} {\Psi(x)}$,  ${\Psi(x)}$ is the annihilation operator at site $x$. The second step is to utilize the Jordan-Wigner transformation to transform the second-quantized operators into Pauli gates which can be done as
\begin{align}
    {\Psi(x)} = \prod_{x'<x}Z_{x'}\left(\frac{X_x - iY_x}{2}\right), 
\end{align}
where $X_x,Y_x,Z_x$ are Pauli operators on site $x$. 

In this formalism, we now treat the spin-resolved fermionic sites as qubits with states $\ket{1}(\ket{0})$, corresponding to the site unoccupied (occupied) with an electron. Introducing the notation $P_x = (X_x, Y_x)$, $P_x \cdot P_{x'} = X_xX_{x'} + Y_xY_{x'}$ and $P_x \wedge P_{x'} = X_xY_{x'} - Y_xX_{x'}$, we get
\begin{subequations}
\label{eqn:tr-ham} 
\begin{align}
 &H_{\rm kin} = -\frac{t_0}{2} \sum_{x \neq -\frac{1}{2}}  P_x \cdot P_{x + 1}, \\
  &H_K = - \frac{J_K}{2}[(P_{-\frac{1}{2}} \cdot P_{\frac{1}{2}})x_0 + (P_{-\frac{1}{2}} \wedge P_{\frac{1}{2}})y_0] \nonumber\\ 
  &\quad \quad+ \frac{J_{z}}{2}(Z_{\frac{1}{2}} - Z_{-\frac{1}{2}})z_0, \,
\end{align}
\end{subequations}
which is the Kondo Hamiltonian in terms of qubits. 

\subsection{Kondo Quantum Circuit}
So far, in Eq.~\eqref{eqn:tr-ham},  we have written the Kondo Hamiltonian in terms of qubits. As a next step we design a quantum circuit to emulate its discrete time dynamics 
in terms of the following Floquet unitary
\begin{equation}
U_F = U_K U_{\rm kin} , \label{eq:Floquet}
\end{equation}
where $U_K$ ($U_{\rm kin}$) denotes the Kondo (free fermion) unitary evolution related to the Hamiltonian $H_K$ ($H_{\rm kin}$) introduced above (details follow)
and we consider its quench dynamics. 

\subsubsection{Initial state(s)}
\label{sec:InitialStates}

We consider three types of initial states. Firstly, we utilize a Fermi sea state $\ket{FS}$ on both spin up and down chains along with having the impurity site in the down spin $\ket{\downarrow} \equiv \ket{1}$ as shown in \figref{fig:main-img} a). Initializing the $\ket{FS}$ requires us to {implement}
\begin{equation}
\label{eqn:fs-state}
\ket{FS} = \prod_{k \in FS} c^\dag_{k}\ket{\underline{0}},
\end{equation}
where $\ket{\underline{0}}$ is the vacuum state, and $c_k$ is the annihilation operator in the
momentum-basis and ${k \in FS}$ denotes the subset of $k$ states which are occupied in the Fermi sea. Fourier transforming to position basis gives us coefficients of the form $\prod_{i,j}e^{ik_ix_j}$ which enter the $\ket{FS}$ state as antisymmetrized tensors contracted with bitstrings of qubits at positions $\{ x_j \}_j$. The initial state of the circuit then becomes (left to right here means top to bottom in \figref{fig:main-img} a))

\begin{equation}
\ket{\psi}_{\rm in} = \ket{FS} \otimes \ket{1} \otimes \ket{FS}.
\end{equation}
In actual quantum chips, it is challenging to initialize a linear superposition state with arbitrary coefficients, as it requires an extensive number of 2-qubit unitaries leading to a high circuit depth. A workaround to this problem may be adiabatic state preparation \cite{AlbashLidar2018}, where one starts with a Hamiltonian whose ground state is easy to prepare (e.g. a product state) and slowly interpolates to a final Hamiltonian whose ground state is the target. Faster, diabatic protocols suitable for the present case of one-dimensional {fermions} have also been proposed recently~\cite{AgarwalSondhi2018}. 

Secondly, and motivated by these challenges, we cross-check the dynamics of the $\ket{FS}$ initial state with two other initial states. We utilize a translationally invariant Greenberger-Horne-Zeilinger state, denoted $\ket{TS}$, defined on $N$ fermionic qubits as:
\begin{align}
 \label{eqn:ts-state}
    \ket{TS} = \frac{1}{\sqrt{2}}(\underbrace{\ket{1010...}}_{\text{N qubits}} + \underbrace{\ket{0101...}}_{\text{N qubits}}).
\end{align}
Both the $N$ up and the $N$ down spin qubits are initialized separately in this $\ket{TS}$ state, while the impurity is initialized in a spin-down state ($\ket{1}$) so that 
\begin{equation}
\ket{\psi}_{\rm in} = \ket{TS} \otimes \ket{1} \otimes \ket{TS}.
\end{equation}

Finally, we look at the randomized tensor product (RTP) state in which the state of every single fermionic qubit is defined as:
\begin{align}
\label{eqn:rtp-state}
    \ket{\psi}_x= \cos\left(\frac{\theta_x}{2}\right)\ket{0} + e^{i\phi_x}\sin\left(\frac{\theta_x}{2}\right)\ket{1}.
\end{align}
Here, $\theta_x$ and $\phi_x$ are local random parameters taken from uniform distributions over $[0, \pi]$ and $[0,2\pi)$ respectively. The initial state is the tensor product of all such $\ket{\psi}_{x}$ along with a spin-down impurity state.

\subsubsection{Free fermion evolution $U_{\rm kin}$}

We first consider the free fermion Floquet evolution, i.e. a simple match-gate circuit, denoted $U_{\rm kin}$ in \eqnref{eq:Floquet}. 
We employ the following {2}-qubit gate, which has been previously implemented experimentally in Refs.
~\cite{MorvanRoushan2022,MiRoushan2022}
\begin{subequations}
\begin{align}
e^{i\frac{\theta}{2}(X \otimes X + Y \otimes Y)} &= fsim(\theta,0,0),\\
 fsim(\theta, \phi, \beta) &= \begin{pmatrix}
        1 & 0 & 0 & 0\\
        0 & \cos(\theta) & ie^{i\beta}\sin(\theta) & 0\\
        0 & ie^{-i\beta}\sin(\theta) & \cos(\theta) & 0\\
        0 & 0 & 0 & e^{i\phi} \\
    \end{pmatrix},
\end{align}
\end{subequations}
where we define the matrix in the computational basis state $\{\ket{00},\ket{01},\ket{10},\ket{11}\}$. 
The free fermion unitary $U_{\rm kin}$ is then defined as
\begin{subequations}
    \label{eqn:final-hop}
\begin{align}
    &U_{\rm kin} = U_{t{2}}U_{t{1}}, \\
    &U_{t1} = e^{i\frac{\theta}{2} \sum_{n=0}(X_{2n}X_{2n+1} + Y_{2n}Y_{2n+1})}, \\
    &U_{t2} = e^{i\frac{\theta}{2} \sum_{n=0}(X_{2n+1}X_{2n+2} + Y_{2n+1}Y_{2n+2})},  
\end{align}
\end{subequations}
which is an approximate version of $e^{-iH_{\rm kin} T/2}$ (cf.~the brick wall structure of Fig.~\ref{fig:main-img} a)), where $T$ is the time period of one Floquet step. In the Trotter limit ($\theta \to 0$), where we can identify $\theta \simeq t_0T$, $U_{\rm kin}$ approaches the continuum evolution. 
In our numerical studies, we consider open boundary conditions of these operators. 

In the Floquet dynamics, the energy of the system is not conserved, but there is a dispersion relation connecting the quasienergy (as inferred from the corresponding eigenvalue of $U_{\rm kin}$) to the momentum in the Brillouin zone.
For $\theta < \pi/2$, this dispersion crosses zero energy twice, and the Fermi velocity at half-filling is
\begin{equation}
    v = \sin(\theta) \label{eq:FermiVelocity}.
\end{equation}
We emphasize that the natural velocity units employed here are $2a/T$, with $a$ the lattice spacing and $T$, as mentioned, the Floquet time. We use Eq.~\eqref{eq:FermiVelocity} as an input for the analytical results obtained in section~\ref{sec:Analytics} and defer the technical details regarding the exact dispersion relation to \appref{sec:Matchgate}.

\subsubsection{Kondo evolution $U_{K}$}
\begin{figure*}
    \includegraphics[scale = 1]{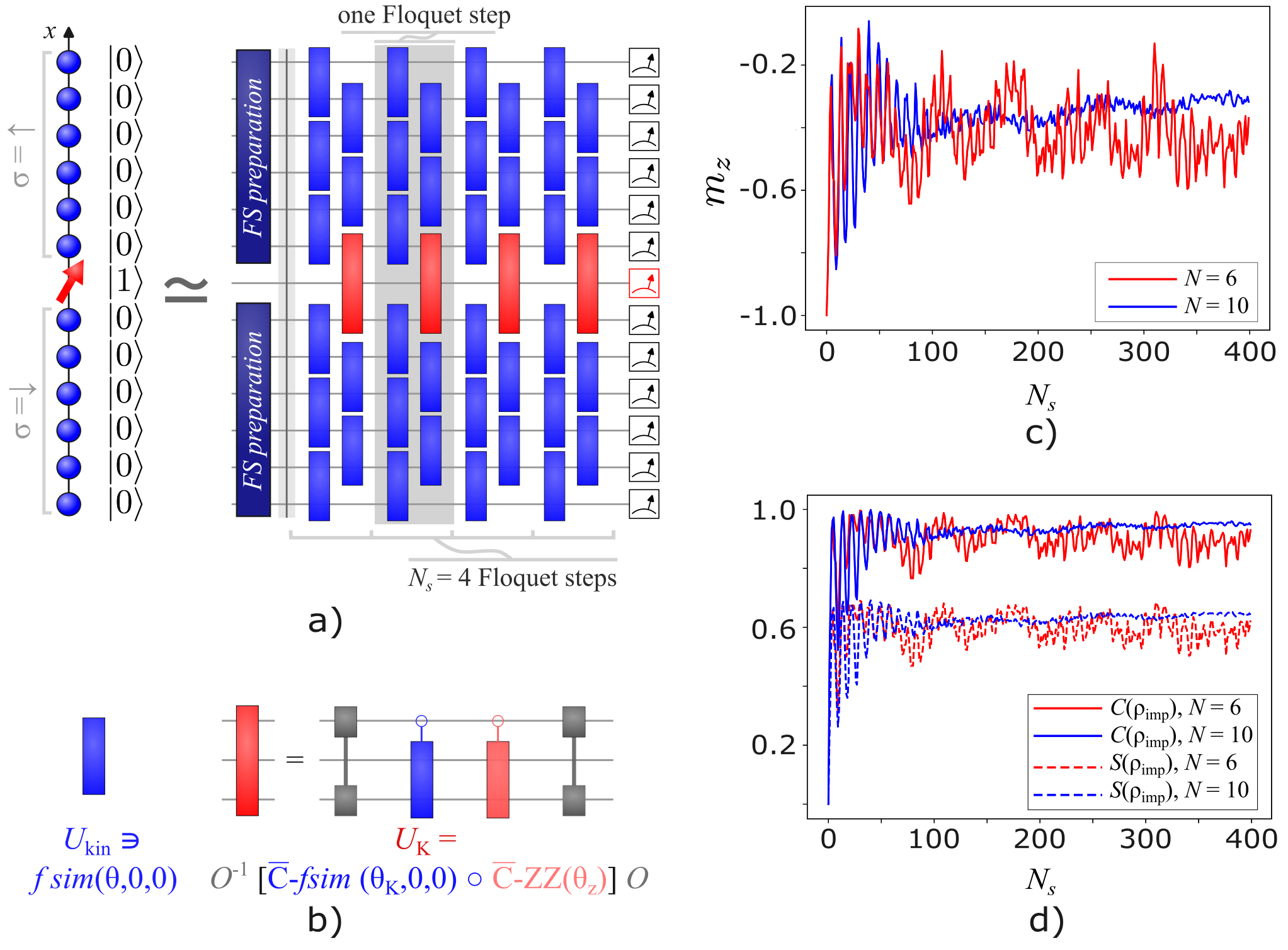}
    \caption{ a) Graphical representation of the unfolded Kondo setup, \eqnsref{eqn:unfolded1}{eqn:tr-ham}{eqn:final-hop} and the main quantum circuit. The qubit labeling the impurity state is initialized as a down spin ($\ket{1} = \ket{\downarrow}$), while the first and last $N$ qubits denote the occupation of electrons in the down and up spins, respectively, where $\ket{0}$ is a vacant site. The dark blue boxes at the start initialize the qubits representing spin-up and down electrons into separate Fermi seas $\ket{FS}$ (for other initializations see Sec.~\ref{sec:InitialStates}). At the end of the evolution, one may locally read-out in the computational basis to obtain the probability distribution of the classical bitstrings, which is used to recreate the final state. b) Legend for the gates {employed in panel a)}: The 2-qubit blue boxes are $fsim(\theta,0,0)$ gates and constitute $U_{\rm kin}$, Eq.~\eqref{eqn:final-hop}. The red boxes are the 3-qubit Kondo interaction gates $U_K$, Eq.~\eqref{eqn:KondoFloq} which can be decomposed into controlled \textit{fsim} and {$R_{ZZ}$ (2-qubit rotation)} gates. c)  Impurity magnetization $m_z = \langle z_0 \rangle$ for $N = 6$ (red curve) and $N = 10$ (blue curve) as a function of time (Floquet steps) at the Toulouse point for $\theta = \pi/3,~ \theta_K = \pi/6$. d) Entanglement measures between impurity and fermionic {bath}: Concurrence (solid curves) and von Neumann entropy ({dashed curve}) 
    for $N = 6$ (red) and $N = 10$ (blue) as a function of time (Floquet steps) at the Toulouse point for $\theta = \pi/3,~ \theta_K = \pi/6$.}
    \label{fig:main-img}
\end{figure*}

The Kondo term in the Hamiltonian Eq.~\eqref{eqn:tr-ham} couples the qubits at $x=\pm 1/2$ to the impurity. 
Accordingly, we define the 3-qubit  unitary gate $U_K$, {depicted as red boxes in Fig.~\ref{fig:main-img} a) and} acting on the central three qubits as an $8 \times 8$ matrix,
\begin{widetext}
 \begin{align}
\label{eqn:KondoFloq}
     U_K = \begin{pmatrix}
        1 & 0 & 0 & 0 & 0 & 0 & 0 & 0\\
        0 & 1 & 0 & 0 & 0 & 0 & 0 & 0\\
        0 & 0 & e^{i\frac{\theta_z}{2}}\cos(\theta_K) & 0 & 0 & ie^{i\frac{\theta_z}{2}}\sin(\theta_K) & 0 & 0\\
        0 & 0 & 0 & e^{-i\frac{\theta_z}{2}} & 0 & 0 & 0 & 0\\
        0 & 0 & 0 & 0 & e^{-i\frac{\theta_z}{2}} & 0 & 0 & 0\\
        0 & 0 & ie^{i\frac{\theta_z}{2}}\sin(\theta_K) & 0 & 0 & e^{i\frac{\theta_z}{2}}\cos(\theta_K) & 0 & 0\\
        0 & 0 & 0 & 0 & 0 & 0 & 1 & 0\\
        0 & 0 & 0 & 0 & 0 & 0 & 0 & 1\\
    \end{pmatrix},
\end{align}   
\end{widetext}
where, in the basis of states $\ket{\sigma_{-1/2}, \sigma_{1/2}, m_z}$, we used the {ordering} $\{\ket{000}, \ket{001}, \ket{010}, \ket{011},\ket{100}, \ket{101}, \ket{110},\\ \ket{111}  \}$. We can identify in the Trotter limit 
the dimensionless parameters $\theta_K \simeq J_K T/2$ and $\theta_z \simeq J_z T/2$, {correspdonding to antiferromagnetic exchange interactions.}

Standard universal gate sets for quantum computation only include 1- and 2-qubit gates. Thus, it is instructive to break this 3-qubit gate into a circuit comprising of 2-qubit and 1-qubit gates as follows. First, notice that in the subspace of fermionic qubits adjacent to the Kondo impurity, $x = \pm 1/2$, $H_K$ only acts non-trivially on the two-dimensional subspace $\ket{\sigma_{-1/2},\sigma_{1/2}} \in \{\ket{01}, \ket{10} \}$, corresponding to one fermion at site $n=0$. We thus first use the 2-qubit change of basis
\begin{align}
    \hat O &= \ket{00}\bra{01} + \ket{01}\bra{10} + \ket{10}\bra{11} + \ket{11}\bra{00}
\end{align}
which is represented by the matrix
\begin{align}  
   O = \begin{pmatrix}
        0 & 1 & 0 & 0\\
        0 & 0 & 1 & 0\\
        0 & 0 & 0 & 1\\
        1 & 0 & 0 & 0 \\
    \end{pmatrix},
\end{align}
to effectively encode fermion states $\ket{01}, \ket{10}$ as $\ket{00}, \ket{01}$ in the new basis, where the left qubit projects to the singly-occupied manifold (when in state $\ket{0}$) while the right qubit encodes the two states. Now one may again use the $fsim$ gate introduced above to encode Kondo interactions, i.e. for qubits at position $x = -1/2,0,1/2$, cf. Fig.~\ref{fig:main-img} b)
\begin{align}
    U_K & = \hat O^\dagger \left({\bar C}-e^{i \theta_K \frac{x_0X_{1/2} + y_0Y_{1/2}}{2}}\right) \left({\bar C}-e^{-i\theta_z \frac{z_{0}Z_{1/2}}{2}}\right) \hat O.
\end{align}
This product of two controlled-unitary gates, $ {\bar C}-fsim(\theta_K,0,0)$ and {$ {\bar C}-R_{ZZ}(\theta_z/2)$} (controlled-ZZ rotation gate, {$\bar C-$ is control activated by $\ket{0}$ instead of the usual $\ket{1}$}) can be decomposed into 1 and 2-qubit gates \cite{NielsenChuang2010}, see \appref{sec:appn1b}. This is just one way this decomposition could be achieved, with the possibility of a more optimal decomposition left for future studies.

We thus conclude that the depth of a single Floquet unitary $U_F$ is 2 if arbitrary 3-qubit gates are available, but 5 if arbitrary 2-qubit unitary gates and controlled-unitaries can be implemented, see Fig.~\ref{fig:main-img} a), b). {Hence, our circuit depth becomes $2N_s$ and the {total} number of two-qubit gates in the circuit (utilizing the decomposition in \appref{sec:appn1b}) becomes $(2N+13)N_s$, where $N_s$ is the number of Floquet steps.}

\section{Computables under consideration}
\label{sec:sec3}

In this section we summarize the computables under consideration, in particular the observables whose expectation values are being studied as a function of time (i.e. circuit depth).

\subsection{Impurity magnetization}
We firstly consider the magnetization 
\begin{equation}
\vec{m} = 2\langle \vec S \rangle
\end{equation}
as a function of Floquet steps. Let us denote the reduced density matrix for the impurity as
 \begin{align}
 \label{eqn:rho_imp}
     \rho_{imp} = \frac{1}{2}(\mathbf{1} + 2\vec{m}\cdot\vec{S}),
 \end{align}
where $\mathbf{1}$ is the $2\times 2$ identity matrix.
Since we always start off in a state where $m_x,m_y = 0$, $x$ and $y$ components remain zero for all evolved states thereafter. To see this, simply note that the total magnetization along the z-axis is conserved in the present circuit, i.e., $U^\dagger_F S_z^{\rm tot} U_F = S_z^{\rm tot}$. The operators $S_{x,y}$ change the total magnetization along the z-axis and therefore 
\begin{align}
    m_{x,y} = 2 \text{Tr}[S_{x,y} 
    \rho_{total}] = 0.
\end{align}
Here $\rho_{total}$ is the density matrix of the system. Thus, $\norm{\vec{m}} = |m_z|$, implying that measuring only the magnetization along the z-axis is sufficient. 

In Secs.~\ref{sec:MagnetizationNum}, \ref{sec:ItCalc}, we evaluate numerically the magnetization, see Figs.~\ref{fig:main-img} c), \ref{fig:sz-plots1}, \ref{fig:sz-plots2} and present a complementary analytical derivation in Eq.~\eqref{eq:AnalyticsDisc}, below.

\subsection{Local entanglement}
We furthermore study the entanglement between the impurity and fermionic sites of the system. Two entanglement measures that are considered in this regard are concurrence $C(\rho)$ and von Neumann entanglement entropy $S(\rho)$. To understand concurrence as a measure of the entanglement between impurity (spin-1/2) and fermionic sites, we use a 
definition as in Ref.~\cite{Zhu2009}
\begin{align}
\label{eq:conc}
     C &= \sqrt{2(1-\text{Tr}(\rho_{B}^2))},
\end{align}
which slightly differs from the standard definition of concurrence between two qubits as in \cite{HillWooters1997}, and is used to define entanglement for multiparticle pure states in arbitrary dimensions. Here $\rho_B$ is the reduced density matrix obtained from a pure state by tracing out complementary degrees of freedom of a bipartition $B$. We partition our system into an impurity site and fermionic sites. Thus our $\rho_B = \rho_{imp}$, and we have  
\begin{equation}
\Tr(\rho_{imp}^2) = (1 + \norm{\vec{m}}^2)/2.
\end{equation}
The standard von Neumann entanglement measure is given as $S(\rho) = -\text{Tr}(\rho \ln \rho)$. For $S(\rho_{imp})$, we have:
\begin{align}
\nonumber
    S(\rho_{imp}) &= -\frac{1}{2} \sum_{\pm} (1 \pm \norm{\vec{m}})\ln\left(\frac{1 \pm \norm{\vec{m}}}{2}\right) \\
    &= -\frac{1}{2}\bigg(\ln\left(\frac{1 - \norm{\vec{m}}^2}{4}\right) + \norm{\vec{m}}\ln \left(\frac{1 + \norm{\vec{m}}}{1 - \norm{\vec{m}}}\right)\bigg).
\end{align}
Thus, the impurity magnetization determines the entanglement measures for the impurity problem. To place the above result in a more general context,
we note that, more accurately speaking, the entanglement entropy is
associated to the 2nd cumulant of the spin distribution
 \begin{align}
  \langle \Delta\vec{S}^2 \rangle &= \langle \vec{S}^2 \rangle - \langle \vec{S} \rangle^2= \frac{3 - \norm{\vec{m}}^2}{4},
 \end{align}
and that for the present spin-1/2 system, higher cumulants vanish. The relationship of entanglement with quantum noise is well known from works on full-counting statistics~\cite{KlichLevitov2009}. More specifically, in a point contact geometry for electrical charge transport, the entanglement entropy is related to linear combinations of the even cumulants of the charge probability distribution. 

A numerical evaluation of entanglement measures discussed here can be found in Secs.~\ref{sec:MagnetizationNum}, \ref{sec:ItCalc}, the results are also plotted as a function of Floquet steps in Figs.~\ref{fig:main-img} d) and \ref{fig:sz-plots1} c), d).

\subsection{Bulk entanglement}

For entanglement measures on the impurity, it is clear that the magnetization along $z$-axis suffices. However, instead of tracing out all fermionic sites from $\rho_{total}$ to form $\rho_{imp}$, we can keep some {fermionic sites up to position $n$ on both up and down chains}, to have a measure of entanglement with respect to distance in the fermionic sites \cite{Affleck2008,WuStoudenmire2022}. Specifically and in terms of the unfolded chain, for a given $n$, we keep qubits from $x = - (2n+1)/2$ to $x = (2n+1)/2$ and trace out the rest. We define this as
\begin{align} \label{eq:Sn}
    S(n) = -\text{Tr}(\rho_n \ln \rho_n),
\end{align}
where $\rho_n$ is the reduced density matrix obtained after tracing out all but the impurity site and fermionic sites up to site with position $n$ from impurity site. 

In the limit of large system size $N$, one expects the Kondo impurity to give rise to a universal boundary entropy contribution~\cite{AffleckLudwig1991b,
ZhouSchollwock2006,Affleck2008} to $S({n})$ in addition to the bulk entanglement entropy of the gapless chain~\cite{CalabreseCardy2004}. 
The same measure has been used to study the spin chain formulation of the  Kondo model~\cite{LaflorencieAffleck2008,
DeschnerSorensen2011,
LaflorencieAffleck2006}. 

We show a numerical evaluation of the bulk entanglement versus Floquet steps for different positions in \figref{fig:ent-dist} and versus position at a fixed time step {(see Appendix~\ref{sec:appn5} for more details regarding the second figure)}.

\subsection{Heating}

In periodically driven systems one generically expects the system to heat up over time. Specifically, starting from a low-energy state,
energy is supplied in each cycle and at small times, we expect to see that the internal energy increases.

However, contrary to the case of a driven solid state system (e.g. a material exposed to light), it is here not a priori clear how to define the internal energy. As a proxy to characterize the heating in the system, we consider the expectation value of the total Hamiltonian 
associated to the Floquet unitary circuit, as defined in \eqnref{eqn:main-ham}. We compute $\langle H \rangle$ as a function of time (Floquet steps). 
This observable is also a means to characterize the many-body state after $N_s$ Floquet steps and to answer if it is close to the ground state of the Kondo Hamiltonian or rather a highly excited state. The eigenstate-thermalization-hypothesis~\cite{AlessioRigol2016,Deutsch2018} states that in non-integrable systems and in regards to local observables such as the magnetization, such a state may be mimicked by thermal distribution with temperature proportional to $\langle H \rangle$.

The numerical evaluation of the internal energy can be found in Sec.~\ref{sec:NumHeat}, below.

\section{Numerical simulations}
\label{sec:sec4}

\begin{figure*}
    \includegraphics[width=2\columnwidth]{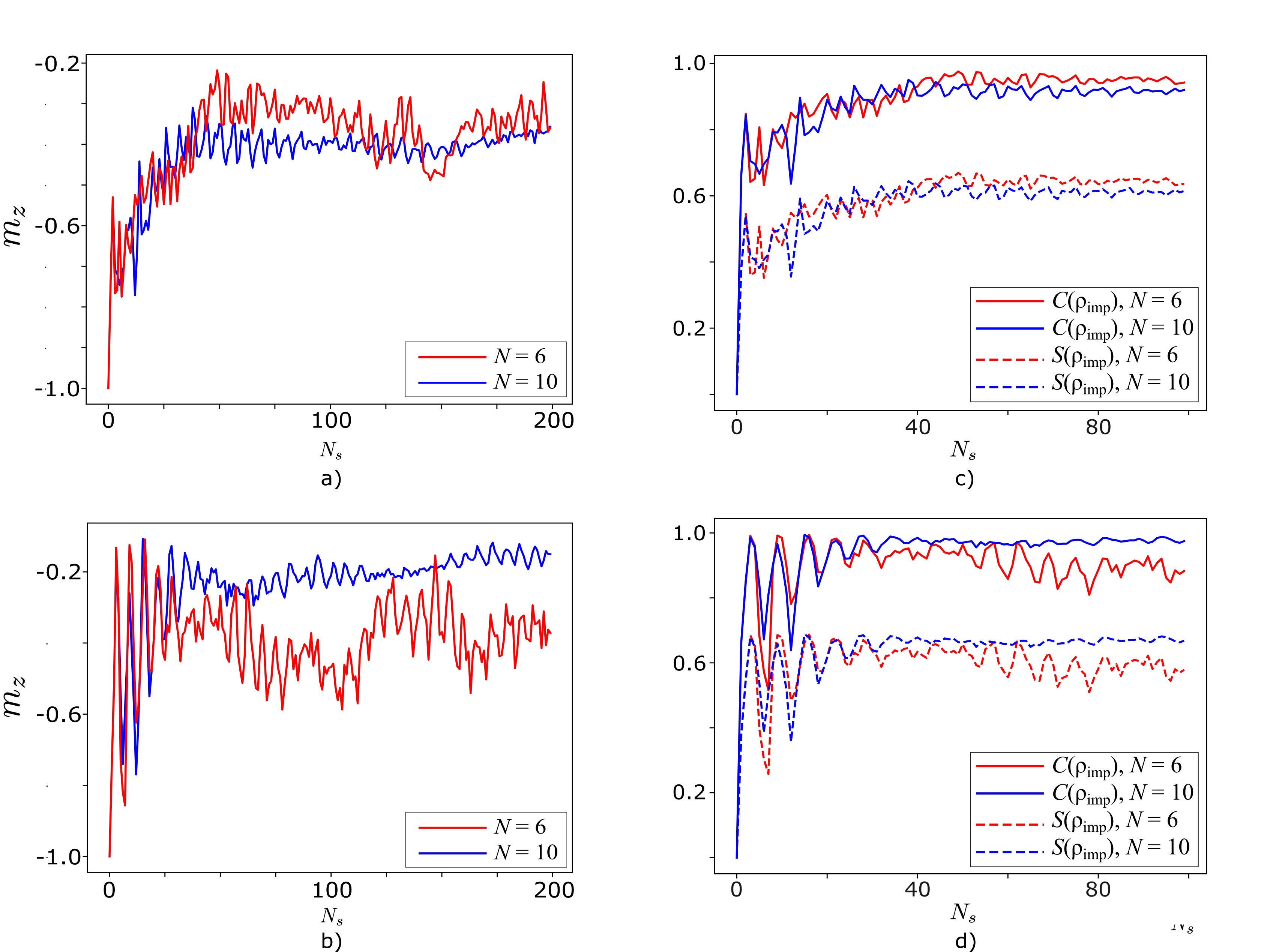}
    \caption{Impurity magnetization expectation values ($m_z = \langle z_0 \rangle$) as a function of time (Floquet steps)
    for $N = 6$ (red) and $N = 10$ (blue) fermionic sites at $\theta = \pi/3,~ \theta_K = \pi/4$ for a) isotropic point,  b) Toulouse point. Entanglement measures of concurrence (solid curve), Eq.~(\ref{eq:conc}), and von Neumann entropy (dashed curve), Eq.~(\ref{eq:Sn}), between subsystems of impurity and fermionic leads {versus Floquet steps} at: c) isotropic point, d) Toulouse point.}
    \label{fig:sz-plots1}
\end{figure*}

This section contains the main numerical simulations. We first carefully analyze magnetization and entanglement dynamics using data obtained by means of exact diagonalization for $N = 6$ (i.e., 13 qubits), Sec.~\ref{sec:EDN6} and for a fermionic bath initialized in a Fermi sea state. Next, in Sec.~\ref{sec:NumChecks}, partly using complementary iterative numerical methods, we numerically demonstrate how these main characteristics persist for larger system sizes, that no additional features appear at longer time scales, and that the asymptotic behavior is independent of the initial bath state.

\subsection{Exact diagonalization for 13 qubits}
\label{sec:EDN6}

In this section, we use exact diagonalization (ED) to obtain all the eigenvalues and eigenvectors of the Floquet unitary ($U_F$) matrix of size $2^{2N + 1} \times 2^{2N + 1}$ in the case of $N = 6$ sites (or $2N + 1 = 13$ qubits). This is done for both the isotropic Kondo ($|\theta_z| = |\theta_K|$) and the anisotropic case at the Toulouse point ($\theta_z =  {\gamma \sin(\theta)/\sqrt{2}}$, $\gamma = \sqrt{2} - 1$),
which is the point in parameter space where the model can be analytically treated using bosonization, see Sec.~\ref{sec:Analytics} below.
All calculations presented in this section are, in contrast, numerical and obtained for the electron bath initialized in the Fermi sea state $\ket{FS}$, see App.~\ref{sec:FermiSea} for details about the numerical implementation thereof.
\subsubsection{Raw data: Magnetization and entanglement curves}

\label{sec:MagnetizationNum}

The typical evolution of impurity magnetization as a function of the Floquet steps is presented in Figs.~\ref{fig:main-img} c), \ref{fig:sz-plots1} a), b). In all of those figures the ED {data} for $N = 6$ is plotted in red. Given the initial conditions, {all} curves start out at $m_z = -1$ but {the values of $|m_z|$ } decay into the long-time asymptote with a smaller magnitude {of oscillation} {by around $40-50$ Floquet steps}. On top of this general trend, we observe oscillatory behavior, in particular a very high frequency leading to oscillations with very small time periods on the order of 10s of steps and much lower frequencies with time periods of the order of 100 steps. At least qualitatively, the behavior of the isotropic Kondo model and the model at the Toulouse {point} are similar.

The entanglement measures plotted in Figs.~\ref{fig:main-img} d), \ref{fig:sz-plots1} c), d) follow similar trends but start out at vanishing entanglement and saturate at a large value. In view of the previously established relationship between quantum noise of the local magnetization and entanglement, this is of course not surprising.

{We also note that in the parameter regime considered here, we should be able to access Kondo physics in a circuit depth of about {100 Floquet steps} {(circuit depth 500 using two-qubit gates). For experimental reference, we highlight that in state-of-the-art} works like~\cite{MiRoushan2022} for the Floquet-Ising chain 
noise
effects destroy bulk features at around 30 steps and topological edge states at about 150 steps.}

    

\subsubsection{Postprocessed data: Characterization of main features}
\label{sec:mainf}

\begin{figure}
    
    \includegraphics[width=\columnwidth]{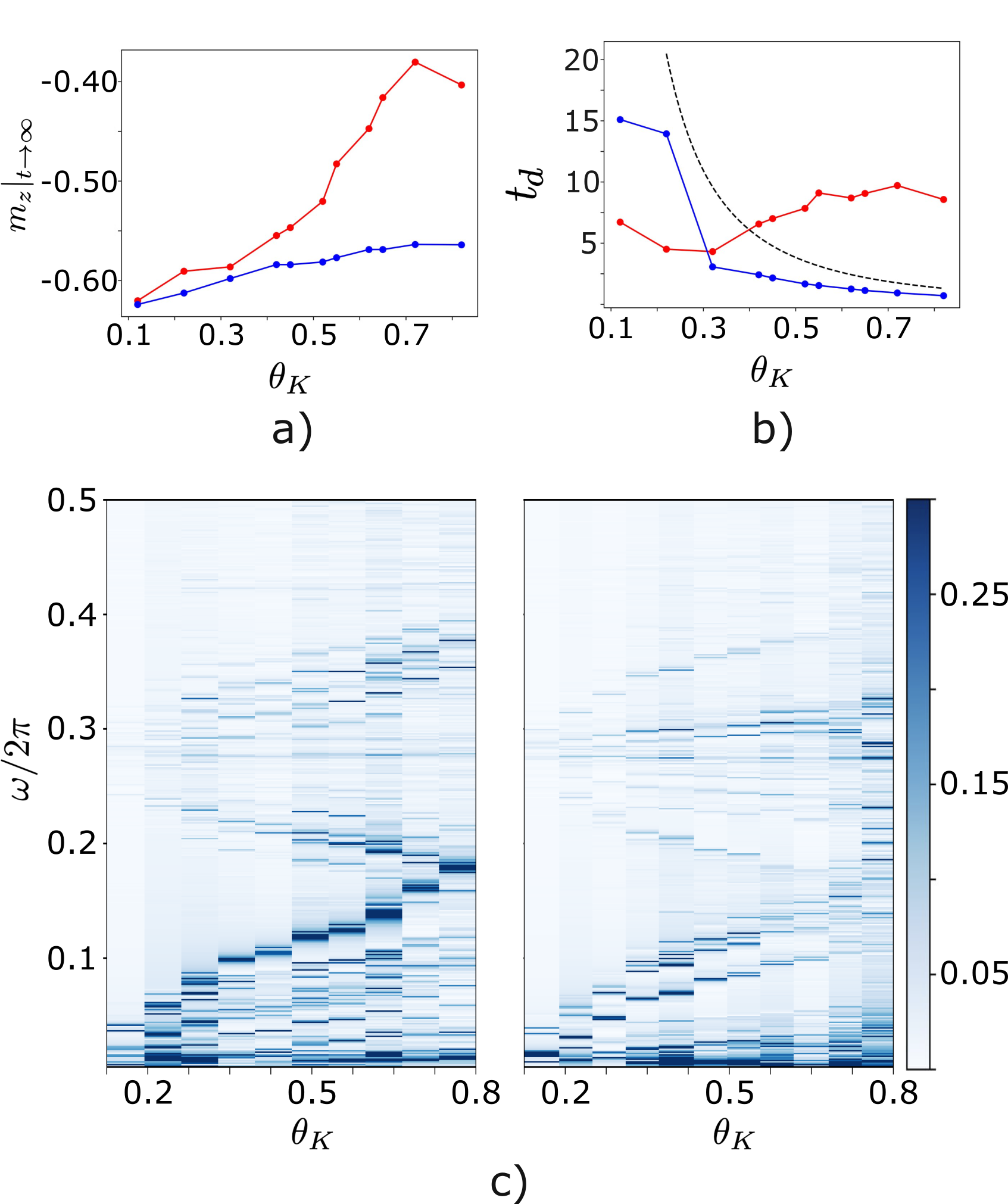}
    \caption{Characteristic properties of the magnetization dynamics versus $\theta_K$ extracted  
    for constant $\theta = \pi/3$ at Toulouse point (blue) and isotropic point (red) for $N = 6$. a) Magnetization asymptote by averaging between 100 and 1000 Floquet steps. b) Decay times ($t_d$) measured utilizing curve fits, as presented in \secref{sec:mainf}. The black dashed curve is the decay time obtained analytically in Sec.~\ref{sec:Analytics} (no fitting involved). c) Color plot of frequencies ($\omega$) of oscillations (left: Toulouse point, right: isotropic Kondo model). The color bar indicates the magnitude of the FFT peak normalized to 1 {(i.e. the sum of the FFT values for all peaks {for each value of $\theta_K$ (i.e. locally)} is 1)}. } 
    \label{fig:prop-plots}
\end{figure}

Given the central role of impurity magnetization, we here quantitatively characterize the features qualitatively described in the previous subsection and extract decay time, asymptote, frequencies of oscillation, and, finally, their {magnitude}. To graphically illustrate their behavior, we compose two types of plots: (i) Keeping $\theta_K$ constant at $\pi/6$, and varying $\theta$, (ii) keeping $\theta$ constant at $\pi/3$ and varying $\theta_K$. In the main text, \figref{fig:prop-plots}, we concentrate on type (ii) plots but discuss the variation with both $\theta$ and $\theta_K$ in words. The type (i) plots are presented in \appref{sec:appn5}. The following is observed: \\

    First, the long-time asymptote is taken to be the average value of the plot between 100 to 1000 steps (i.e. when the magnetizations seems to oscillate about an average). For the isotropic case, the asymptote 
    {increases} towards 0 with increasing $\theta$ or $\theta_K$ and the same trend is observed for increasing $\theta_K$ and $\theta$ in the Toulouse point case as highlighted in \figref{fig:prop-plots} a) {for increasing $\theta_K$.} \\
    
    Second, we take the decay time as the value of $t_d$ obtained from the curve-fit of $m_z(t) = (-1 - m_z \vert_{t \rightarrow \infty})e^{-\frac{t}{t_d}} + m_z \vert_{t \rightarrow \infty}$ to the magnetization plots, where $m_z \vert_{t \rightarrow \infty}$ is the long-time asymptote as described above and the decay rate is $t_d^{-1}$. In the isotropic case, the decay times seem to decrease then increase for increasing $\theta_K$ {at fixed $\theta$}, while for increasing $\theta$ {at fixed $\theta_K$}, it is constant around a very low value of the order of $10^{-2}$. At the Toulouse point the decay times seem to decrease monotonically with increasing $\theta_K$ {at fixed $\theta$}  (see \figref{fig:prop-plots} b)), while they are (mostly) constant for increasing $\theta$ {at fixed $\theta_K$}
    In regards to the $\theta_K$ dependence, it should be emphasized that magnetization curves at very low $\theta_K$ 
    are fundamentally different from curves at other $\theta_K$. We attribute this to a finite size effect (the Kondo length scale $\sim v/T_K$ exceeds {the} system size at smallest $\theta_K$). \\


    Third, we measure the oscillation frequencies ($\omega$) with the use of Fast Fourier Transform (FFT) of the impurity magnetization data from 0 to 1000 Floquet steps and plot it as a color plot in Fig.~\ref{fig:prop-plots} c). For both Toulouse point and isotropic limit, the main low-frequency features appear to be essentially $\theta_K$ independent, while there is also a characteristic frequency which monotonously increases with $\theta_K$. Another note-worthy point is that the magnitude of frequencies for the Toulouse point case is about an order of magnitude higher than the isotropic case at around $\theta_K > 0.4$.\\

    Fourth, we measure the oscillation range by calculating the minimum and maximum values of the magnetization from a 100 to a 1000 steps, and then subtracting them to find the range. For the isotropic case, this range increases first, then decreases with increasing $\theta_K$ {at fixed $\theta$}, while it goes like a sinusoid for increasing $\theta$ {at fixed $\theta_K$}. (see Appendix~\ref{sec:appn5}). In the Toulouse point case, the trend with increasing $\theta_K$ {at fixed $\theta$}  remains that same, while with increasing $\theta$ {at fixed $\theta_K$}, {no clear} %
    trend is observed.  
    The magnitude of ranges is of the same order in both cases. \\

\begin{figure}
    
    \includegraphics[width=\columnwidth]{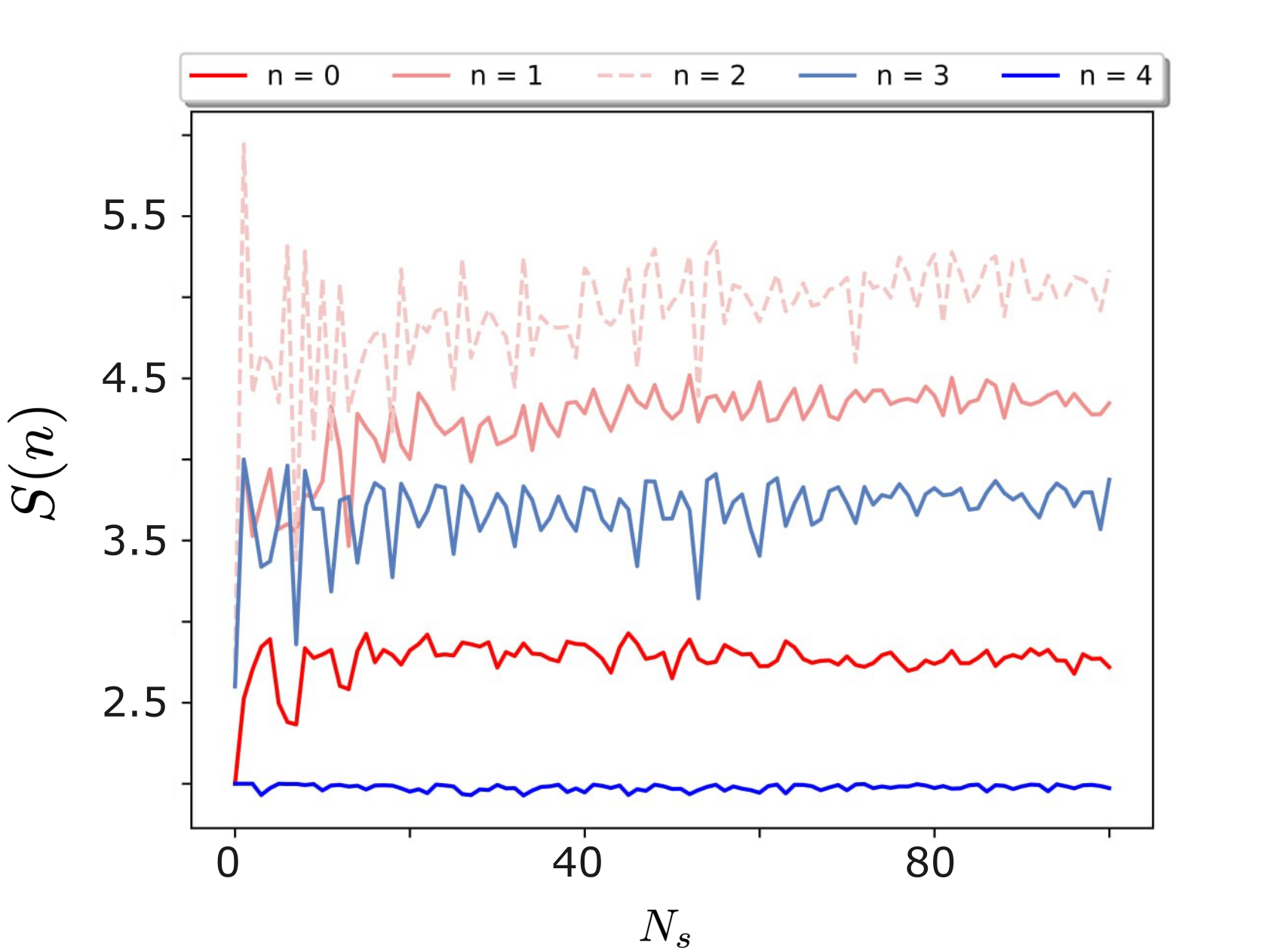}
    \caption{Von Neumann entropy, Eq.~(\ref{eq:Sn}),  for $N = 6$ sites at the Toulouse point as a function of time (Floquet steps) for various distances from the impurity.
    See also Fig.~\ref{fig:ent-dist2}. 
    }
    \label{fig:ent-dist}
\end{figure}    

Finally, \figref{fig:ent-dist} shows the von Neumann entanglement entropy as a function of time for $N = 6$ for different distance values. It is pretty clear from the graph that as a function of distance, the entanglement increases up to $n = 2$ and then decreases rapidly, which is also shown  
{in Appendix~\ref{sec:appn5}}. These results can be taken as evidence for the {formation of a} Kondo cloud, i.e. 
the extent to 
which 
the impurity spin is entangled with the conduction electrons, {analogously to the}
equilibrium {case}~\cite{Affleck2009,PixleyIngersent2015, WuStoudenmire2022}.
While we see an extension up till the 3rd site in a 6-site model, the numerical data is, of course, too inconclusive to make detailed statements.

\subsection{Numerical cross-checks}
\label{sec:NumChecks}

This section is intended to supplement Sec.~\ref{sec:EDN6} in such a way, that it reinforces the physical significance of the aforementioned phenomena but it does not contain new physics beyond Sec.~\ref{sec:EDN6}. Readers purely interested in the main results may thus skip this section.

Using an iterative calculation summarized in Sec.~\ref{sec:ItCalc}, which is numerically amenable for $N = 6$ and $N = 10$, we discuss qualitatively finite size effects. Resorting again to the ED solution of Sec.~\ref{sec:EDN6}, we demonstrate in Sec.~\ref{sec:LongTime} that the asymptotic behavior discussed previously persists at least for the first $10^5$ Floquet steps, i.e.~far beyond the coherence times of NISQ machines. Additionally, we probe the independence of the asymptotic long-time dynamics on the initial state, Sec.~\ref{sec:IniState}, using again the iterative solution.

\subsubsection{Comparing system sizes $N = 6$ and $N = 10$ using iterative solution}
\label{sec:ItCalc}

To cross-check our simulations at larger $N$, we utilized IBM Qiskit's \cite{WilleNaveh2019}
circuit classes called \textit{Estimator} and \textit{SparsePauliOp} to estimate the observables after running the circuit for a certain number of Floquet steps. {We note that we use statevector simulation under the \textit{Statevector} class method which efficiently evolves quantum states by applying unitary matrices directly to a statevector of size $2^{2N+1}$, avoiding the need to store and diagonalize the full Hamiltonian matrix, as required in ED. In statevector simulation, a quantum state is represented as a complex vector, and quantum gates act as sparse matrix operations that update only the relevant amplitudes, unlike ED, which requires constructing and storing the entire $2^{2N+1} \times 2^{2N+1}$ Hamiltonian and computing its full spectrum (scaling as $O(2^{6N})$). Statevector simulation only requires matrix-vector multiplications, leading to a significantly lower memory footprint and computational cost of $O(2^{2N})$. Additionally, Qiskit employs parallelization, look-up tables, and merging consecutive gates into a single matrix, making it feasible to simulate circuits with dozens of qubits, while ED quickly becomes generically intractable beyond 16 qubits. Hence, for $N > 6$ sites, we have only done the Qiskit simulations because straightforward ED is too {memory}-intensive and impractical.} For now, again, the bath may be initialized in a Fermi sea state $\ket{FS}$. We double-checked that the 
impurity magnetization 
obtained from
ED and Qiskit are virtually identical
for $N = 6$ sites.

The magnetization dynamics for $N = 10$ are plotted in blue in Figures~\ref{fig:main-img} c), \ref{fig:sz-plots1}.
On comparing the resulting impurity magnetization curves obtained from the $N = 10$ and $N = 6$ sites, we observe that the long-time
dynamics in both scenarios remains similar and features
oscillations
with various time periods 
and {magnitudes} about a long-time asymptotic value. The value of this long-time asymptote has a smaller magnitude for $N = 10$ sites at the Toulouse point, while it is comparable 
in the isotropic case. We also observe that the amplitudes for oscillations are much more damped for $N  = 10$ sites, and mostly 
smaller frequencies (time periods of the order of nearly 100 steps) are dominant. Both the $N = 10$ and $N = 6$ site curves seem to reach their asymptotic values in roughly the same number of Floquet steps.

Since the entanglement measures, Figs.~\ref{fig:sz-plots1} c) and d), are directly dependent on the impurity magnetization, we have the same observations there as well, explaining why in the $N = 10$ curve the entanglement measures seem to oscillate much less than the $N = 6$ site case after longer times. 

\subsubsection{Heating}
\label{sec:NumHeat}

\begin{figure}
    
    \includegraphics[width=\columnwidth]{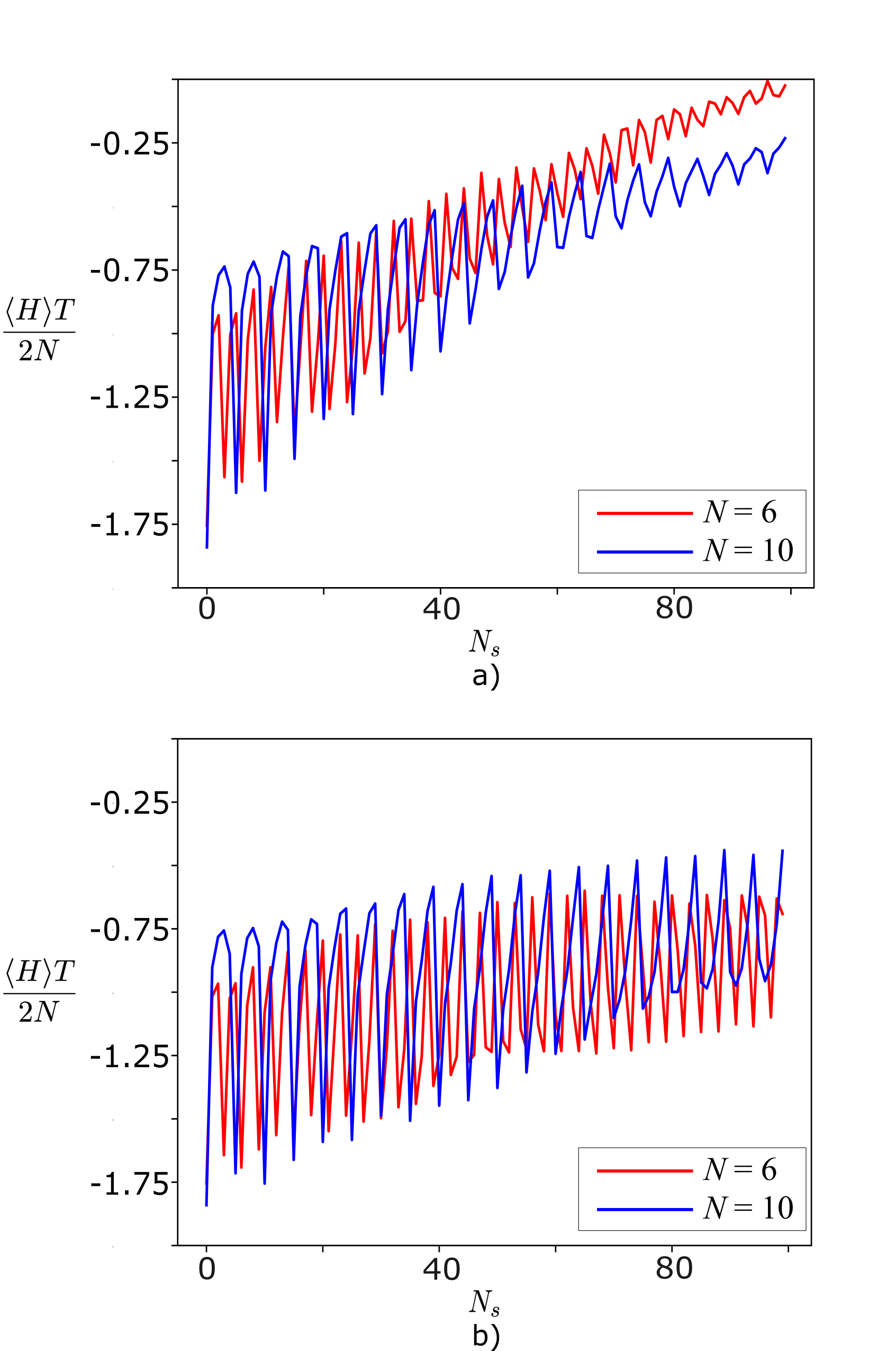}
    \caption{Expectation of total Hamiltonian ($\langle H \rangle  = \langle H_{\rm kin} + H_K \rangle$) as a function of time (Floquet steps) {starting with Fermi sea state $\ket{FS}$} for different parameters for $N = 6$ (red) and $N = 10$ (blue) fermionic sites at Toulouse point. a) $\theta = \pi/4,~ \theta_K = \pi/4$. b) $\theta = \pi/4,~ \theta_K = \pi/6$. }
    \label{fig:h-plots2}
\end{figure}

We now also discuss the average internal energy density, 
see Fig.~\ref{fig:h-plots2}. 
We observe that the energy density starts from a negative value $\mathcal O(-\theta)$, corresponding to the filled Fermi sea and subsequently heats up in an oscillatory fashion and more strongly for larger $\theta_K$. As for magnetization, oscillation  
frequencies {de}crease with circuit width $N$
for both the isotropic and Toulouse point. 
This is consistent with the expectation that
the fast frequency of oscillation is
determined by the single-particle level spacing, which goes as $1/N$.

\subsubsection{Long-time dynamics}
\label{sec:LongTime}

\begin{figure}
    
    \includegraphics[width=\columnwidth]{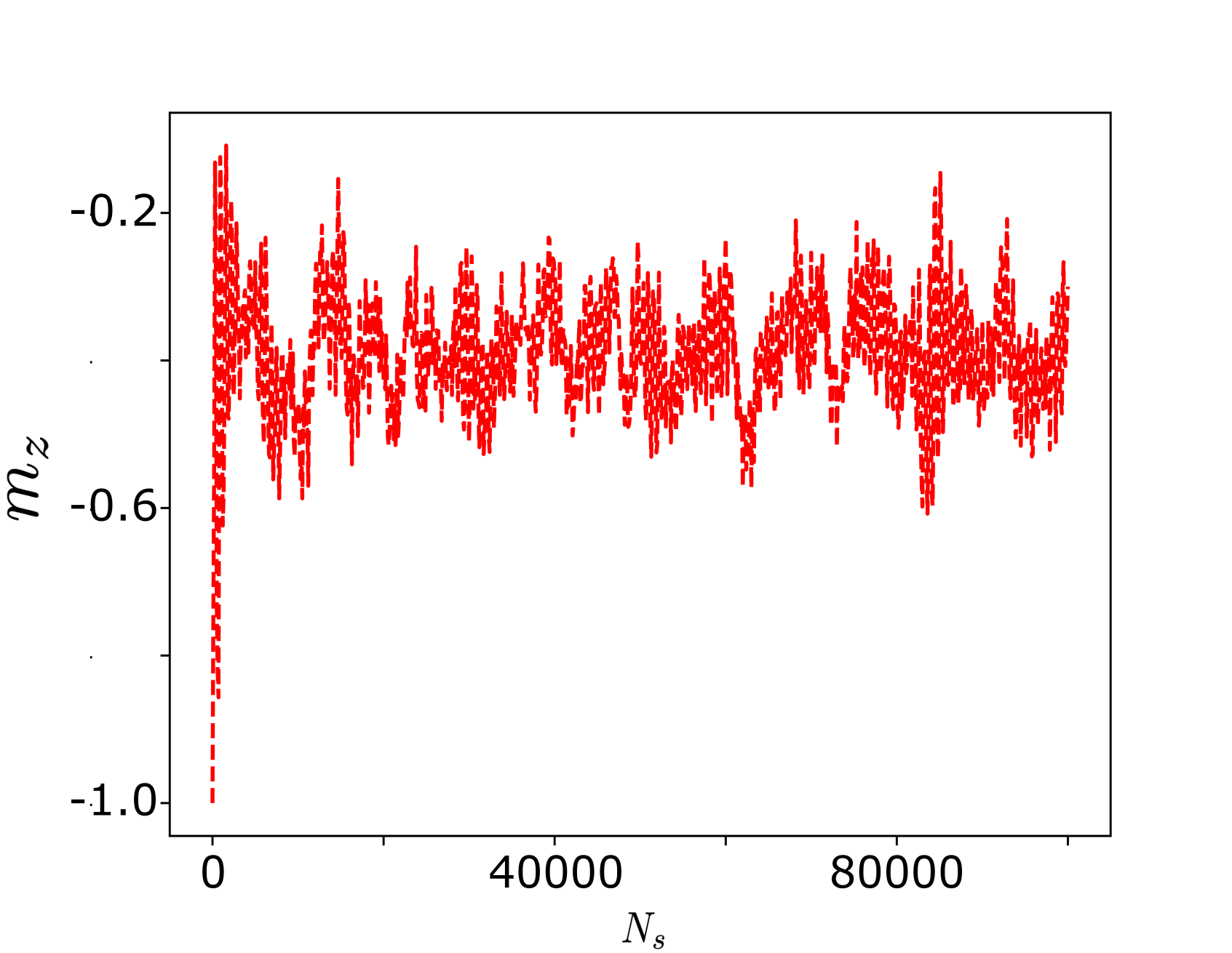}
    \caption{Impurity magnetization expectation values ($m_z = \langle z_0 \rangle$) as a function of time (Floquet steps) for $N = 6$ at the Toulouse point, $\theta = \pi/3,~ \theta_K = \pi/4$ till $T = 10^5$ steps, spaced in intervals of 100 steps.}
    \label{fig:sz-plots2}
\end{figure}

One may wonder whether the asymptotic long-time behavior discussed so far is only transient or corresponds to an asymptotic steady state as would be obtained by first sending system size $N \rightarrow \infty$ and subsequently $N_s \rightarrow \infty$. To numerically address this question,
\figref{fig:sz-plots2} highlights the evolution of the impurity magnetization for $N = 6$ sites at the Toulouse point up until $10^5$ steps. We are not able to observe any new emergent physical phenomena, e.g. new frequencies of oscillation or any change in the long-time magnetization asymptote at such times. 

Given that, however, our simulation occurs at fixed (rather small) $N$, it is imposed by unitarity that the asymptotic steady state behavior is eventually lost and the system reexplores its initial conditions after a time scale called the recurrence time. The latter can be estimated by the inverse many-body level spacing. {In Fig.~\ref{fig:dos-plot} of the Appendix we present the} numerical (ED) {many-body Floquet spectrum, the corresponding density of states appears to contain quasi-continuous bands as well as gaps. The level spacing within the quasi-continuous bands sets} 
an upper bound on the lowest eigenenergy spacing, which is machine precision $10^{-16}$, so we expect recurrence times of at least $10^{16}$ steps and accordingly do not observe any recurrence here as expected.

\subsubsection{(In-)dependence on initial state}
\label{sec:IniState}

\begin{figure*}
    
    \includegraphics[width=2\columnwidth]{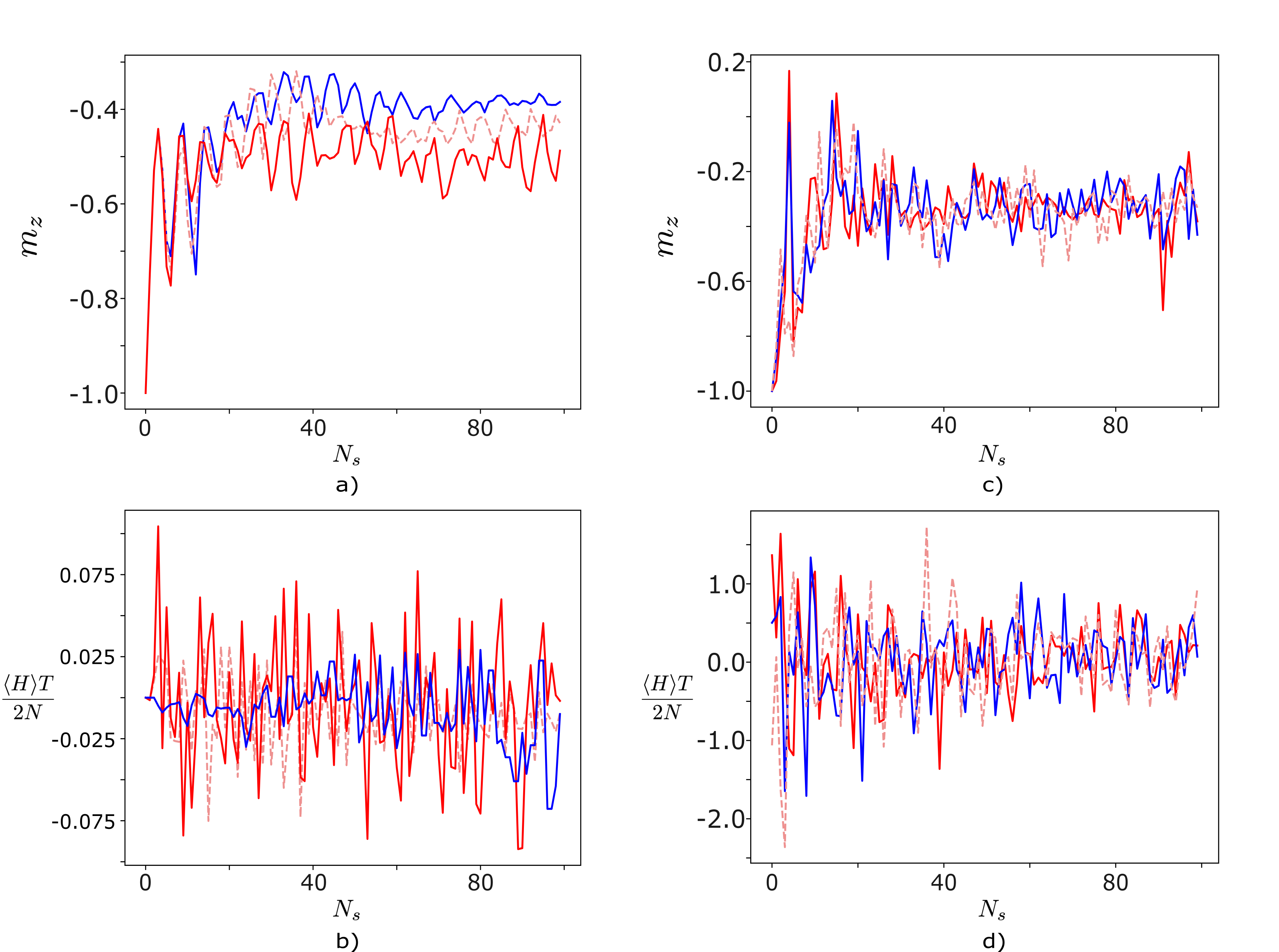}
    \caption{Plots ({as a function of Floquet steps}) at 
    the isotropic point $\theta_z = \theta_K$ for the TS starting state, Eq.~\eqref{eqn:ts-state}, at $\theta = \pi/3$ and $\theta_K = \pi/4$ for $N = 6$ (red, solid), $N = 8$ (light red, dashed) and $N = 10$ (blue, solid). a)  Impurity magnetization $m_z = \langle z_0 \rangle$. b) Hamiltonian expectation ($\langle H \rangle$). Plots at the isotropic point $\theta_z = \theta_K$ for randomized tensor product (RTP) (Eq.~\eqref{eqn:rtp-state}) states at $N = 6$, $\theta = \pi/4$ and $\theta_K = \pi/6$ for three randomized states labelled by red solid, light red dashed and blue solid curves. c) Impurity magnetization $m_z = \langle z_0 \rangle$. d)  Hamiltonian expectation ($\langle H \rangle$).}
    \label{fig:ts-plots}
\end{figure*}

    

So far the entire discussion was devoted to quench dynamics for Fermi sea initial states. This section provides evidence that crucial long-time properties of magnetization and entanglement are independent of the starting state. {We note that the previously chosen} system sizes $N = 6$ and $N = 10$ are chosen for $\ket{FS}$ starting states, because they both host non-degenerate Fermi sea states at half filling and periodic boundary conditions, but $N = 8$ is also considered for $\ket{TS}$ and RTP initial states.

Figs.~\ref{fig:ts-plots} {a) to d)} highlight the evolution of some observables with the starting states $\ket{TS}$ (see \eqnref{eqn:ts-state}) and RTP (see \eqnref{eqn:rtp-state}) respectively. It is observed that while the impurity magnetization and concurrence measure as a function of Floquet steps is very similar to the one observed when the $\ket{FS}$ state was used, average energy  
curves differ. Particularly, with the $\ket{TS}$ state, $\langle H \rangle$ starts at zero and subsequently seems to fluctuate randomly about 0, and has no fixed visible trend. It also varies on a much smaller scale as compared to using $\ket{FS}$ as a starting state.\\

Also for the RTP state, the impurity magnetization 
seems to follow the 
same trend, leading to a not insignificant overlap between the curves (save from some outlier peaks by one of the curves) with different randomized starting states, cf. Fig.~\ref{fig:ts-plots} c). We also compared (not shown) these curves with an average over 100 configurations of the RTP starting state, where each configuration consisted of $2N + 1$ sets of \{$\theta_x, \phi_x$\} for each qubit chosen randomly over uniform distributions as defined in Eq.~\eqref{eqn:rtp-state}. These average curves where comparable to the curves with $\ket{TS}$ starting state.

The trend for the three curves itself is close to the $\ket{TS}$ initial state configuration with the same values of {$\theta,\theta_K$ and $\theta_z$}, where we can observe the magnetization oscillating about an asymptotic value. The internal energy
curves for the RTP state is also very similar to that of the $\ket{TS}$ state, with the values fluctuating randomly about 0. If the $\ket{TS}$ and RTP states are taken as some excited states, with $\ket{FS}$ as the ground state, then it makes sense as to why the trend of the energy curve is the way it is.

\section{Analytical results near the thermodynamic limit}
\label{sec:Analytics}

Here we present the analytical calculations for the impurity magnetization as a function of time {at a specific point in parameter space, the \textit{Toulouse point}.} We mostly focus on the quench dynamics assuming discrete time steps (after $N_s$ Floquet steps, each of time $T$) but also compare it to continuum time evolution. {All of our analytical calculations assume a Fermi sea initial state and we represent the problem close to the thermodynamic limit as a Floquet field theory, Sec.~\ref{sec:floqth}.}

{Extending standard steps, as in the literature for Hamiltonian Kondo models~\cite{EmeryKivelson1992}, w}e first transform the Kondo-Floquet unitary 
via bosonization and refermionization to a Floquet unitary for an interacting resonant level model{, Secs.~\ref{sec:bandr}, \ref{eq:EffFloquetUnity}. In the Hamiltonian case, }
the Toulouse point corresponds to the values in parameter space where the {anisotropic} Kondo impurity {Hamiltonian} can be mapped onto a non-interacting resonant level model. {Based on the irrelevance of anisotropies under renormalization group, this special point captures the same physics as the isotropic Kondo problem.}
Contrary to the Hamiltonian situation, interactions do not completely vanish at the Toulouse point in the Floquet case, yet {the residual terms} can be shown to {have} strongly
irrelevant {scaling dimensions}. {We drop} 
these residual interactions, which is expected to be accurate for
the long-time limit.

{Consequently, the next step is to solve the Floquet quench dynamics for a specific initial state, Sec.~\ref{sec:StartingDM}, of a non-interacting resonant level model. To this end, in Sec.~\ref{eq:FloqEvolve}, w}e closely follow a construction introduced in \cite{HeylKehrein2010b} to solve the non-interacting {Floquet impurity} problem. All {of} the calculations are done near the thermodynamic ($N \to \infty$) limit, keeping leading $1/N$ corrections~\cite{ZarandVonDelft2000}. The main results are highlighted here, with supporting calculations presented in \appref{sec:appn4}.

{We briefly summarize the justification and motivation for our approach, see also the discussion in Sec.~\ref{sec:DiscussionAnalytics}.} Effectively, the evaluation of the Floquet evolution of the magnetization is a multi-point correlator with respect to {the initial state, which we assume to be the Fermi sea state throughout this section. We further}
{
linearize
{the free fermion Floquet dispersion} term around the Fermi edge keeping in mind that} 
{
the mapping of the Kondo lattice model to the {Dirac} field theory is believed to be accurate if only states near the Fermi surface are involved, effectively requiring $J_{K,z} \ll t_0$.} {In the regime of its applicability, the} multi-point correlator of 
Dirac fermions in one dimension 
can 
be treated using bosonization (even in the present non-equilibrium, i.e., quench, set-up).
Of course, a direct connection to the lattice model is no longer justified {in the Floquet problem} once heating effects lead to an effective temperature of the order of the free fermion bandwidth on the lattice. 


\subsection{Floquet field theory}
\label{sec:floqth}
We consider an anisotropic Kondo model for a continuum spatial variable $x$, so that the analogue to Eqs.~\eqref{eqn:main-ham} is
\begin{subequations}
\begin{align} \label{eq:lin}
    &\mathcal H_{\rm kin} = v \sum_{k,\sigma} k :c^\dag_{k,\sigma}c_{k,\sigma}:,\\ 
    &\mathcal H_z = \mathcal{I}_z S_z \Psi^\dag(0) \sigma_z \Psi(0), \\ 
    &\mathcal H_\perp = \mathcal{I}_K (S_+ \Psi^\dag(0) \sigma_- \Psi(0) + H.c.).
\end{align}
\label{eq:InitialHams}
\end{subequations}
Here ${\Psi}(x)$ is the fermionic field operator
\begin{align}
    \Psi_\sigma(x) = \sqrt{\frac{2\pi}{L}}\sum_k e^{-ikx}c_{k,\sigma}, 
    \label{eq:Psispinful}
\end{align}
and $v = 2\sin(\theta)a/T$ is the Fermi velocity, where we used Eq.~\eqref{eq:FermiVelocity} to make connection to the lattice model studied numerically, but for convenience physical units of lattice spacing $a$ and Floquet time $T$ are restored here. Note that $\mathcal{I}_z,\mathcal{I}_K$ have units of velocity, i.e. they also have absorbed a factor of $a$ as compared to their counterparts $J_z, J_K$ in Eq.~\eqref{eqn:main-ham}. The fermionic operators $c_{k, {\sigma}}$ are dimensionless.  The spacing of free fermion momenta $k$ and energies is $\Delta_L = 2\pi/L, \Delta_E =  v \Delta_L$, where will later identify $L = Na$. 
In this section we only consider antiperiodic boundary conditions, such that $k \in \Delta_L [\mathbb Z + 1/2]$ and a uniquely defined Fermi sea $\ket{\mathcal{FS}}$, which is a spin singlet and
\begin{subequations}
\begin{align}
   c_{k,\sigma} \ket{\mathcal{FS}} &= 0, \forall k>0, \\
      c_{k,\sigma}^\dagger \ket{\mathcal{FS}} &= 0, \forall k<0.
\end{align}
\end{subequations}
Normal ordering, denoted by $: \dots :$, is defined with respect to the filled Fermi sea by expansion of a given operator in fermionic modes and ordering in $c_{k, \sigma}\vert_{k>0}$ as well $c_{k, \sigma}^\dagger \vert_{k<0}$ to the right of all other operators.
To make connection to the lattice model studied numerically, we followed the standard procedure of mapping the fermions of the Kondo problem onto a chiral field~\cite{Affleck1995} and assumed fermions to live on a ring with circumference $L$, i.e. $x \in [{-L/2, L/2})$ to capture finite size effects~\cite{ZarandVonDelft2000}.

In full analogy to the numerical study, one may want to consider $U_F = U_K U_{\rm kin}$,
\begin{subequations}
\begin{align}
    U_{\rm kin} & = e^{- i \mathcal H_{\rm kin} T/2},\\
    U_K & = e^{- i (\mathcal H_z + \mathcal H_\perp) T/2}.
\end{align}
\end{subequations}

\subsection{Bosonization and Refermionization}
\label{sec:bandr}
We here apply well known steps of bosonization 
and refermionization \cite{EmeryKivelson1992,ZarandVonDelft2000, HeylKehrein2010b} to the present case of the Floquet unitary on a finite ring. The main bosonization identities are 
\begin{subequations}
\begin{align}
    \Psi_\sigma(x) &= \frac{1}{\sqrt{a}}F_\sigma e^{-i\Delta_L x(\hat{N}_\sigma - P_0/2)} e^{-i\phi_\sigma(x)}, \\
    :\Psi_\sigma^\dagger(x) \Psi_\sigma(x): & = \partial_x \phi_\sigma (x) + \Delta_L \hat N_\sigma, 
\end{align}
\end{subequations}
where $F_\sigma$ are Klein factors which act as fermionic ladder operators on eigenstates of $\hat N_\sigma$ with fixed number of electrons of a given spin~\cite{ZarandVonDelft2000}. {Here $\phi_\sigma(x)$ is the chiral bosonic field~\cite{EmeryKivelson1992,Heyl2012} with the following commutation relations in the thermodynamic limit:
\begin{subequations}
\begin{align}
    [\phi_\sigma(x), \partial_{x'}\phi_{\sigma'}(x')] &\to 2\pi i \delta_{\sigma,\sigma'}\left[\frac{a/\pi}{(x-x')^2 + a^2} - \frac{1}{L}\right] \\
    [\phi_\sigma(x), \phi_{\sigma'}(x')] &\to -2i\delta_{\sigma,\sigma'}\arctan\left[\frac{x-x'}{a}\right]
\end{align}
\end{subequations}
where our lattice spacing $a$ also acts as a ultra-violet cutoff in the regularization scheme.}

In the present case of antiperiodic boundary conditions, $P_0 = 1$, but for periodic boundary conditions $P_0 = 0$. 
In bosonic language, the free Hamiltonian becomes
\begin{align}
    \mathcal H_{\rm kin} &= \sum_\sigma \frac{\Delta_E}{2}\hat{N}_\sigma (\hat{N}_\sigma + 1 - P_0) \notag \\ 
    &+ \sum_\sigma \frac{v}{4\pi} \int_{-L/2}^{L/2} dx :(\partial_x\phi_\sigma(x))^2:.
\end{align}
Next we bosonize Kondo terms and convert to spin (s) and charge degrees of freedom $\phi_{c/s} = (\phi_\uparrow \pm \phi_\downarrow)/\sqrt{2}$, $\mathcal N_{c/s} = (N_\uparrow \pm N_\downarrow)/\sqrt{2}$. Our Kondo impurity only couples to the spin mode, so we henceforth drop the charge mode and the subscript $_s$ for spin mode. Then we have the kinetic Hamiltonian as
\begin{subequations}
\begin{align}
    \mathcal H_{\rm kin} =  \frac{\Delta_E}{2}\mathcal N^2 + \frac{v}{4\pi} \int_{-L/2}^{L/2} dx:(\partial_x\phi(x))^2:,
\end{align}
while the Kondo Hamiltonian becomes
\begin{align}
    \mathcal H_z &= \frac{\mathcal{I}_z}{\sqrt{2}} S_z (\partial_x \phi(0)+ \Delta_L \mathcal N), \\ 
    \mathcal H_\perp &= \frac{\mathcal{I}_K}{2a} (S_+F^\dag_{\downarrow} F_{\uparrow}e^{-i\sqrt{2}\phi(0)} + H.c.). 
\end{align}
\end{subequations}
Note that the kinetic term corresponds to a Hamiltonian of bosonic spin modes in anti-periodic boundary conditions, i.e. with a unique ground state, just as before the rotation to charge and spin modes. 

We now perform a standard Emery-Kivelson~\cite{EmeryKivelson1992} transformation using $U = e^{i\gamma S_z \phi(0)}$, with $\gamma = \sqrt{2} - 1$ under which we have the transformed Hamiltonians as
\begin{subequations}
\begin{align}
    \mathcal H_{\rm kin} \rightarrow \tilde {\mathcal H}_{\rm kin} &\equiv U \mathcal H_{\rm kin} U^\dag 
     = [{\mathcal H}_{\rm kin} + \frac{\Delta_E}{2} \mathcal N] \notag \\ 
     &- \gamma v [\partial_x \phi (0) + \Delta_L \mathcal N] S_z + \Delta_E \mathcal N [\gamma S_z - \frac{1}{2}] , \label{eq:HkinEff}\\
    \mathcal H_z \rightarrow \tilde {\mathcal H}_{z} &\equiv U \mathcal H_{z} U^\dag = \frac{\mathcal{I}_z}{\sqrt{2}} S_z (\partial_x \phi(0) + \Delta_L \mathcal N), \\
    \mathcal H_\perp \rightarrow \tilde {\mathcal H}_{\perp} &\equiv U \mathcal H_{\perp} U^\dag  = \frac{\mathcal{I}_K}{2a} (S_+F^\dag_{\downarrow} F_{\uparrow}e^{-i \phi(0)} + H.c.).
\end{align}
\end{subequations}
Constant shifts of the Hamiltonian have been dropped throughout. Here and in the following, the tilda denotes operators in the rotated basis. 

Note that the first square brackets in Eq.~\eqref{eq:HkinEff} correspond to Dirac fermions on a ring with periodic boundary conditions, while the last term of the same equation can be simplified using that the total magnetization along z-axis of the system is conserved, and that $\mathcal N/\sqrt{2} = (N_{\uparrow} - N_{\downarrow})/2$ is the magnetization of the fermionic bath 
\begin{subequations}
\begin{align}
    \mathcal N/\sqrt{2} + S_z &= S_z^{tot} = -1/2,\\ 
    \Rightarrow  \gamma S_z \mathcal N- \mathcal N/2 &= \gamma S_z.
\end{align}
\end{subequations}
For refermionization, we use a transformation with a second unitary $U_{f} = e^{i\pi \frac{\mathcal N}{\sqrt{2}}S_z}$ to ensure the correct statistics such that $S_+ \rightarrow U_{f} S_+ {U_{f}}^\dagger = e^{i \pi \mathcal N/\sqrt{2}} S_+$. We now use the bosonization identity, along with the observation that $F^\dag_{\downarrow} F_{\uparrow}$ acts as a ladder operator (Klein factor) for $\mathcal N$, to introduce new spinless pseudofermions ${\tilde{ \Psi}}(x)$ and the corresponding modes as
\begin{subequations}
\begin{align}
  {\tilde{ \Psi}}(x) &= \frac{1}{\sqrt{a}}F^\dag_{\downarrow} F_{\uparrow}e^{-i\Delta_L x (\mathcal N - P_0/2)}e^{-i\phi(x)}, \\
 c_k &= \frac{1}{\sqrt{2\pi L}} \int_{-L/2}^{L/2} dx e^{ikx} \Psi(x),
\end{align}
\end{subequations}
where $P_0 = 0$, i.e.~periodic boundary conditions, such that $k \in \Delta_L  \mathbb{Z}$. In particular, $k=0$ is now allowed, contrary to Eq.~\eqref{eq:Psispinful}. 
We further use a fermionic representation of the spin operators and absorb a phase $i$ 
\begin{equation}
d = e^{-i\pi \mathcal N}S_-,
\end{equation}
to get $S_z = d^\dag d - 1/2$.
With all this, we have the following refermionized operators 
\begin{subequations}
\label{eqn:final-hams}
\begin{align} 
    \tilde{\mathcal H}_{\rm kin} &= v \sum_k k :c^\dag_k c_k : - v  \gamma [d^\dagger d - 1/2] : {\tilde{ \Psi}}^\dagger(0) {\tilde{ \Psi}}(0) :  \notag \\ & 
    - \mu d^\dagger d, \\ 
    \tilde{\mathcal H}_z &= \frac{\mathcal{I}_z}{\sqrt{2}} [d^\dagger d - 1/2] : {\tilde{ \Psi}}^\dagger(0) {\tilde{ \Psi}}(0) : \\
    \tilde{\mathcal H}_\perp &=  \frac{\mathcal{I}_K}{2 \sqrt{a}} (d^\dag {\tilde{ \Psi}}(0) + H.c.),
\end{align}
\end{subequations}
where $\mu = -\Delta_E \gamma$ is a finite size induced chemical potential for $d$ fermions and we have made use of the fact that $\partial_x\phi(0) +\Delta_L \mathcal N= \colon{\tilde{ \Psi}}^\dag(0){\tilde{ \Psi}}(0)\colon$.

Several comments are in order: First, we emphasize that all the steps leading from Eq.~\eqref{eq:InitialHams} to \eqref{eqn:final-hams} are just a basis transformation and a rewriting. But of course, these steps are designed in a way, that the new degrees of freedom are close to the collective excitations of the many-body system.

Second, we recall that the Fermi sea which is the ground state of $\mathcal H_{\rm kin}$ is non-degenerate, while the impurity spin degree added a groundstate degeneracy of two (we initialize the system in one of these two states). In contrast, the refermionized Hamiltonian $\tilde{\mathcal H}_{\rm kin}$ has a doubly degenerate ground state, depending on whether the $k=0$  mode $c_0$ is occupied or not. At the same time the  
$\mu$ term acts like an external field which, for $\gamma > 0$, tends to polarize $S_z$ to $S_z = -1/2$. So the degeneracy has been transferred from impurity to bath electrons, but, of course, remains. On the side, we remind the reader that the Kondo fixed point can be interpreted as a state with newly emergent, altered boundary conditions, as encoded in a universal phase shift. 

Third, the initial state corresponding to total magnetization $-1/2$ is the one with $c_0$ mode occupied, $c_0^\dagger c_0 = 1$ (in addition to $d^\dagger d = 0$).

Finally, we employ a point-splitting regularization, such that $\colon{\tilde{ \Psi}}^\dag(0) {\tilde{ \Psi}}(0)\colon = \Delta_L (\sum_{kk'}c^\dag_k c_{k'} - L/2a)$. A less field theoretical regularization scheme {with the same effect} is to simply truncate the momentum spectrum keeping only a finite odd integer $L/a$ states and occupying the first $L/2a + 1/2$ of them. In this case
\begin{subequations}
\begin{align}
{\tilde{ \Psi}}(x) \rightarrow C_x &= \frac{1}{\sqrt{L/a}} \sum_k e^{i k x} c_k, \\
c_k &= \frac{1}{\sqrt{L/a}} \sum_x e^{-i k x} C_x,
\end{align}
\end{subequations}
and $C_x$ fulfill standard fermionic exchange statistics.

\subsection{Effective Floquet unitary}
\label{eq:EffFloquetUnity}

It is convenient to express the terms entering the exponential in the Floquet unitaries as
\begin{subequations}
\begin{align}
\tilde{\mathcal H}_0 & = \sum_k v k : c^\dagger_k c_k:, \\
\tilde{\mathcal H}_z & = \frac{\mathcal{I}_z \sqrt{2} \pi}{a}(d^\dagger d - 1/2)(C_0^\dagger C_0 - 1/2), \\ 
\tilde{\mathcal H}_\perp & = V[d^\dagger C_0 + H.c.],\\
\delta \tilde{\mathcal H}_L & = -\mu (d^\dagger d - 1/2),
\label{eq:ChemPot}
\end{align}
where $V = {\mathcal{I}_K \sqrt{(\pi/2)}}/{a}$ and the last term, {Eq.~\eqref{eq:ChemPot},} is $\mathcal O(1/L)$, while all others are finite in the thermodynamic limit. It is {seen} that
$\delta \tilde {\mathcal H}_L$ commutes with $\tilde{\mathcal H}_0$ and $  \tilde{\mathcal H}_z$, and $\tilde{\mathcal H}_z$ additionally also commutes with $\tilde{\mathcal H}_\perp$. 
The relevant unitaries at the Toulouse point ($\mathcal{I}_z = \sqrt{2} v \gamma$) have the form
\begin{align}
\tilde U_{\rm kin} & = \text{exp}(- i T [\tilde{\mathcal H}_0 - \tilde{\mathcal H}_z - \delta \tilde{\mathcal H}_L]/2) \notag \\
& = \underbrace{\text{exp}(- i T [\tilde{\mathcal H}_0 - \tilde{\mathcal H}_z]/2)}_{\equiv \tilde U_{\rm kin,0}} \underbrace{\text{exp}( i T \delta \tilde{\mathcal H}_L/2)}_{\equiv \tilde U_L}, \\
\tilde U_{\rm K} & = \text{exp}(- i T [\tilde{\mathcal H}_z + \tilde{\mathcal H}_\perp]/2) \notag \\
& =  \underbrace{\text{exp}(- i T \tilde{\mathcal H}_\perp/2)}_{\equiv \tilde U_{K,0}} \underbrace{\text{exp}(- i T \tilde{\mathcal H}_z/2)}_{\equiv \tilde U_z}.
\end{align}
\label{eq:EffProblem}
\end{subequations}
The evolution to the time $N_sT$  {is then} 
\begin{align}
\tilde U(N_sT) & \equiv U_{f} U (\tilde U_K \tilde U_{\rm kin} )^{N_s} [U_{f} U]^\dagger  \notag\\
& = [ \tilde U_{K,0} \tilde U_z \tilde U_L^{1/2} \tilde U_{\rm kin,0} \tilde U_L^{1/2}]^{N_s} \notag \\
&= \tilde U_L^{-1/2} [\underbrace{U_{2}U_{1}}_{\equiv \tilde{U}_F} ]^{N_s} \tilde U_L^{1/2}.
\end{align}
Here we introduced the Floquet unitary $\tilde U_F$
split into a half sequence which is pure kinetic energy $U_1$ and a second term $U_2$ which only mixes $d$ and $C_0$, i.e. 
\begin{subequations}
\begin{align}
\label{eq:rgscaled}
U_{1} & = \tilde U_z \tilde U_{\rm kin,0}  \simeq e^{- i \sum_{k} \bar v k :c^\dagger_k c_k:}, \\
U_{2} & = (\tilde U_L^{1/2} \tilde U_{K,0} \tilde U_L^{1/2}) \notag  \\
&\simeq \exp\left \{- i (d^\dagger C_0^\dagger) \left ( \begin{array}{cc}
    \frac{-\bar \mu - \Delta \bar \mu}{2} & \bar V  \\
    \bar V & \frac{-\bar \mu + \Delta \bar \mu}{2}
\end{array}\right ) \left (\begin{array}{c}
     d  \\
     C_0 
\end{array} \right) \right \}.\label{eq:CBHUnitaries}
\end{align}
\label{eq:FloquetCycle12}
\end{subequations}
We generally use a bar to denote the dimensionful quantities multiplied by half a Floquet time, e.g. $\bar v = v T/2$ and emphasize that, contrary to the case of Hamiltonian time evolution, the Toulouse point does not imply perfect cancellation of the interaction term $\tilde{\mathcal H}_z$ in the resonant level model. Instead, Campbell-Baker-Hausdorff multiplication of $U_{\rm kin,0}$ and $U_z$ leads to an exponential which contains the quoted $ \sum_k {\bar v} k : c^\dagger_k c_k :$, but also local interaction terms multiplied by momenta $k$ {(i.e. with additional derivatives)}. Such terms
{have} however {a} highly 
irrelevant {scaling dimension}
and {are} thus expected to be unimportant for the long-time evolution. We thus drop these terms at the $\simeq$ sign.

In the equation for $U_2$ the same sign indicates that 
terms $\mathcal O(1/L^2)$ are dropped, see App.~\ref{sec:appn4-2}. We have the quantities:
\begin{subequations}
\begin{align}
    \bar V & \equiv VT/2 = \frac{\mathcal{I}_KT \sqrt{\pi}}{\sqrt{8} a}, \\
    \bar \mu & = \mu T/2 = - \gamma \bar \Delta_E ,\\
    \Delta \bar \mu & = \bar \mu \bar V \cot(\bar V) .
\end{align}    
\end{subequations}
In this limit the Kondo problem becomes a matchgate (i.e. free fermion) circuit.

\subsection{Starting density matrix}
\label{sec:StartingDM}

The density matrix at time $0^-$ is the filled Fermi sea of scattering states in the presence of a potential scatterer. We see this from \eqnref{eqn:final-hams}, where we set {$d^\dag d$} in $\tilde{\mathcal H}_{\rm kin}$ as {$0$}, and all the constant terms drop out. We then have
\begin{align}
\label{eqn:dmH}
   \mathcal H = \sum_k vk :c_k^\dagger c_k : + V_0 \sum_{k,k'} c^\dagger_k c_{k'},
\end{align}
with $V_0 = \Delta_E \gamma/2$, and $v$ as the Fermi velocity.

\subsubsection{Exact treatment}

We find that the single-particle density matrix $\rho_{kk'}$ is
\begin{align}
\label{eqn:dmfin}
    \rho_{kk'} \equiv \langle c^\dagger_k c_{k'}  \rangle &= \sin^2\left (\frac{\pi \delta}{\Delta_E} \right)  \frac{\psi^{(0)}(\frac{vk - \delta}{\Delta_E}) - \psi^{(0)}(\frac{vk' - \delta}{\Delta_E} )}{\pi^2\frac{vk - vk'}{\Delta_E}}.
\end{align}
Here, $\psi^{(0)}$ is the digamma function and the quantities introduced here are:
\begin{subequations}
\begin{align}
\bar V_0 & = \pi V_0/\Delta_E = \gamma \pi/2,\\
\delta & = \frac{\Delta_E}{\pi} \arccot\left ( \frac{1}{\bar V_0} \right) \approx 0.18 \Delta_E.
\end{align}    
\end{subequations}

The details of this derivation are presented in \appref{sec:appn4-1-1}, where we use the Matsubara technique to find the Green's functions.

\subsubsection{Approximations in the density matrix}
Now, we introduce a useful approximation of $\delta/\Delta_E \ll 1$, which gives us (keeping only leading order terms, see \appref{sec:appn4-1-2}):
\begin{align}
\label{eqn:dmapprox}
    \rho_{kk'} = \theta(-k) \delta_{kk'} -\delta \frac{\theta(-k k')}{\vert vk - vk' \vert } + \mathcal O(\delta^2/\Delta_E^2).
\end{align}
Here we notice that this expression is valid for $k,k' \neq 0$. For $k = k' = 0$, we have $\rho_{kk'} = 1$ as $c^\dagger_0 c_0 =1$. For $k(k') = 0$ and $k'(k) \neq 0$, we take $\theta(0) = 1$. Note that $\rho_{kk'}$ is a symmetric matrix. We write
$\rho_{kk'} = \rho^{(s)}_{kk'} + \rho^{(a)}_{kk'}$, where 
%
\begin{subequations}
\begin{align}
    \label{eqn:rho-sym}
    \rho_{kk'}^{(s)} & = \underbrace{\frac{\delta_{kk'}}{2}}_{\rho_{kk'}^{(s,0)}} -\underbrace{\frac{\delta \theta(-k k')}{\vert vk - vk' \vert }}_{\rho_{kk'}^{(s,1)}} + \mathcal O(\delta^2/\Delta_E^2),  \\
    \rho_{kk'}^{(a)} & = - \text{sign}(k)\frac{\delta_{kk'}}{2} + \mathcal O(\delta^2/\Delta_E^2),
\end{align}
\end{subequations}
{where $\text{sign}(0) = -1$}.

\subsection{Evolution under Floquet cycles}
\label{eq:FloqEvolve}

\subsubsection{Single cycle dynamics}
For  
our Floquet cycle 
$U_F = U_2 U_1$, Eq.~\eqref{eq:FloquetCycle12},
we follow the technique which Heyl and Kehrein \cite{HeylKehrein2010b} used in a slightly different Kondo-Floquet problem. We introduce fermionic operators $a_l$ with $a_l = d$ for $l = d$ and $a_l = c_k$ for $l = k$ and compactly write a formula for the continuous time evolution within a given stretch
\begin{equation}
    a_l(t) \equiv U^\dagger(t) a_l U(t) =  G^R_{ll'}(t) a_{l'},
\end{equation}
where, 
\begin{align}
    G^R_{ll'}(t) & = \theta(t) \langle \lbrace a_l (t), a_{l'}^\dagger \rbrace \rangle.
\end{align} 
Within one Floquet cycle this corresponds to 
\begin{subequations}
 \begin{align}
    a_l & = \mathcal M_{ll'} a_{l'}, \\
    \mathcal M_{ll'} & = G_{ll''}^{R,2}(T/2) G_{l''l'}^{R,1}(T/2), \label{eq:Mdef}
\end{align}   
\end{subequations}
where $G_{ll''}^{R,i}$ is the retarded Green's function corresponding to the unitary $U_i$, {cf. Eq.~\eqref{eq:FloquetCycle12},} which are given in \appref{sec:appn4-3-1}. {Regarding the regularization of the retarded Green's function ("$+i 0$" in the denominator of the Green's function) we highlight the following mathematical detail of important physical consequences: Throughout, the regulator $+ i 0$ is assumed to have a larger absolute value than the finite size level space $\Delta_E$ allowing to replace momentum sums by integrals.}

\subsubsection{Iterative solution for $N_s$ Floquet cycles}

It is obvious that 
\begin{align}
    a_l(N_s T) = [\mathcal M^{N_s}]_{ll'} a_{l'}.
\end{align}
To calculate the magnetization we specifically need $\mathcal M^{N_s}_{dk}$. Since we assume $\langle d^\dagger d \rangle (N_s = 0) = 0$, we have 
\begin{equation}
    m_z (N_s T) = 2[\mathcal M^{N_s}]^*_{dk} [\mathcal M^{N_s}]_{dk'} \underbrace{\langle c_k^\dagger c_{k'} \rangle (N_s = 0)}_{\rho_{kk'}} - 1. \label{eq:mzMainText}
\end{equation}
We find that (cf. \cite{HeylKehrein2010b} and \appref{sec:appn4-3-2}):
\begin{align} \label{eq:MdkMainMain}
    [\mathcal M^{N_s}]_{dk} = \frac{\mathcal M_{dd}^{N_s} - e^{- i N_s \bar v k}}{\mathcal M_{dd} - e^{- i \bar v k}} \mathcal M_{dk}.
\end{align}
This result, Eq.~\eqref{eq:MdkMainMain}, has a simple physical interpretation. It describes the transition amplitude from starting state $k$ to final state $d$ after $N_s$ steps. A crucial equation to derive the result within the formalism~\cite{HeylKehrein2010b} is $(\lambda >0$)
    \begin{equation} \label{eq:MdkMain}
        \sum_k e^{- i \lambda k}\mathcal M_{kd} \mathcal M_{dk} = 0,
    \end{equation}
due to the analytical structure of retarded Green's functions. It means that a particle that has jumped to the impurity orbital $d$ is stuck at $d$ for eternity. Thus the transition amplitude is just the sum over all processes where at step ${\ell}$ the particle makes the transition. As long as the particle sits at $k$ it will accumulate phase $e^{-i \bar v k}$ at each time step $1, \dots, {\ell}$. To make the transition it needs $\mathcal M_{dk}$. Once it reached the final state it will 
slowly incoherently decay with rate $- \ln[\mathcal M_{dd}]$ for the remaining $N_s - 1 - {\ell}$ steps. Summing over all possible ${\ell}$ leads to Eq.~\eqref{eq:MdkMainMain}. 

Inserting this matrix structure into the definition of $m_z$ we thus get:
\begin{align}
    m_z(N_s T)  = &- \cos^{2N_s}(\bar V) + \frac{ \bar \mu a}{\pi \bar v  \alpha} [1- 2  \cos^{N_s}(\bar V) \cos(N_s \bar \mu)\notag\\ 
    &+ {\cos^{2N_s}(\bar V)}], \label{eq:AnalyticsDisc}
\end{align}
where $\alpha \simeq [{1 + \cos(\bar V)}]/[{1 - \cos(\bar V)}]$. 

This can be compared to the continuum time evolution where we have analogous expressions with the replacement $[\mathcal M^{N_s}]_{dk} \rightarrow \mathcal G^R_{dk}$. Leaving details to \appref{sec:appn4-4} we find
\begin{align}
    m_z(t) \simeq -e^{-2 \Delta t} + \frac{4}{\pi} \left [\frac{{\mu}}{\Delta} - \sin({\mu} t) \frac{e^{- \Delta t}}{\Delta t}\right ]. \label{eq:AnalyticsCont}
\end{align}
Parts of this result have been reported before~\cite{NordlanderLangreth1999,LobaskinKehrein2005,HeylKehrein2010b}.
Here, $\Delta = \pi \rho V^2$, $\rho = a/(2\pi v)$, i.e. the density of states (DOS). Both continuum and discrete results are valid to leading order in $1/L$.

\begin{figure}
    
    \includegraphics[width=\columnwidth]{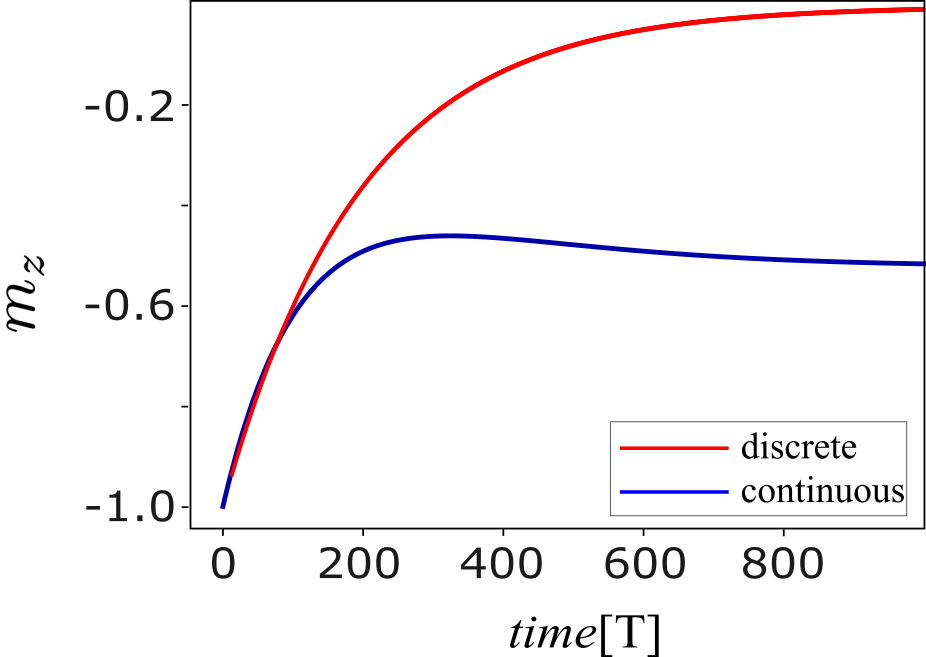}
    \caption{Comparison of discrete and continuum time evolution from the analytical solution Eq.~\eqref{eq:AnalyticsDisc},\eqref{eq:AnalyticsCont}. Here, $\rho = T/5, V = 1/7T, \mu = -5/(3000 T)$. We use the units of Floquet time $T$ even for continuous time $t$. }
    \label{fig:Analytics}
\end{figure}

\subsection{Discussion of analytical solution}
\label{sec:DiscussionAnalytics}

We here discuss physical implications of the analytical continuum, Eq.~\eqref{eq:AnalyticsCont}, and discrete time, Eq.~\eqref{eq:AnalyticsDisc}, evolutions and comment on their comparison to numerics, see also Fig.~\ref{fig:Analytics}. {We note that due to {a} conjectured of universality ({which is well-known for the Hamiltonian} Kondo {model}
), we expect the same physics to be followed for the isotropic point as well, even if we have done all our calculations for the Toulouse point. We remind the reader of the following approximations used: (i) We assume the broadening of bath states 
is greater than finite size momentum spacing $v/L$, without which Kondo effect does not occur,{ see App.~\ref{sec:appn4-3-3} for details.} (ii) We drop the residual interaction term  in Eq.~\eqref{eq:rgscaled} which has highly irrelevant scaling dimensions, unimportant for long-time evolution.}

{First, we remark that only the $\delta = 0$ contribution of the starting density matrix Eq.~\eqref{eqn:dmapprox} enters the magnetization dynamics Eq.~\eqref{eq:AnalyticsDisc}, \eqref{eq:AnalyticsCont}. We show this explicitly for the discrete case in App.~\ref{sec:appn4-4} (for the continuum case we only check that $\delta$ terms do not add qualitatively new features). As a corollary, this corroborates the independence of the long-time magnetization dynamics on the initial state.}

{Second,} we {highlight}  
that both {discrete and continuum magnetization dynamics} starts at $m_z = {-1}$ and have a decay towards a long-time asymptote, along with damped oscillations. The decay rate is 
\begin{subequations}
\begin{align}
t_d^{-1} &= 2\Delta \sim V^2 \rho, \quad \text{continuum evolution,} \label{eq:tdCont}\\
t_d^{-1} &= 2\ln(\sec(\bar V))
/T \stackrel{\bar V \ll 1}{\sim} V^2 T, \quad \text{Floquet evolution.} \label{eq:tdDisc}
\end{align}
\end{subequations}
For the latter, we find {good} agreement with the numerics, especially at higher $\theta_K$ values in \figref{fig:prop-plots} b) when the decay time (in units of Floquet times) is parametrically smaller than the system size (in units of the lattice constant). Note the subtle difference in the Trotter limit $\bar V \equiv VT \rightarrow 0$, where the decay time for the discrete time evolution, Eq.~\eqref{eq:tdDisc} does not recover the continuum time result, Eq.~\eqref{eq:tdCont}! Technically, this is related to the identity Eq.~\eqref{eq:MdkMain}, previously derived in~\cite{HeylKehrein2010b} on the basis of the analytical structure of retarded Green's functions. Physically it implies that, within the present alternating Floquet sequence $U_{K} U_{\rm kin}$, a fermion transferred from bath to the impurity site will stay there forever (without hopping back to the bath) and thus no level broadening $\Delta$ can be generated self-consistently by resumming T-matrix diagrams. However, this is only valid so long as the Floquet time $T$ exceeds the time scale of dissipation. Thus, the order of limits of Floquet time to zero (``Trotter limit") and dissipation to zero (as encoded in ``$+i 0$" in the denominator of the retarded Green's functions) do not commute, explaining the different results. Similarly, the value of the asymptote is different for continuum ($\sim \mu/\Delta$) and discrete cases ($\sim \mu \Delta T^2$ in the Trotter limit), {as clearly illustrated in Fig.~\ref{fig:Analytics}}. The frequency of the oscillations 
are proportional to $\mu$ in the continuum and discrete cases which is independent of $\theta_K$ and only depends on $\theta$. {Our analytical solution suggest the following origin of oscillations: 
while the impurity spin keeps flipping between $\ket{\uparrow}$ and $\ket{\downarrow}$ due to interaction with the bath, the total magnetization is fixed to $-\hbar/2$, leading to an effective Zeeman field $\sim \hbar/N$ which creates an energy split between the $\ket{\uparrow}$ and $\ket{\downarrow}$ states, 
{and ultimately} oscillations.}

Qualitatively, the asymptotic value trend matches the numerics, yet quantitative agreement is not observed due to too small system sizes in the numerics. Still, comparing the two numerical system sizes suggests that the asymptotic value, as well as oscillation {magnitude}, vanish as $L \to \infty$, as predicted by the analytical solution at the Toulouse point.  For frequencies, we see that the analytical value remains constant and numerical values increase at higher $\theta_K$ values, see lower panel of \figref{fig:thetak-1}. The numerical trend is most likely due to finite size effects. The oscillations in the continuum case are damped as $e^{-\Delta t}/(\Delta t)$ while they are damped as $\cos^{N_s}(\bar V)$ in the discrete time evolution. However, overdamping of oscillations is not seen in the numerics, again potentially a finite size effect {
{cf.} Figs.~\ref{fig:sz-plots1} a), b).} 
{Oscillation frequencies in the numerics possibly also deviate from analytics due to dropping irrelevant residual interaction terms in the analytical solution (not justified for small systems)
as highlighted in the first paragraph of this subsection.}

\section{Discussion and Outlook \label{sec:Discussion}}
In summary, this study showcases the 
emulation of the quench dynamics of a periodically driven Kondo model using quantum circuits, supported by both numerical simulation and analytical approaches at the Toulouse point. Starting from a polarized spin state, the impurity magnetization exponentially relaxes to much smaller {absolute} values and subsequently exhibits oscillations about a much smaller long-time asymptote. Analytical calculations perturbing about the thermodynamic limit reveal that the asymptotic magnetization and the frequency of oscillations about it scale inversely with system size ($\sim -1/N$), that the decay rate is proportional to $\ln({\sec}(\theta_K))$  in units of the Floquet period. These analytical trends align {qualitatively} well with numerical simulations at small $N$ (see Fig.~\ref{fig:prop-plots}). A minor difference is the analytical expectation of exponentially decaying oscillatory behavior, {while the oscillations appear} to be persistent in numerics. This can be attributed to finite size effects by comparing $N = 6$ with $N = 10$ data.
The long-time behavior is independent of the three types of initial states we studied, and corresponds to finite energy states of the continuum-time Kondo Hamiltonian. This is conceptually interesting, because it implies that aspects {of} Kondo physics are observable despite heating effects, and it is a practically important conclusion, as the state preparation of a random product state is much more amenable on real quantum devices than the state preparation of the Fermi sea state.

Entanglement measures, including concurrence and von Neumann entropy, mirror the trends observed in magnetization dynamics, confirming their dependence on impurity spin fluctuations. 
These results at small $N$ also qualitatively delineate the spatial extent of the Kondo cloud, 
where entanglement peaks at intermediate distances and diminishes beyond the effective screening length.

This work is designed to open up the study of Floquet dynamics of correlated fermionic models, in particular impurity models, on quantum chips of the NISQ era. A first relevant future step would be an actual simulation on a presently available platform. Our numerical simulations demonstrate that 13 noiseless qubits are able to catch the most salient physical properties predicted by the analytical solution near the thermodynamic limit. Future work can leverage the analytical framework and numerical methods presented here to explore experimental implementations of Kondo physics on quantum hardware or study extensions like multi-impurity or multi-channel systems {as an avenue for studying emergent correlated states. A practically important question is a careful analysis of noise.} {Some more open questions remain, particularly regarding the impact of periodic driving on Kondo screening, entanglement growth, and dynamical phase transitions. Additionally, the breakdown of Kondo screening under strong driving, the possibility of Floquet-induced exotic phases, and the role of impurity dynamics in topological and 2D materials remain largely unexplored. 
Quantum computing platforms offer a promising route to address these challenges, providing a scalable and controllable framework for simulating impurity models beyond classical computational limits}. {The unique possibilities of qubit-by-qubit readout and quantum state tomography promise an amount of experimental information about the above mentioned properties of the many-body state which by far exceed the limitations of any solid state platform.} Refining circuit designs and adopting advanced Trotterization techniques~\cite{BertrandAyral2024} will improve accuracy and scalability, enabling deeper insights into non-equilibrium dynamics in quantum impurity systems.

\section{Acknowledgments}
It is a pleasure to thank Thomas Ayral, Po-Yao Chang, Thomas Sch\"afer, Maxim Vavilov and Angkun Wu for useful discussions about this project. JIV and SS thank the Max Planck Institute for Solid State Research for hospitality. SS thanks the German Academic Exchange Service (DAAD) for support via the Working Internships in Science and Engineering (WISE) program. Support for this research was provided by the Office of the Vice Chancellor for Research and Graduate Education at the University of Wisconsin–Madison with funding from the Wisconsin Alumni Research Foundation. This research was supported in part by grant NSF PHY-2309135 to the Kavli Institute for Theoretical Physics (KITP). EJK acknowledges hospitality by the KITP. {All codes and data used in this paper can be found in the public Github profile \url{https://github.com/lo568los/DAAD_project}.}

\appendix

\section{Free Fermion model}
\label{sec:appn1}

This appendix elaborates on the free fermion hopping part of the Floquet  
Hamiltonian and describes its analytical solution, Sec.~\ref{sec:Matchgate}, as well as the numerical procedure to implement a filled Fermi state as a starting state, Sec.~\ref{sec:FermiSea}.

\subsection{Solution of free fermion circuit}

\label{sec:Matchgate}

The results presented in this appendix are based on a circuit with only the hopping 2-qubit $fsim(\theta,0,0)$ gates and with periodic boundary conditions and its exact solution.

The first step is to divide the lattice into unit cells (labelled by $n'$, different from $n$ in the main text) comprising of two sites. Also notice that $n' = n/2$ for even $n$. In a single unit cell, we label the sites as $A$ (left) and $B$ (right). We label the second quantized operators as {${\Psi}_A(n')$ and ${\Psi}_B(n')$ acting on the $A$ and $B$ sites of a unit cell $n'$}. Fourier transform into the momentum space gives us
\begin{subequations}
\begin{align}
\label{eqn:ft}
    c_A(p) &= \frac{1}{2N} \sum_{n'} e^{-ipn'}{\Psi}_A(n'),\\ 
    c_B(p) &= \frac{1}{2N} \sum_{n'} e^{-ip(n' + 1/2)}{\Psi}_B(n').
\end{align}
\end{subequations}
We take $C_p = (c_A(p),c_B(p))^T$ as a two-dimensional vector. Applying \eqnref{eqn:ft} to Eq.~\eqref{eqn:final-hop}
we have:
\begin{align}
\label{eqn:transformed}
    &U_{\rm kin} = \prod_p e^{i\frac{\theta}{2} C_p^\dag T_1(p) C_p}e^{i\frac{\theta}{2} C_p^\dag T_2(p) C_p},\\
    &T_1(p) = \begin{pmatrix}
        0 & e^{ip/2} \\
        e^{-ip/2} & 0 \\
    \end{pmatrix}, \quad 
    T_2(p) = T_1(p)^{*}.
\end{align}
We next use that, for any two $2 \times 2$ matrices, $M_1$ and $M_2$:
\begin{align*}
    [C_p^\dag M_1 C_p, C_p^\dag M_2 C_p] = C_p^\dag [M_1,M_2]C_p.
\end{align*}
We thus have
\begin{align}
   &e^{i\frac{\theta}{2} C_p^\dag T_1(p) C_p}e^{i\frac{\theta}{2} C_p^\dag T_2(p) C_p} = e^{iC_p^\dag H_{\rm eff} C_p}, \\ \nonumber
   &e^{i\frac{\theta}{2} T_1(p)}e^{i\frac{\theta}{2} T_2(p)} = e^{i H_{\rm eff}}, \\ \nonumber
   &H_{\rm eff} = \theta_{\rm eff} T_{\rm eff}.
\end{align}
Since $T_1(p)$ and $T_2(p)$ are $2 \times 2$ matrices, we can decompose it in terms of Pauli matrices $T_1 = \hat{n}_1 \cdot \vec{\sigma}$, $\hat{n}_1 = (\cos(p/2),-\sin(p/2),0)$, and similarly for $T_2$. And since
\begin{align*}
    e^{i \theta \hat{n} \cdot \vec{\sigma}} = cos(\theta)\mathcal{I} + isin(\theta)(\hat{n} \cdot \vec{\sigma})
\end{align*}
we can simplify our expressions and thus find that:
\begin{align}
    &\theta_{\rm eff} = \cos^{-1}(\cos^2(\theta/2) - \cos(p)\sin^2(\theta/2)), \\
    &T_{\rm eff} = \hat{n}' \cdot \vec{\sigma}, \\ \nonumber
    &\hat{n}' = \frac{1}{\sin(\theta_{\rm eff})}(\cos(p/2)sin(\theta), 0, -\sin(p)\sin^2(\theta/2)).
\end{align}
To find the dispersion relation, we diagonalize the $H_{\rm eff}$ matrix and we get that:
\begin{align}
\label{eqn:dispersion}
    E(p) = \pm 2\frac{\theta_{\rm eff}}{s(\theta_{\rm eff})} \left|c\left(\frac{p}{2}\right) s(\theta)\right| \sqrt{c^2(\theta) + s^2\left(\frac{p}{2}\right)s^2(\theta)}.
\end{align}
Here, $c(\theta) = \cos(\theta)$ and $s(\theta) = \sin(\theta)$. 
\begin{figure}
    
    \includegraphics[width=\columnwidth]{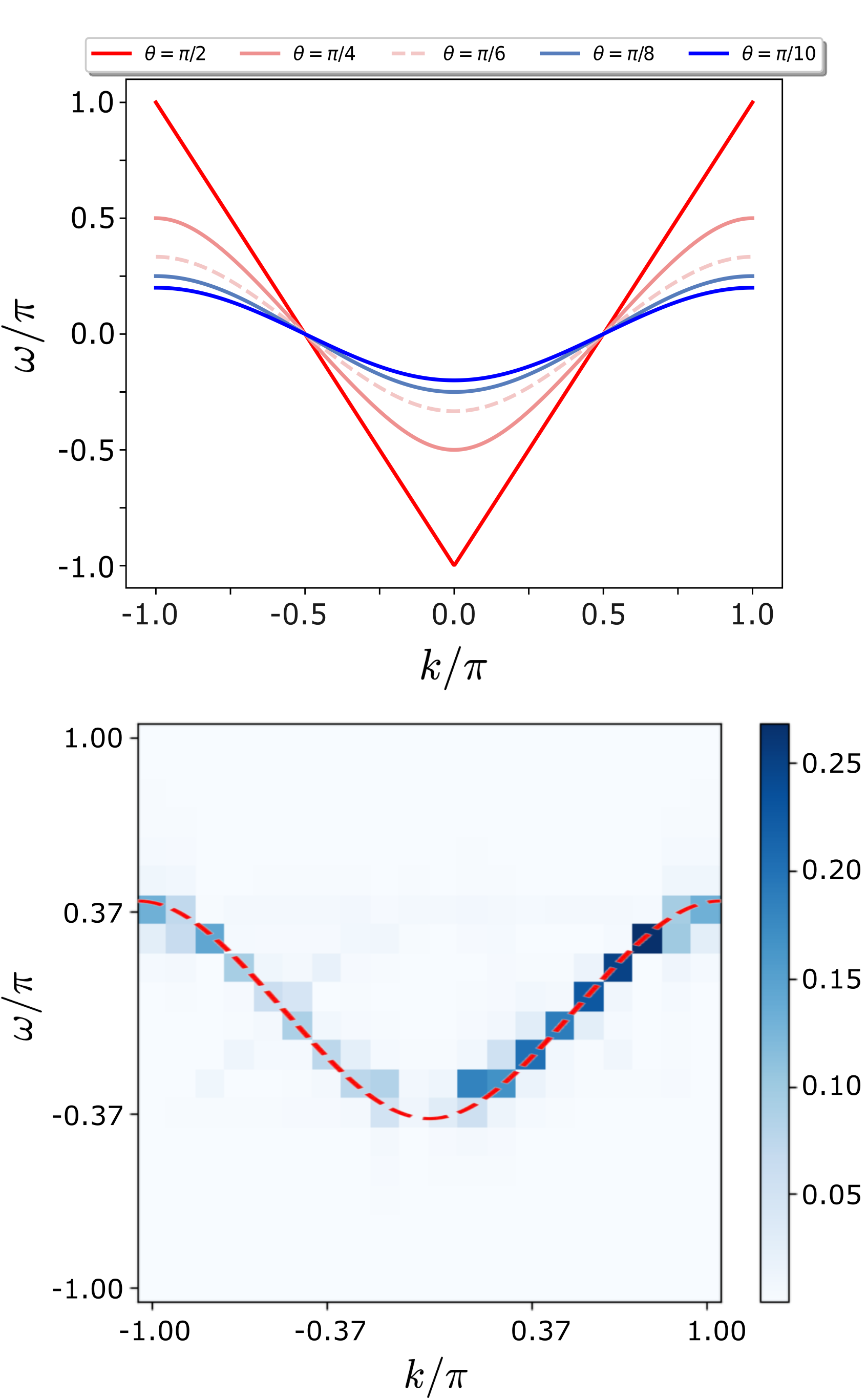}
    \caption{
         Plots for dispersion curves in the reduced Brillouin zone scheme obtained from \eqnref{eqn:dispersion} for various values of hopping parameter $\theta$ (upper panel) and a comparision between the numerically obtained (via direct Fourier transform) dispersion curve and the analytical dispersion relation for $\theta = \pi/6$ (lower panel)}
        \label{fig:disp-curv}
\end{figure}

We see the results in \figref{fig:disp-curv} (upper panel), which highlights the various dispersion curves {(i.e. we plot $\omega$ versus $k$ as per Eq.~\eqref{eqn:dispersion})} in the reduced Brillouin zone scheme for different values of the hopping parameter $\theta$. \figref{fig:disp-curv} (lower panel) is a {numerical check} {of} the analytical result with the corresponding dispersion curves obtained using {$2D$ Fourier transform of the expectation value of the two-body correlator with two adjacent qubits of the circuit prepared in the $+1-\text{eigenstate}$ of X Pauli matrix, as also used in \cite{MorvanRoushan2022}.}

\subsection{Numerical implementation of Fermi sea}
\label{sec:FermiSea}

We utilize the real space description of the Fermi Sea State (cf. Eq.~\eqref{eqn:fs-state}) which we arrive at by writing $c_{k_j} = (1/\sqrt{2 \pi N}) \sum_{x_i} e^{ix_ik_j} $. This gives us coefficients of the form $\prod_{i,j}e^{ix_ik_j}$, which are then calculated recursively for every possible classical bitstring describing a basis state in the $2^{2N+1}$ dimensional Hilbert space with $N/2$ qubits as $\ket{1}$. The linear superposition of these basis states multiplied with the calculated coefficients is our desired Fermi Sea state. We directly pass on these coefficients to Qiskit which does the superposition in the backend under the StateVector class.

\section{Decomposition of $U_K$ into 1 and 2-qubit gates}
\label{sec:appn1b}

\begin{figure}
    
    \includegraphics[width=\columnwidth]{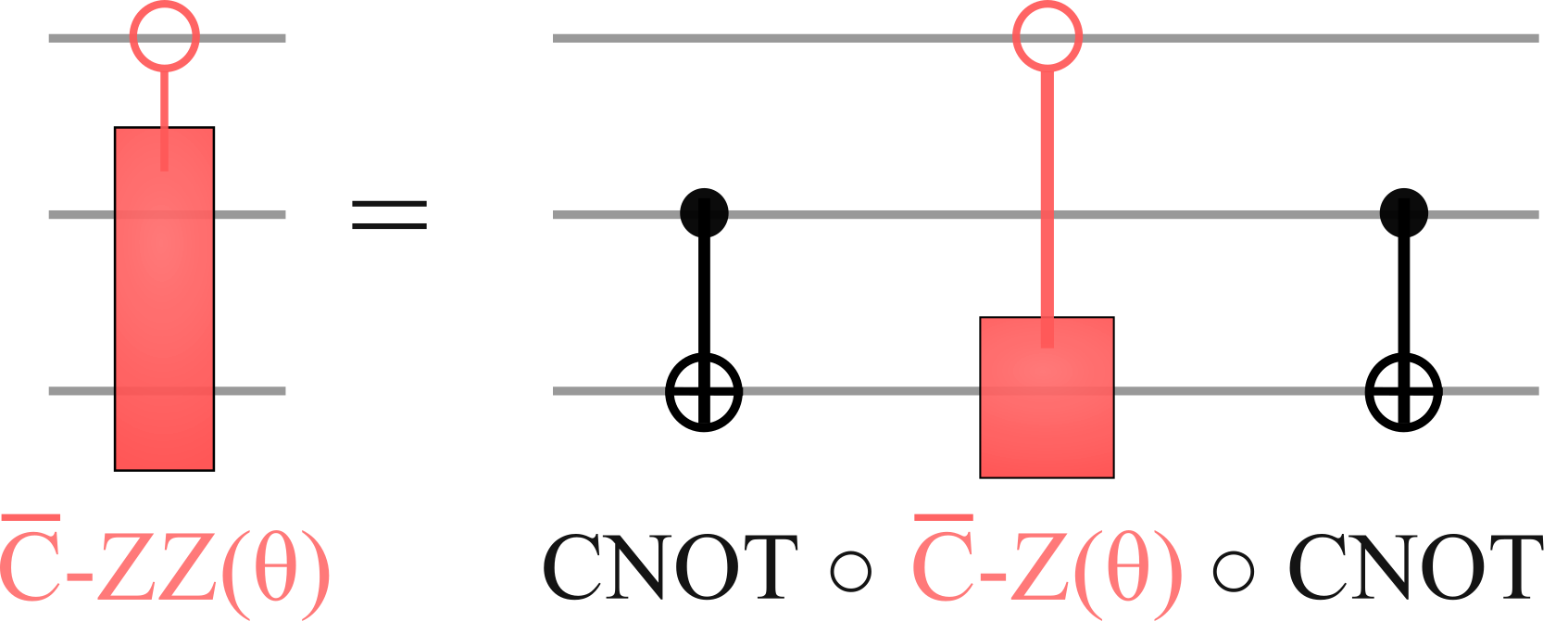}
    \caption{{A graphical description of the decomposition highlighted in \eqnref{eqn:decomp}. The top qubit is the control qubit, while the bottom two are the target qubits.}}
    \label{fig:rzz}
\end{figure}

Here we describe the decomposition of the Kondo unitary $U_K$ into 1 and 2-qubit rotation and $CNOT$ gates. Firstly, note that the cyclic permutation unitary can be decomposed into $X$ and $CNOT$ gates as:
\begin{align}
    \hat O = X_{-1/2} CNOT_{-1/2,1/2} CNOT_{1/2,-1/2} X_{1/2}.
\end{align}
We move on to the two controlled-unitaries $\bar C-\exp(i\theta_K(XX + YY)/2)$ and $ \bar C-\exp(i\theta_zZZ/2)$. The former can be written as a product of $\bar C-R_{XX}(\theta_K/2)$ and $\bar C-R_{YY}(\theta_K/2)$, which can be just derived from $\bar C-R_{ZZ}(\theta_K)$ by rotating the basis, i.e. by applying $H \otimes H$ and $R_X(\pi/2) \otimes R_X(\pi/2)$ before and after the controlled-$ZZ$ rotation on the target qubits. We first note that:
\begin{align}
\label{eqn:zzrot}
 e^{i\theta_K ZZ/2} =  CNOT \cdot(\text{Id} \otimes R_Z(\theta_K))\cdot CNOT.
\end{align}
The controlled version of \eqnref{eqn:zzrot} can be implemented by replacing $R_Z(\theta_K)$ with $\bar C-R_Z(\theta_K)$ while all the other gates still act on the target qubits only, i.e. 
\begin{align}
\label{eqn:decomp}
   \bar C-e^{i\theta_K ZZ/2} =  CNOT \cdot(\text{Id} \otimes \bar C-R_Z(\theta_K)) \cdot CNOT .
\end{align}
Since $\bar C-R_Z(\theta_K)$ is a controlled single qubit gate, we are done. A graphical description of the decomposition in \eqnref{eqn:decomp} is given in \figref{fig:rzz}. {Thus, we note that we utilize $2~(\text{change of basis gates}) + 3~(\text{from } \bar C-\exp(i\theta_zZZ/2)) + 10~(5 \text{ each from } \bar C-R_{XX}(\theta_K/2) \text{ and } \bar C-R_{YY}(\theta_K/2)) = 15$ 2-qubit gates for decomposing $U_K$.}

\section{Technical details of analytics}
\label{sec:appn4}

{In this appendix we summarize the key technical aspects of the analytical calculation as presented in Sec.~\ref{sec:Analytics} of the main text. In \secref{sec:appn4-1}, we describe how the exact and approximate single particle density matrix for the starting state is arrived at. \secref{sec:appn4-2} explains the derivation of $U_2$ (cf. Eq.~\eqref{eq:CBHUnitaries}) via a miniature Jordan-Wigner transformation. In \secref{sec:appn4-3}, we 
{present details of the solution of the free fermion Floquet quench}
to arrive at the time-evolved second quantized operators and the forms of the 
{impurity level} occupation numbers, where the latter are explicitly solved in \secref{sec:appn4-4}.}

\subsection{Derivation of single particle density matrix}
\label{sec:appn4-1}

{This section describes how we arrive at the exact (cf. Eq.~\eqref{eqn:dmfin}) and approximate (cf. Eq.~\eqref{eqn:dmapprox}) starting state density matrices in Sec.~\ref{sec:appn4-1-1} and Sec.~\ref{sec:appn4-1-2} respectively. Starting from the Hamiltonian in Eq.~\eqref{eqn:dmH}, we first use the Matsubara technique by resumming diagrams and then a Wick's transform to arrive at the exact density matrix, after which we keep only leading terms in a $\delta/\Delta_E \ll 1$ limit to arrive at the approximate density matrix.}

\subsubsection{Exact initial density matrix}
\label{sec:appn4-1-1}
We use the Matsubara technique and find that the Green's function has the form
\begin{equation} \label{eq:FullGf}
\mathbf G(i \epsilon, k,k') = \frac{\delta_{k,k'}}{i \epsilon - vk} + \frac{1}{i \epsilon - vk} T(i \epsilon) \frac{1}{i \epsilon - vk'}
\end{equation}
where for the $T$ matrix we need
\begin{align}
\sum_{k} G(i \epsilon, k)  &= \sum_{n} - \frac{i \epsilon}{\epsilon^2 + \Delta_E^2 n^2 }\notag \\
& = - i \frac{\pi}{\Delta_E } \coth \left ( \frac{\pi \epsilon}{\Delta_E} \right) ,
\end{align}
hence
\begin{equation}
T(i\epsilon) = \frac{V_0}{1 + i \frac{\pi V_0}{\Delta_E} \coth (\frac{\pi \epsilon}{\Delta_E} )}.
\end{equation}
We assumed a regularization scheme such as the point-splitting scheme employed in Sec.~\ref{sec:bandr} of the main text which respects particle hole symmetry. Note that the $T$ matrix $T(i\epsilon)$ has zeros at $i \epsilon = v p$ for all momenta, i.e.
\begin{equation}
T(i\epsilon)\vert_{\epsilon = - i v p + \delta \epsilon} = -i \delta \epsilon.
\end{equation}
Hence it cancels up the poles at $vk, vk'$ from the surrounding Green's functions of Eq.~\eqref{eq:FullGf} if $k \neq k'$. So their poles {do not} contribute. 
Moreover, if $k = k'$ we can study the full Green's function near $i \epsilon = vk$ and find
\begin{equation}
\mathbf G (vk+ i \delta \epsilon, k,k) = \frac{1}{i \delta \epsilon} + \frac{1}{(i \delta \epsilon)^2} (- i \delta \epsilon) = 0,
\end{equation}
so also in the case $k = k'$ the poles at $i \epsilon = vk$ are cancelled. {That is} ultimately no surprise, plane waves are simply no eigenstates of the problem.

There are also poles of $T(i \epsilon)$ which are positioned at shifted momenta $i \epsilon = v p + \delta \equiv v \bar p$, which leads to the condition
$1 = \frac{\pi V_0}{\Delta_E} \cot{\pi \delta/\Delta_E}$ and hence
\begin{equation}
\delta = \frac{\Delta_E}{\pi} \arccot \left ( \frac{\Delta_E}{ \pi V_0} \right),
\end{equation}
as quoted in the main text. The behavior near the pole is:
\begin{equation}
T(v \bar p + i \delta \epsilon)  =  -i \frac{1}{1+(\frac{\Delta_E}{\pi  V_0})^2} \frac{\Delta_E^2}{\delta \epsilon \pi^2}.
\end{equation}
Hence we can Fourier transform back:
\begin{align}
\mathbf G(\tau = 0^{-}, k,k') & = \int \frac{d \epsilon}{2\pi} e^{i \eta \epsilon} \mathbf G(i \epsilon, k,k') \notag \\
& =  - \sum_{\bar p < 0} \frac{1}{v \bar p - v k} \frac{1}{v \bar p - v k'} \frac{\Delta_E^2/\pi^2}{1+(\frac{\Delta_E}{\pi  V_0})^2} \label{eq:Sumpbar}
\end{align}
which obviously only captures poles in the lower half-plane of the complex variable $\epsilon$. 
We obtain:
\begin{align}
\mathbf G(\tau = 0^{-}, k,k') & =-  \frac{\Delta_E^2/\pi^2}{1 + ( \Delta_E/\pi V_0)^2} \notag\\
& \times \frac{\psi^{(0)}(\frac{vk - \delta}{\Delta_E}) - \psi^{(0)}(\frac{vk' - \delta}{\Delta_E} )}{(vk - vk')\Delta_E}.
\end{align}
This essentially concludes the derivation of the single particle density matrix for the starting state, as quoted in Eq.~\eqref{eqn:dmfin} of the main text.

\subsubsection{Approximations in the density matrix}
\label{sec:appn4-1-2}

Starting from Eq.~\eqref{eq:Sumpbar}, we proceed with the expansion of the starting density matrix in the limit of small $\delta/\Delta_E \ll 1$ 
\begin{subequations}
 \begin{align}
    \rho_{kk} & \simeq \delta^2 \sum_{\bar p < 0} \frac{1}{(v p + \delta - v k)^2} \notag \\
    & \simeq  \delta^2 \left [ \frac{\theta(-k {+0})}{\delta^2} + \sum_{\bar p < 0, p \neq k} \frac{1}{(v p + \delta - v k)^2} \right ] , \\
    \rho_{k k'} &\stackrel{k \neq k'}{\simeq} \delta^2 \theta(-k {+0})\theta(-k'{+ 0}) \Big[  \frac{1}{\delta} \frac{1}{v k - vk'} + \frac{1}{\delta} \frac{1}{v k' - vk} \notag \\
    &+ \sum_{\bar p < 0, p \neq k,k'} \frac{1}{v \bar p - v k}\frac{1}{v \bar p - v k'} \Big ] \notag \\
    &+ \delta^2 \theta(k{-0})\theta(-k'{+ 0}) \notag \\ & \times \Big[  \frac{1}{\delta} \frac{1}{v k - vk'} + \sum_{\bar p < 0, p \neq k} \frac{1}{v \bar p  - v k}\frac{1}{v \bar p  - v k'} \Big ] \notag \\
    &+ \delta^2 \theta(-k {+0})\theta(k' {-0})\notag \\ &\times \Big[  \frac{1}{\delta} \frac{1}{v k' - vk} + \sum_{\bar p < 0, p \neq k'} \frac{1}{v \bar p  - v k}\frac{1}{v \bar p  - v k'}\Big ] \notag
    \\
    &+ \delta^2 \theta(k{-0})\theta(k'{-0}) \sum_{\bar p < 0} \frac{1}{v \bar p  - v k}\frac{1}{v \bar p  - v k'}.
\end{align}   
\end{subequations}
{The additions "$\pm 0$" regulate the Heaviside function at $k = 0$.}
Keeping leading terms we thus find the approximate starting density matrix as Eq.~\eqref{eqn:dmapprox}.

\subsection{Derivation of $U_2$}
\label{sec:appn4-2}
We consider:
\begin{subequations}
 \begin{align}
    U(B_x,B_z) &= e^{-i B_x [C^\dagger_0 d + H.c.] - i 2B_z(d^\dagger d - 1/2)} \notag \\
       & = e^{- i \frac{B_x}{2} [Xx+ Yy] - i B_z z}, \\
       & = \frac{1+Zz}{2} e^{- i B_z \sigma_z} \frac{1 + Zz}{2} \notag\\
       & + \frac{1-Zz}{2} e^{- i [B_x \tau_x + B_z \tau_z]} \frac{1 - Zz}{2}.
\end{align}   
\end{subequations}
Here we used the miniature Jordan-Wigner transformation $C_0 = [X-iY]/2, 2C_0^\dagger C_{0} - 1 = Z, d = Z [x-iy]/2, 2 d^\dagger d -1 = z$ and the Pauli matrices $\sigma_{z}$ ($\tau_{x,z}$) act in the 2x2 subspace of $Zz$ eigenvalue +1 (-1) spanned by $\ket{\Uparrow, \uparrow},\ket{\Downarrow, \downarrow}$ ($\ket{\Downarrow, \uparrow},\ket{\Uparrow, \downarrow}$). For small $A,B_z \ll 1$ we calculate to subleading order in the exponent:
\begin{subequations}
 \begin{align}
    &U(0,A/2)U(B_x,B_z) U(0,A/2)  \\
     &= \frac{1+Zz}{2} e^{- i (A+B_z) \sigma_z} \frac{1 + Zz}{2} \notag\\
       & + \frac{1-Zz}{2} e^{- i \tilde B_x \tau_x   -i  [A + \tilde B_z] \tau_z } \frac{1 - Zz}{2} \\
       & = \text{exp} \Big \lbrace - i\frac{1+Zz}{2} (A + B_z z) \frac{1+Zz}{2} \notag \\
       & - i \frac{1-Zz}{2} [\frac{\tilde B_x}{2} [Xx +Yy]   +  [A + \tilde B_z] z ] \frac{1-Zz}{2} \Big \rbrace \\
       & = e^{ - i \tilde B_x [Xx + Yy] - i [A + \frac{B_z + \tilde B_z}{2}] z - i  \frac{B_z - \tilde B_z}{2} Z}.
\end{align}   
\end{subequations}
We here introduced the quantities:
\begin{subequations}
\begin{align}
    \tilde B_x & = B_x + \text{sign}(B_x) A [1- B_x \cot (B_x)], \\
    \tilde B_z & = \frac{B_x}{\sin(B_x)}[B_z + A \cos(B_x)] - A.
\end{align}   
\end{subequations}
The expansions correspond to the leading order expression in small $\Delta_E T \sim v T/L \ll 1$. For us:  
\begin{subequations}
 \begin{align}
&A  = -\gamma \Delta_E\frac{T}{2} \equiv -\gamma \bar \Delta_E = \bar \mu,\\
    &B_x =  \frac{\mathcal{I}_K \sqrt{2\pi}}{2a}\frac{T}{2}, \\
    &B_z  =  0,
\end{align}   
\end{subequations}
so that:
\begin{subequations}
\begin{align}
    \tilde B_x & =  \frac{\mathcal{I}_K \sqrt{2\pi}}{2a} \frac{T}{2} + \mathcal O({1/N}){\equiv \bar V}, \\
    \tilde B_z & = -\left \{\frac{\bar V}{\sin(\bar V)}[\gamma \cos(\bar V)]   - \gamma \right \} \bar \Delta_E.
\end{align}    
\end{subequations}
This concludes the derivation of Eq.~\eqref{eq:CBHUnitaries} of the main text.

\subsection{Solution of free fermion Floquet quench}
\label{sec:appn4-3}
{In this section, we utilize the idea of periodic $\mathcal{M}$ matrices, as first used in \cite{HeylKehrein2010b}, which essentially utilize key mathematical properties of the product of Greens function matrices (cf. Sec.~\ref{sec:appn4-3-3}). We first find the matrix forms of the Greens functions (cf. Sec.~\ref{sec:appn4-3-1}) corresponding to our unitaries $U_1$ and $U_2$, which are then multiplied to obtain $[\mathcal M]_{ll'}$ for a single step. These $\mathcal M$ matrices are then multiplied $N_s$ times to obtain the desired matrix elements which evolve our second quantized operators}

\comment{\begin{figure}
    
    \includegraphics[width=\columnwidth]{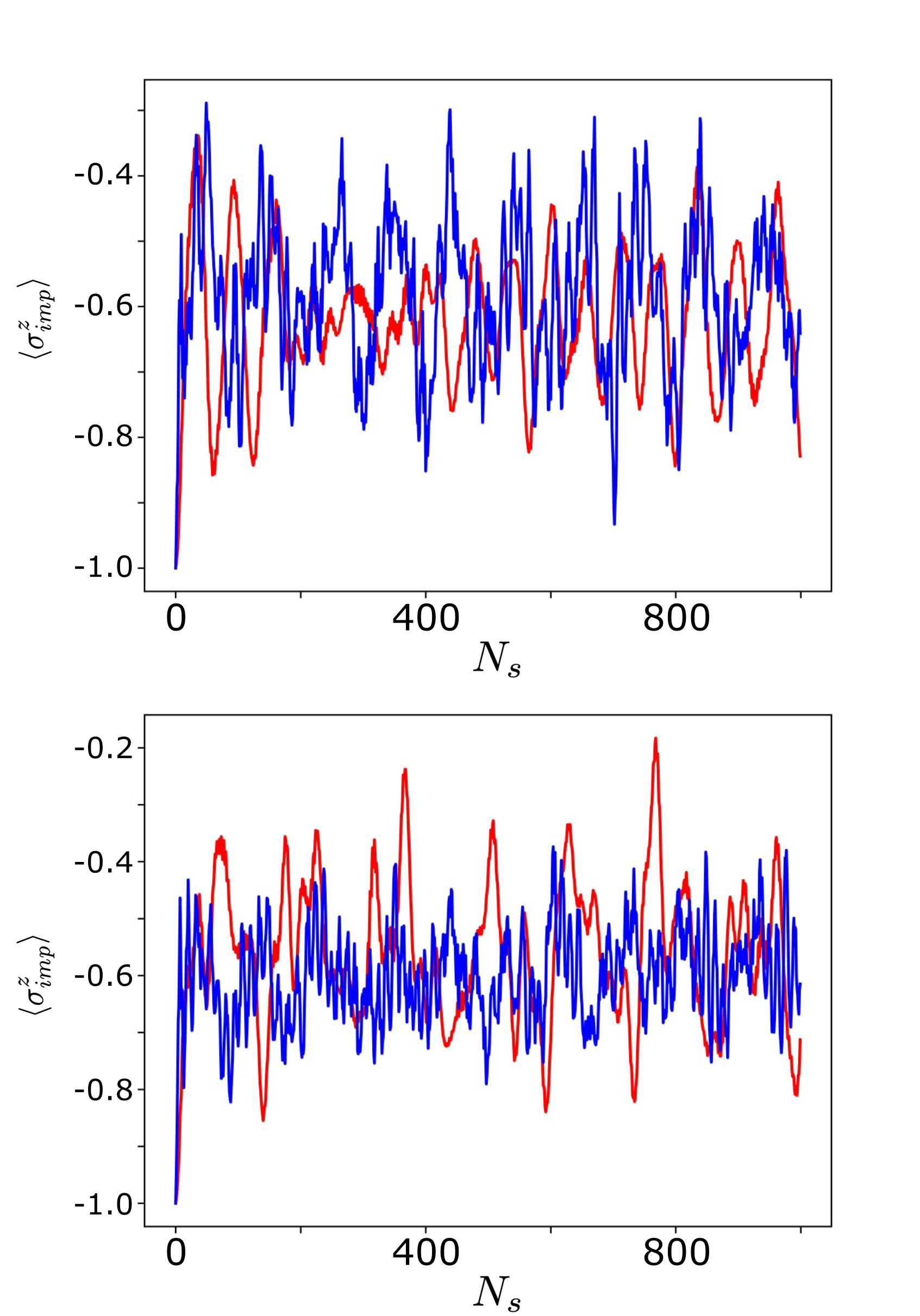}
    \caption{Impurity magnetization for $\theta_K = 0.12$ (red) and $\theta_K = 0.32$ (blue) v/s Floquet steps ($N_s$). Upper figure is for isotropic point, while the lower figure is for the Toulouse point {with Fermi sea starting state (cf. Eq.~\eqref{eqn:fs-state}) and $N = 6$}. }
    \label{fig:comp-low}
\end{figure}}

\subsubsection{Green's functions {approach}}
\label{sec:appn4-3-1}
We see that the form of the Hamiltonian in the exponent of Eq.~\eqref{eq:FloquetCycle12} is $\mathcal H_{{1,2}} = {a_l^\dag h^{{(1,2)}}_{ll'} a_{l'}}$, where {$h^{(1,2)}$ has a $2 \times 2$ block structure in the space spanned by}
$a_l = (C_0, c_k)$.
 We {further} use that for any Hamiltonian $h_{ll'}$ which is diagonalized by $u_{l \lambda}$ one can write
{\allowdisplaybreaks

\begin{align}
     a_l(t) & = e^{ i t a_l^\dagger h_{ll'}
 a_{l'}}  a_l e^{- i t a_l^\dagger h_{ll'}
 a_{l'}} \notag\\
     & = \sum_\lambda u_{l \lambda} e^{i t \epsilon a^\dagger_\lambda  a_\lambda} a_\lambda e^{ -i t \epsilon a^\dagger_\lambda  a_\lambda} \notag\\
     & = \sum_\lambda u_{l \lambda} e^{- i t \epsilon_\lambda} (u^\dagger)_{\lambda l'} a_{l '} \notag \\
     & = (e^{- i t h})_{ll'} a_{l'}.
\end{align}
}
Thus, for our model, we find that:
\begin{widetext}
\begin{subequations}
\begin{align}
    G_{ll'}^{R,1}(T/2) &= \left (\begin{array}{c c}
        1 & 0  \\
        0 & e^{-i  \bar v k}  \delta_{k,k'} 
        \end{array} \right) \\
    G_{dd}^{R,2}(T/2) &= e^{i \frac{\bar \mu}{2}} \left( \cos(\sqrt{\bar V^2 + (\Delta \bar \mu /2)^2}) + i \frac{(\Delta \bar \mu /2)}{\sqrt{\bar V^2 + (\Delta \bar \mu /2)^2}} \sin(\sqrt{\bar V^2 + (\Delta \bar \mu /2)^2})\right) \\
    G_{dk}^{R,2}(T/2) &= G_{kd}^{R,2}(T/2) = e^{i \frac{\bar \mu}{2}}\left(- i \sqrt{\frac{a}{L}}\frac{\bar V}{\sqrt{\bar V^2 + (\Delta \bar \mu /2)^2}} \sin(\sqrt{\bar V^2 + (\Delta \bar \mu /2)^2})\right) \\
    G_{kk'}^{R,2}(T/2) &= e^{i \frac{\bar \mu}{2}}\left(\delta_{k,k'} + \frac{a}{L} \left ( \cos(\sqrt{\bar V^2 + (\Delta \bar \mu /2)^2})-1 - i \frac{(\Delta \bar \mu /2)}{\sqrt{\bar V^2 + (\Delta \bar \mu /2)^2}} \sin(\sqrt{\bar V^2 + (\Delta \bar \mu /2)^2}) \right ) \right)
\end{align}    
\end{subequations}
\end{widetext}

Combining these two expressions leads to explict expressions for $\mathcal M_{ll'}$, cf. Eq.~\eqref{eq:Mdef} of the main text.

\subsubsection{Basic identities for $\mathcal{M}$ matrices}
\label{sec:appn4-3-3}
We use the following identity:
\begin{equation}
    \sum_k e^{- i m \bar v k} = 0, \forall m \in \mathbb N^+. 
\end{equation}
This is obvious for all $m \bar \Delta_E \equiv m \Delta_E T/2 \notin  \mathbb Z$ as $\sum_k e^{-ix k} = \delta(x)$.

As per the appendix in \cite{HeylKehrein2010b}, this identity should be read in Fourier space (for simplicity here for continuous time and $\lambda > 0$):
\begin{align}
    \sum_k \theta(t) e^{- i t v k} e^{-i \lambda k}  & \equiv i\sum_k\int (d\omega) \frac{e^{- i \omega t } e^{- i \lambda k }}{\omega+i \eta - v k} \notag \\
    & = - i \int (d\omega) \sum_k \frac{e^{- i \omega t} e^{- i \lambda k }}{ v k - \omega - i \eta } 
\end{align}
We will work in the limit $\eta \gg \Delta_E$ so that we can replace the sum by an integral (the $\eta$ smears the function under the sum more than can be resolved by discrete steps). Assuming an appropriate regularization scheme ({e.g. the particle-hole symmetric scheme exposed in the main text}) we follow arguments by Heyl and Kehrein to deform the contour of integration into the lower half plane for $\lambda \neq 0$, where there are no poles, while for $\lambda = 0$, the inner sum over momenta gives $-i \pi /\Delta_E$, but the integral over $\omega$ gives a delta function which is again 0 for $t > 0$. Hence the sum vanishes.

From the vanishing of internal $\sum_k e^{-i m \bar v k}$ sums we find \cite{HeylKehrein2010b}
\begin{subequations}
\begin{align}
    \sum_k e^{- i \lambda k} \mathcal M_{dk} \mathcal M_{kd} &= 0, \label{eq:MdkMkd}\\
        \sum_k e^{- i \lambda k} \mathcal M_{k_1k} \mathcal M_{kd} &= e^{- i (\bar v + \lambda ) k_1} \mathcal M_{kd}, \\
         \sum_k e^{- i \lambda k} \mathcal M_{dk} \mathcal M_{kk_1} &= \mathcal M_{dk} e^{- i (\bar v + \lambda ) k_1},\\
         \sum_k e^{- i \lambda k} \mathcal M_{k_1k} \mathcal M_{kk_2} & =  [\mathcal M_{k_1,k_2} + \delta \mathcal M e^{- i \bar v k_1} ]e^{- i (\lambda + \bar v) k_2}.
\end{align}
\end{subequations}
Again, positive $\lambda$ is assumed and incommensuration would be needed for small $\eta$. We introduced:
\begin{equation}
    \mathcal M_{kk'}  = e^{- i \bar v k} \delta_{kk'} + \mathcal \delta \mathcal M e^{- i \bar v k'}. 
\end{equation}
\subsubsection{Finding $[\mathcal{M}^{N_s}]_{dk}$}
\label{sec:appn4-3-2}

\begin{figure}
    
    \includegraphics[width=\columnwidth]{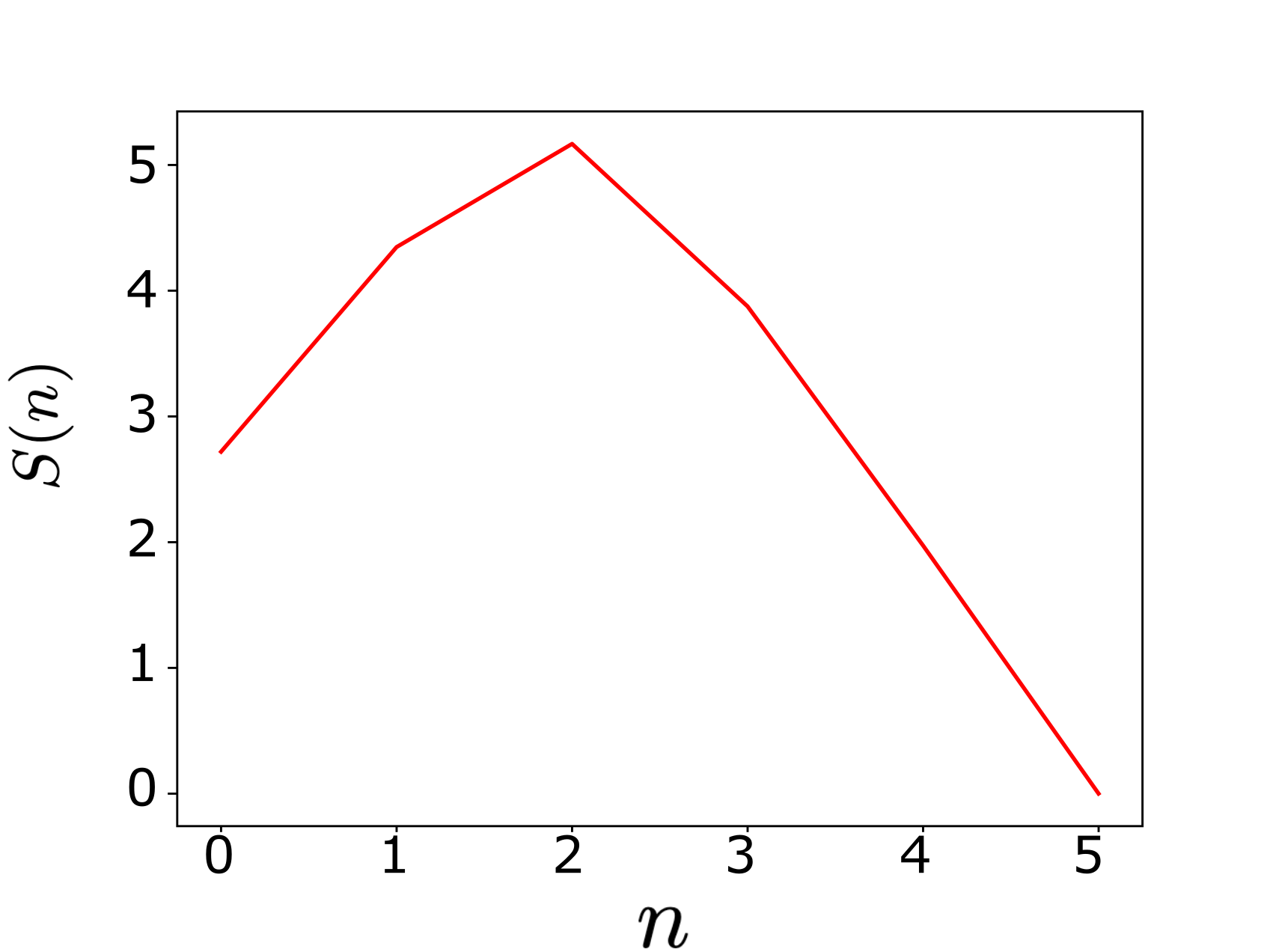}
    \caption{von Neumann entanglement ($S(n)$) as a function of site position ($n$) for $N = 6$ sites {with Fermi sea starting state (cf. Eq.~\eqref{eqn:fs-state})} after $N_s = 100$ Floquet steps. We see that the entanglement increases up to $n = 3$ after which it decreases monotonically.}
    \label{fig:ent-dist2}
\end{figure}

\begin{figure}
 \includegraphics[width=\columnwidth]{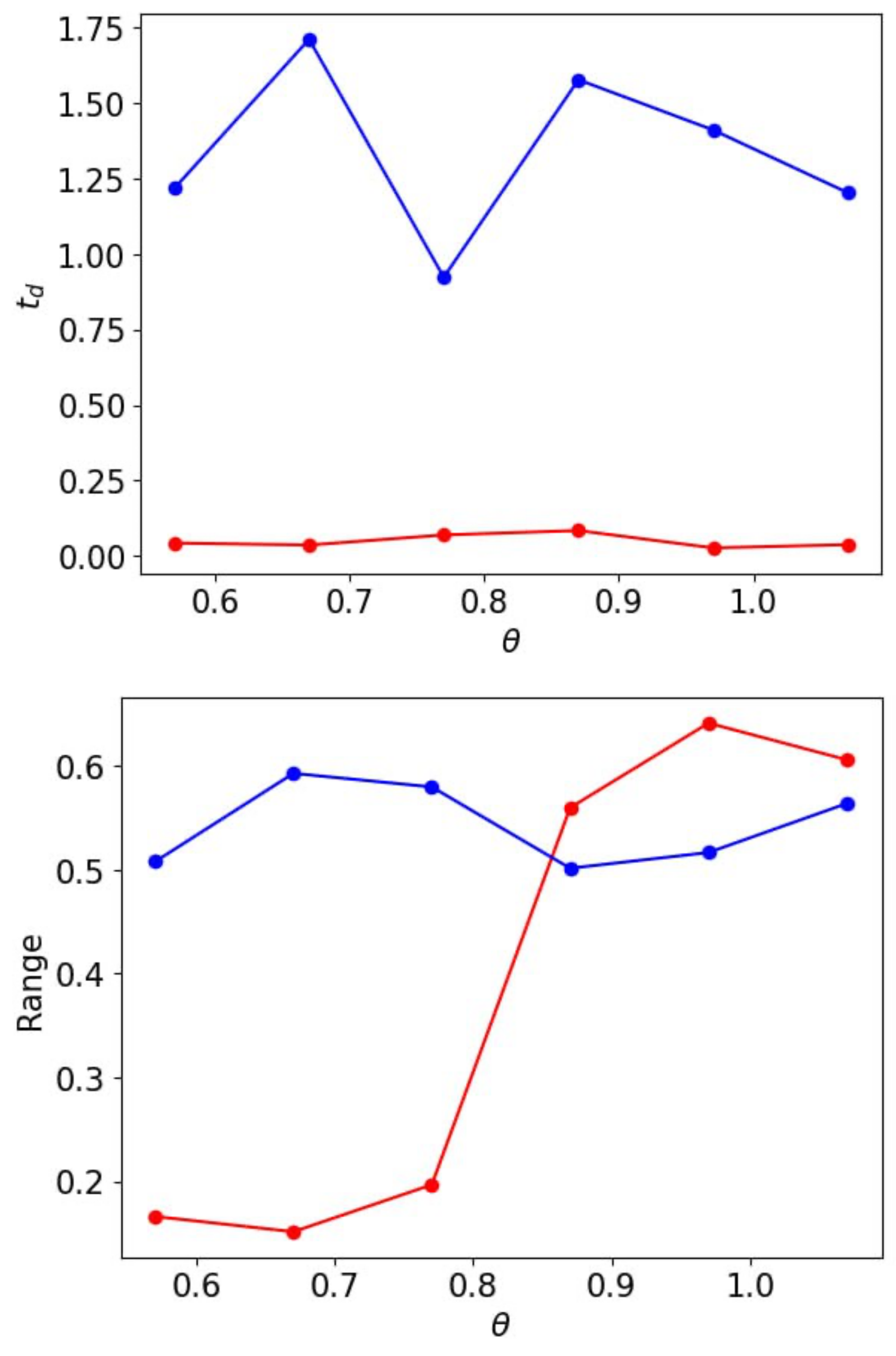}
        \caption{Plots for properties varying as a function of $\theta$ keeping $\theta_K = \pi/6$ fixed at Toulouse (blue) and isotropic (red) points {with Fermi sea starting state (cf. Eq.~\eqref{eqn:fs-state}) and $N=6$}. Upper panel: Decay times. Lower panel: Oscillation range between 100 and 1000 steps.}
        \label{fig:theta-1}
\end{figure}
Hence, we find for the most relevant matrix elements (Einstein sums assumed here)
\begin{subequations}
 \begin{align}
    [\mathcal M^{N_s}]_{dk} & =  [\mathcal M^{N_s-1}]_{dd}  \mathcal M_{dk} + [\mathcal M^{N_s-1}]_{dp}  \mathcal M_{pk} \notag\\
    & = [\mathcal M^{N_s-1}]_{dd}  \mathcal M_{dk} \notag\\
    &+ [\mathcal M^{N_s-2}]_{dd}  \mathcal M_{dp} \mathcal M_{pk} + [\mathcal M^{N_s-2}]_{dp}  \mathcal M_{pp'} \mathcal M_{p'k} \\
    & = [\mathcal M^{N_s-1}]_{dd}  \mathcal M_{dk}  + [\mathcal M^{N_s-2}]_{dd} \mathcal M_{dk} e^{- i \bar v k}  \notag \\
    &+[\mathcal M^{N_s-2}]_{dp}  \mathcal M_{pk} e^{- i \bar v k} + [\mathcal M^{N_s-2}]_{dp}  \delta \mathcal M e^{- i \bar v k- i \bar v p} \\
    & = [\mathcal M_{dd}]^{N_s-1}  \mathcal M_{dk}  + [\mathcal M^{N_s-1}]_{dk}e^{- i \bar v k} \notag \\
    & + [\mathcal M^{N_s-2}]_{dp}  \delta \mathcal M e^{- i \bar v p- i \bar v k}. \label{eq:itResult}
\end{align}   
\end{subequations}
We solve this with the Ansatz {
(cf. Eq.~\eqref{eq:MdkMainMain} of the main text)
}
\begin{align}
  &[\mathcal M^{N_s}]_{dk}  =  \sum_{{\ell} = 0}^{N_s - 1} [\mathcal M_{dd}]^{N_s-1-{\ell}} e^{- i {\ell} \bar v k} \mathcal M_{dk} \notag \\  
  &= \frac{\mathcal M_{dd}^{N_s} - e^{- i N_s \bar v k}}{\mathcal M_{dd} - e^{- i \bar v k}} \mathcal M_{dk}.
  \label{eq:Mdk}
\end{align}
By construction, for $N_s = 1,2$ this is the correct answer. {Let us} now check arbitrary $N_s \geq 2$ by plugging it into the previously found iterative result, Eq.~\eqref{eq:itResult}, where we readily see that the $p$ sum vanishes.  Hence we just need to check
\begin{align}
    [\mathcal M^{N_s}]_{dk} &= [\mathcal M_{dd}]^{N_s - 1} \mathcal M_{dk}  \notag \\
    &+ [\mathcal M_{dd}]^{N_s - 2} \sum_{{\ell} = 0}^{N_s - 2} \left ( \frac{e^{- i \bar v k}}{\mathcal M_{dd}} \right )^{{\ell}} e^{- i \bar v k} \mathcal M_{dk},
\end{align}
which is indeed correct and concludes the proof by induction. Note that $[\mathcal M^{N_s}]_{dk}$ describes the transition amplitude from starting state $k$ to final state $d$ after $N_s$ steps. It is also important to note that:
\begin{align}
    [\mathcal M^{N_s}]_{dk}^* = -[\mathcal M^{N_s}]_{d,-k} \vert_{(\bar \mu, \Delta \bar \mu) \rightarrow - (\bar \mu,\Delta \bar \mu)}.
\end{align}

\subsubsection{$d$-level occupation}

The magnetization takes the form $m_z(t)/2 = \langle S_z \rangle(t) = n(t) -1/2$, where $n (t)$ is the $d$ level occupation (in principle $t$ can be continuous or $t = N_s T$ discrete). We symmetrize/antisymmetrize the ladder with respect to the chemical potentials $\bar \mu_{d(C)} = (\bar \mu \pm \Delta \bar \mu)/2$.
\begin{equation}
    m_z(t) = {2}(n^{(s)}(t) + n^{(a)}(t)) -1,
\end{equation}
where
\begin{align}
    n^{(s/a)}(N_sT)& = {\frac{1}{2}}\{[\mathcal M^{N_s}]_{dk}^* [\mathcal M^{N_s}]_{dk'} \notag \\
    &\pm [\mathcal M^{N_s}]_{dk}^* [\mathcal M^{N_s}]_{dk'} \vert_{(\bar \mu,\Delta \bar \mu) \rightarrow - (\bar \mu, \Delta\bar \mu)} \} \rho_{kk'}, \notag \\
    & = [\mathcal M^{N_s}]_{dk}^* [\mathcal M^{N_s}]_{dk'} \underbrace{\frac{\rho_{kk'} \pm \rho_{-k',-k}}{2}}_{\rho_{kk'}^{(s/a)}}.
\end{align}
Using \eqnref{eqn:rho-sym}, we thus can write to leading order $n^{(s)}(t) \simeq n^{(s)}_0(t) + n^{(s)}_1(t)$ with: 
{\allowdisplaybreaks
\begin{subequations}
\begin{align}
    n^{(s)}_0(N_sT) & = \frac{1}{2} \sum_{k} [\mathcal M^{N_s}]_{dk}^* [\mathcal M^{N_s}]_{dk}, \\
    n^{(s)}_1(N_sT) & = - \sum_{k,k'} [\mathcal M^{N_s}]_{dk}^* [\mathcal M^{N_s}]_{dk'}  \frac{\delta \theta(-kk')}{\vert v k- v k' \vert},\\
    n^{(a)}(N_sT) & = -\frac{1}{2} \sum_{k} \text{sign}(k)[\mathcal M^{N_s}]_{dk}^* [\mathcal M^{N_s}]_{dk},
\end{align}
\label{eqn:n}
\end{subequations}
}
with $\text{sign}(0) =-1$.
\subsection{Calculation of $n^{(s)}$ and $n^{(a)}$}
\label{sec:appn4-4}

\begin{figure}
    
    \includegraphics[width=\columnwidth]{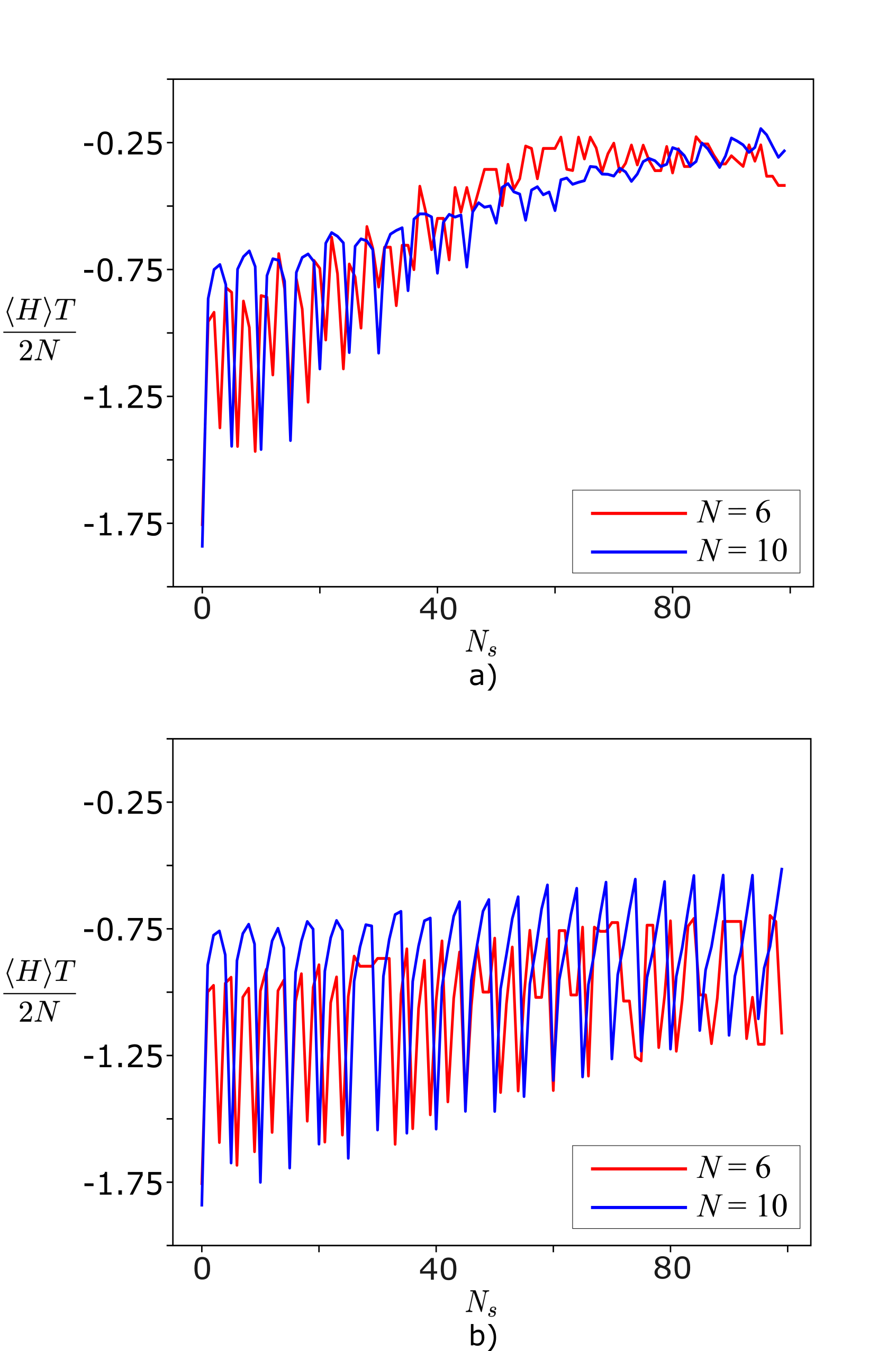}
    \caption{Expectation of total Hamiltonian ($\langle H \rangle  = \langle H_{\rm kin} + H_K \rangle$) as a function of time (Floquet steps) for different parameters for $N = 6$ (red) and $N = 10$ (blue) fermionic sites at isotropic point {with Fermi sea starting state (cf. Eq.~\eqref{eqn:fs-state})}. a) $\theta = \pi/4, \theta_K = \pi/4$. b) $\theta = \pi/4, \theta_K = \pi/6$ }
    \label{fig:h-plots1}
\end{figure}

\begin{figure}
\includegraphics[width=\columnwidth]{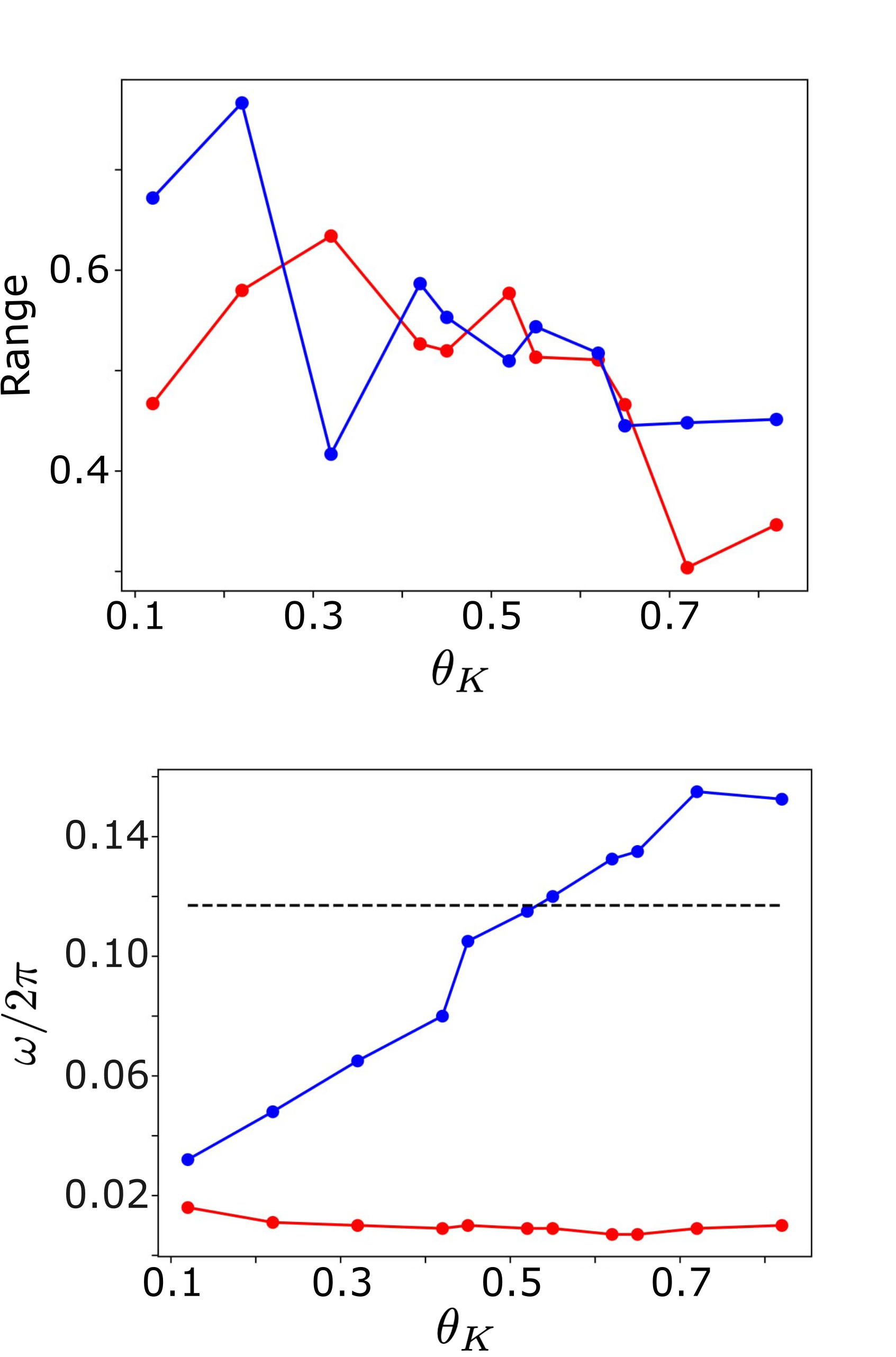}
        \caption{Plots for Oscillation range 
        (upper panel) and frequencies (lower panel) at the Toulouse point (blue) and isotropic point (red) as a function of $\theta_K$ keeping $\theta = \pi/3$ {with Fermi sea starting state (cf. Eq.~\eqref{eqn:fs-state}) and $N = 6$}. The black dashed curve in the lower panel is for the analytical frequencies obtained at the Toulouse point.}
        \label{fig:thetak-1}
\end{figure}

\begin{figure}
 \includegraphics[width=\columnwidth]{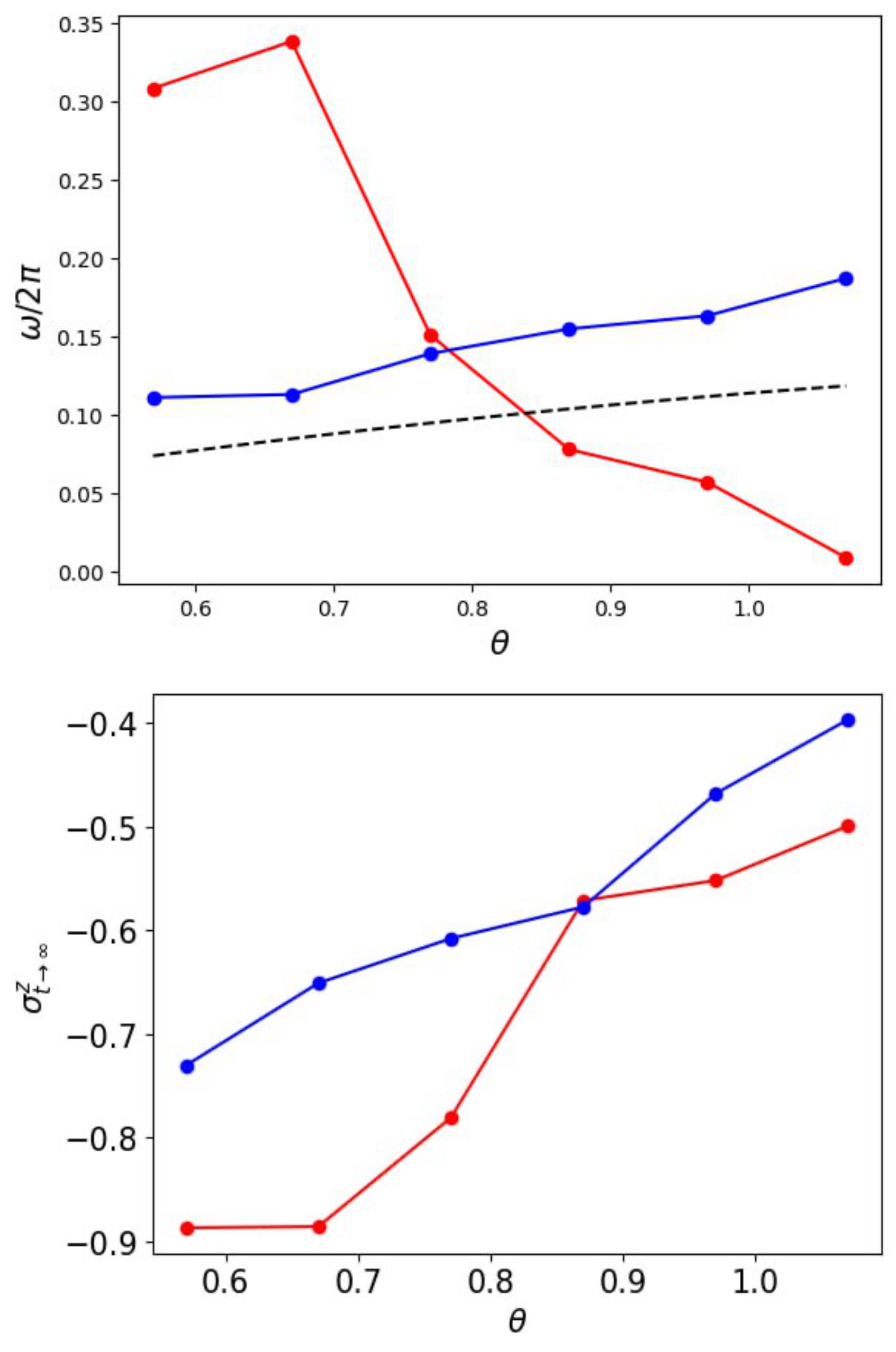}
        \caption{Plots for properties varying as a function of $\theta$ keeping $\theta_K = \pi/6$ fixed at Toulouse (blue) and isotropic (red) points {with Fermi sea starting state (cf. Eq.~\eqref{eqn:fs-state})}. Upper panel: Frequencies corresponding to the highest/second-highest FFT peaks. The black dashed curve is for the frequencies obtained analytically. Lower panel: Magnetization asymptote by averaging between 100 and 1000 steps.}
        \label{fig:theta-2}
\end{figure}
\subsubsection{Continuum time evolution}
It is instructive to first have a look at the quench dynamics for the case of continuous time evolution and for simplicty we set $\mu_C = (\mu - \Delta\mu)/{2} =0$, $\mu_d = (\mu + \Delta\mu)/{2} = \mu$. 
Then utilizing the standard  retarded Green's function definition as $\mathbf G^{R}_{ll'} = -i\theta(t)\langle \{c_l(t), c_{l'}^\dag \} \rangle$, we have:
\begin{subequations}
 \begin{align}
     \mathbf G_{dd}^R(\omega) & = \frac{1}{\omega + \mu_d + i \Delta}, \\
    \mathbf G_{kk'}^R(\omega) & = \frac{\delta_{kk'}}{\omega +  i \eta - v k}, \\
    \mathbf G_{dk}^R(\omega) & = \frac{V}{[\omega + \mu_d + i \Delta][\omega +  i \eta - v k]}.
 \end{align}   
\end{subequations}
In this case:
{\allowdisplaybreaks
\begin{align}
     n(t) &= \langle d^\dagger(t) d(t) \rangle \notag\\
     & = \frac{a}{L}\mathbf G_{dk}^{R,*}(t) \mathbf G_{dk'}^R(t) \rho_{kk'} \notag \\
     & = \frac{a}{L} \int (d\omega ) (d\omega') e^{i \omega t - i\omega' t} \rho_{kk'} \notag \\
     &  \times \frac{V}{[\omega + \mu_d - i \Delta][\omega -  i \eta - v k]} \notag \\
     & \times \frac{V}{[\omega' + \mu_d + i \Delta][\omega' +  i \eta - v k']}.
\end{align}
}
For $n_0^{(s)}$ we need:
\begin{align}
    \frac{V^2a}{2L}\sum_k \frac{1}{\omega^- - v k} \frac{1}{{\omega'}^+ - v k} \simeq i \Delta \frac{1}{\omega' - \omega + i 2 \eta},
\end{align}
where we used that for {$L \to \infty$} we can replace the sum by an integral. We also used that $ \frac{V^2a}{2L} \frac{2\pi}{\Delta_E} = \pi (aV^2/2\pi v) = \pi V^2 \rho = \Delta$ is the decay rate. Thus:
\begin{align}
    n_0^{(s)}(t) & = \int (d\omega) (d\omega') \frac{i \Delta e^{i \omega t} e^{- i \omega' t}}{\omega' - \omega + i 2 \eta} \frac{1}{\omega + \mu_d - i \Delta}\frac{1}{\omega' + \mu_d + i \Delta}, \notag\\
    & = \Delta \int (d\omega) \frac{1- e^{i \omega t + i \mu_d t - \Delta t}}{[\omega + \mu_d - i \Delta][\omega + \mu_d + i \Delta]} \notag \\
    & = \frac{1 - e^{-2 \Delta t}}{2}.
\end{align}
It is probably unsurprising that there are no oscillations in the limit where only particle-hole symmetric contributions are kept. The result coincides with what is typically quoted in the literature as in \cite{HeylKehrein2010b}. 

For $n^{(a)}$ we directly perform calculations in time domain and replace $\sum_k \rightarrow \int dx/\Delta_E$:
{\allowdisplaybreaks
\begin{align}
    n^{(a)} (t) & = - \underbrace{\frac{V^2 a}{2L \Delta_E}}_{= \Delta/2\pi} \int dx \text{sign}(x) \notag \\
    & \frac{e^{i x t} - e^{-i \mu_d t - \Delta t}}{x+ \mu_d - i \Delta } \frac{e^{-i x t} - e^{i \mu_d t - \Delta t}}{x+ \mu_d + i \Delta }  \notag \\
    & = -\frac{\Delta}{\pi} \int d x \text{sign}(x)\frac{1 - \cos((x + \mu_d)t) e^{- \Delta t}}{(x+ \mu_d)^2 + \Delta^2} \notag \\
    & \stackrel{y = x + \mu_d}= -\frac{\Delta}{\pi} \int_0^\infty d y [\text{sign}(y-\mu_d) - \text{sign}(y+\mu_d)]\notag \\ &\times \frac{1 - \cos(yt) e^{- \Delta t}}{y^2 + \Delta^2}  \notag\\
    & = 2 \text{sign}(\mu_d) \frac{\Delta}{\pi} \int_0^{\vert \mu_d \vert} dy \frac{1 - \cos(yt) e^{- \Delta t}}{y^2 + \Delta^2} 
\end{align}
}
We next use that $\vert \mu_d \vert \ll \Delta$ since $\mu_d \sim 1/L$ but $\Delta$ is constant. Then 
\begin{equation}
    n^{(a)}(t) =  \frac{2}{\pi} \left [\frac{\mu_d}{\Delta} - \sin(\mu_d t)\frac{e^{- \Delta t}}{\Delta t} \right].
\end{equation}
Thus we find a contribution to the asymptote, which is $\mathcal O(\mu_d/\Delta)$ and an oscillation with frequency $\mu_d$ which decays as $e^{-\Delta t}$ (i.e., slower than the non-oscillating term). It also makes sense that for positive $\mu_d$, the level tends to acquire a positive asymptotic value (recall that $\mu_d$ enters as $-\mu_d$ in the Hamiltonian, so no occupation is favored).

For the calculation of $n_1^{(s)}$ we first need to re-express the $\delta$ induced correction to the distribution function
\begin{align}
 \rho_{kk'}^{(1)} & = - \delta \frac{\theta(k) \theta (-k')-\theta(-k) \theta (k')}{vk - vk'} \notag \\
 & = - i \delta \int (d \Omega) \frac{\theta(k) \theta (-k')-\theta(-k) \theta (k')}{(\Omega^+ - vk)(\Omega^- - vk')} \notag\\
 & = -  i \frac{\delta}{2} \int (d \Omega) \frac{\text{sign}(k)-\text{sign} (k')}{(\Omega^+ - vk)(\Omega^- - vk')}.
\end{align}
We next define (we use $z = - \mu_d - i\Delta$)
\begin{subequations}
 \begin{align}
    \mathbf G^R (t, \Omega) & = \sum_k \frac{\mathbf G^R_{dk} (t)}{v k - \Omega^-} \notag \\
    & = - 2 \pi i \frac{V}{\Delta_E} \frac{e^{- i \Omega t} - e^{- i z t}}{\Omega-z}\\
    \tilde{\mathbf G}^R (t, \Omega) & = \sum_k \text{sign}(k) \frac{\mathbf G^R_{dk} (t)}{v k - \Omega^-} \notag \\
    & = 2\frac{V}{\Delta_E} \frac{1}{z- \Omega} \Big ( e^{- i z t} [\ln(-z) - \ln(-\Omega^{-})] \notag\\
    &+e^{- i t \Omega} [Ci(t\Omega^{-}) +i Si(t\Omega^{-}) + i \pi/2] \notag \\
    & -e^{- i z t} [Ci(tz) +i Si(tz) + i \pi/2]  \Big ).
\end{align}   
\end{subequations}
Here $Ci$ is the cosine integral and analogously $Si$ the sin integral function. 

Thus
\begin{align}
    n_1^{(s)}(t) & = - i \frac{a}{2L} \delta \int (d\Omega) [\mathbf G^{R,*} (t, \Omega) \tilde {\mathbf G}^R (t, \Omega) - c.c.] \notag \\
    & = 2\frac{\delta \Delta}{\Delta_E} \int (d\Omega) \Big [ \frac{e^{i \Omega t}- e^{i z^* t}}{(\Omega - z^*)(z- \Omega)} \notag \\
    & [\ln(-z) - \ln(-\Omega^{-})] \notag\\
    &+e^{- i t \Omega} [Ci(t\Omega^{-}) +i Si(t\Omega^{-}) + i \pi/2] \notag \\
    & -e^{- i z t} [Ci(tz) +i Si(tz) + i \pi/2] +c.c.\Big].
\end{align}
We will not seek additional correction to terms decaying as $e^{- \Delta t}$ (we only want to look for corrections to the asymptote and slower decaying oscillations).  Thus we drop all terms with $z$ in the exponent. The only term which has a chance of not falling off exponentially is thus:
\begin{align}
    n_1^{(s)}(t) \approx - \frac{4 \Delta \delta}{\Delta_E} \int_{\rm P.V.}(d\Omega) \frac{1}{(\Omega + \mu_d)^2 + \Delta^2} Ci(t \vert \Omega \vert).
\end{align}
However, it is an easy check that
\begin{align}
    \frac{d n_1^{(s)}}{dt}  &= - \frac{4 \Delta \delta }{\Delta_E t} \int (d\Omega) \frac{\cos(\Omega t)}{(\Omega + \mu_d)^2 + \Delta^2} \propto e^{- \Delta t}.
\end{align}
Thus for sure all time dependence is exponential decay. There might be constant, but it can not occur to first order in $\mu_d$ for symmetry reasons. One may perform one partial integral to find at zero $\mu$
\begin{align}
    n_1^{(s)}(t) = \frac{4 \Delta \delta }{\Delta_E } \int (d\Omega) \arctan (\Omega/\Delta) \frac{\cos(\Omega t)}{\Omega}.
\end{align}
It is not at all obvious from looking at this expression that it is not oscillating and instead decays exponentially, but numerical checks confirm that it decays to zero.  Thus, the fact that the starting state is not a simple Fermi sea plays no role at large times.

\subsubsection{Discrete time evolution}

Equipped with this understanding, we now switch to the Floquet case. For the dominant term $n^{(s)}_0$ we find
{\allowdisplaybreaks
\begin{align}
    n_0^{(s)} (N_sT) & = \frac{1}{2}\sum_{{\ell},{\ell}' =1}^{N_s} \mathcal M_{dd}^{*,N-{\ell}}\mathcal M_{dd}^{N-{\ell}'} \vert \mathcal M_{dk} \vert^2 \underbrace{\sum_k e^{i {\ell} \bar v k}e^{-i {\ell}' \bar v k}}_{= \delta_{{\ell}{\ell}'}L/a} \notag \\
    & = \frac{1}{2}\sum_{{\ell} = 1}^{N_s} \vert \mathcal M_{dd} \vert^{2N-2{\ell}} \vert \mathcal M_{dk} \vert^2 L/a \notag \\
    & = \frac{1}{2}\underbrace{\frac{ \vert \mathcal M_{dk} \vert^2 L/a}{1 - \vert \mathcal M_{dd} \vert^2}}_{ = 1} \left [ 1 - \vert \mathcal M_{dd} \vert^{2N_s} \right ] \notag \\
    & = \frac{1 - \left ( \frac{\bar V^2}{\bar V^2 + \bar \mu^2} \cos^2(\sqrt{\bar V^2 + \bar \mu^2}) - \frac{\bar \mu^2}{\bar V^2 + \bar \mu^2} \right)^{N_s}}{2} \notag \\
    & \simeq \frac{1 - \left ( \cos^2(\bar V) \right)^{N_s}}{2}.
\end{align}
}

In the last line, we drop terms that are $\mathcal O(1/L)$. To calculate $n^{(a)}$, we obtain:

{\allowdisplaybreaks
\begin{widetext}
\begin{align}
    n^{(a)} & = -\frac{1}{2} \sum_k \text{sign}(k) \frac{\mathcal M_{dd}^{*,N_s}- e^{ i  N_s \bar v k}}{\mathcal M_{dd}^{*}- e^{ i  \bar v k}}\frac{\mathcal M_{dd}^{N_s}- e^{- i  N_s \bar v k}}{\mathcal M_{dd}- e^{ - i  \bar v k}} \vert \mathcal M_{dk} \vert^2 \notag \\
    &=-\frac{1}{2} \frac{1}{\bar \Delta_E} \vert \mathcal M_{dk} \vert^2 \int dx \text{sign}(x) \frac{\vert \mathcal M_{dd} \vert^{2N_s} + 1 - 2 \vert \mathcal M_{dd} \vert^{N_s} \cos[N_s(\varphi+ x)]}{\vert \mathcal M_{dd} \vert^{2} + 1 - 2 \vert \mathcal M_{dd} \vert \cos(\varphi+ x)} \notag \\
    & = \frac{1}{2} \frac{\vert \mathcal M_{dk} \vert^2}{\bar \Delta_E}  \int dy \frac{\text{sign}(y+\varphi)-\text{sign}(y-\varphi)}{2} \frac{\vert \mathcal M_{dd} \vert^{2N_s} + 1 - 2 \vert \mathcal M_{dd} \vert^{N_s} \cos(N_sy)}{\vert \mathcal M_{dd} \vert^{2} + 1 - 2 \vert \mathcal M_{dd} \vert \cos(y)} \notag\\
    & = \frac{\vert \mathcal M_{dk} \vert^2}{\bar \Delta_E} \text{sign}(\varphi) \int_0^{\vert \varphi \vert} dy \frac{\vert \mathcal M_{dd} \vert^{2N_s} + 1 - 2 \vert \mathcal M_{dd} \vert^{N_s} \cos(N_sy)}{\vert \mathcal M_{dd} \vert^{2} + 1 - 2 \vert \mathcal M_{dd} \vert \cos(y)}.
\end{align}
\end{widetext}
}
We used $\mathcal M_{dd} =  \vert\mathcal M_{dd} \vert e^{i \varphi}$, where $\varphi$ is an odd function of the chemical potentials $\bar \mu_{d(C)} = {(\bar \mu \pm \Delta \bar \mu)/2}$ and $\bar \Delta_E = \Delta_E T/2$. Note that the only oscillating terms as a function of $N_s$ are exponentially damped, as in the continuum time evolution. Since we only want the leading order in $\varphi$ we can obtain that by Taylor expansion (differentiating the integral w.r.t. to $\vert \varphi\vert$). We further use (taking $\bar \mu \ll \bar V$):
\begin{subequations}
\begin{align}
    &\vert \mathcal M_{ dd} \vert^2  =  \cos^2(\bar V), \\
    &\vert \mathcal M_{ dk} \vert^2  = \frac{a}{L} \sin^2(\bar V)  = \frac{a}{L} [1 - \vert \mathcal M_{ dd} \vert^2],\\
    &\varphi  = \bar \mu + \arctan \left (\frac{\Delta \bar \mu}{\bar V} \tan (\bar V) \right ) \simeq \bar \mu + \mathcal O\left(\frac{1}{L}\right), \\
    &\frac{a/L}{\bar \Delta_E }  = \frac{a}{\pi v T} = \frac{2\rho}{T},
\end{align}    
\end{subequations}
where $\rho$ is the DOS and we obtain:
\begin{align}
\label{eqn:nafull}
    n^{(a)} & = \frac{2 \varphi \rho}{T \alpha}
    \left [ 1- 2 \vert \mathcal M_{dd} \vert^{N_s} \cos(N_s \varphi) + \vert \mathcal M_{dd} \vert^{2N_s} \right ]. \notag \\
    \alpha = &\frac{\sin^2(\bar V)}{\cos^2(\bar V) + 1 - 2\vert \cos(\bar V)\vert \cos(\varphi)} 
\end{align}
{We now solve $n_1^{(s)}$. To this end,} we briefly copy-paste the expression 
\begin{align}
    [\mathcal M^{N_s}]_{dk}  &=  \sum_{{\ell} = 0}^{N_s - 1} [\mathcal M_{dd}]^{N_s-1-{\ell}} e^{- i {\ell} \bar v k} \mathcal M_{dk}  \notag \\
    & =  \sum_{{\ell} = 1}^{N_s} [\mathcal M_{dd}]^{N_s-{\ell}} e^{- i {\ell} \bar v k} [e^{i {\ell} \bar v k} \mathcal M_{dk} ].
\end{align}
The expression in the last square brackets is $k$ independent. We will also use the Fourier transform of the signum function
\begin{align}
    \int_{P.V.} dy \frac{i e^{- i yx}}{ y\pi} = \text{sign}(x),
\end{align}
where $\int_{P.V.}$ is the principal value of the integral. As in the continuum calculation we also need:
\begin{widetext}
\begin{align}
    \mathcal M(\Omega) & = \sum_k \frac{[\mathcal M^{N_s}]_{dk}}{vk - \Omega^{-}}  = \frac{1}{\bar \Delta_E}\sum_{{\ell} = 1}^{N_s} [\mathcal M_{dd}]^{N_s-{\ell}} e^{- i l \Omega} [e^{i {\ell} \bar v k} \mathcal M_{dk} ] \notag \\
   \tilde{ \mathcal M}(\Omega) & = \sum_k \text{sign}(k) \frac{[\mathcal M^{N_s}]_{dk}}{vk - \Omega^{-}} =  \int  \frac{dx}{\bar \Delta_E}\frac{\text{sign}(x) }{x- \Omega^-}\sum_{{\ell} = 1}^{N_s} [\mathcal M_{dd}]^{N_s-{\ell}} e^{- i {\ell} x} [e^{i l \bar v k} \mathcal M_{dk} ] \notag \\
   & = \frac{e^{i {\ell} \bar v k} \mathcal M_{dk}}{\bar \Delta_E} \int_{P.V.} \frac{dy i}{y \pi} \int \frac{dx}{x- \Omega^{-}}\sum_{{\ell} = 1}^{N_s} [\mathcal M_{dd}]^{N_s-l} e^{- i ({\ell}+y) x} \notag \\
   &  = \frac{e^{i {\ell} \bar v k} \mathcal M_{dk}}{\bar \Delta_E} \int_{-{\ell}}^\infty dy \int_{P.V.} \frac{ i}{y \pi} \sum_{{\ell} = 1}^{N_s} [\mathcal M_{dd}]^{N_s-{\ell}} e^{- i ({\ell}+y) \Omega}.
\end{align}

In full analogy to the continuum case:
\begin{align}
    n_1^{(s)}(t) & = - \frac{i}{2} \delta \int (d\Omega) [\mathcal M^* ( \Omega) \tilde {\mathcal M} ( \Omega) - c.c.] \notag \\
    & = \frac{\delta \vert \mathcal M_{dk} \vert^2}{2 \pi \bar \Delta_E^2} \int_{-{\ell}'}^\infty dy \int_{P.V.} \frac{1}{y}\int (d\Omega) \sum_{{\ell}{\ell}' = 1}^{N_s} \mathcal M_{dd}^{*,N_s-{\ell}}\mathcal M_{dd}^{N_s-{\ell}'} e^{i [{\ell}-{\ell}'-y] \Omega} + c.c. \notag\\
     & = \frac{\delta \vert \mathcal M_{dk} \vert^2}{2 \pi \bar \Delta_E^2}   \sum_{{\ell}{\ell}' = 1}^{N_s} \mathcal M_{dd}^{*,N_s-{\ell}}\mathcal M_{dd}^{N_s-{\ell}'} \frac{1- \delta_{{\ell}{\ell}'}}{{\ell} - {\ell}'}+ c.c. \notag \\
      & = \frac{\delta \vert \mathcal M_{dk} \vert^2}{2 \pi \bar \Delta_E^2}   \sum_{{\ell}{\ell}' = 1}^{N_s} \left (\mathcal M_{dd}^{*,N_s-{\ell}}\mathcal M_{dd}^{N_s-{\ell}'} + \mathcal M_{dd}^{N_s-{\ell}}\mathcal M_{dd}^{*,N_s-{\ell}'} \right)\frac{1- \delta_{{\ell}{\ell}'}}{{\ell} -{\ell}'} \notag \\
      & = 0.
\end{align}
\end{widetext}

In the continuum case we found that this contribution is at most a non-oscillating, quickly decay function. This calculation says it is zero in the Floquet case. This wraps up the calculation of $n^{(s)}$ and $n^{(a)}$ for the discrete time evolution.

\section{Complementary numerical data}
\label{sec:appn5}
\begin{figure}
    \includegraphics[width=\columnwidth]{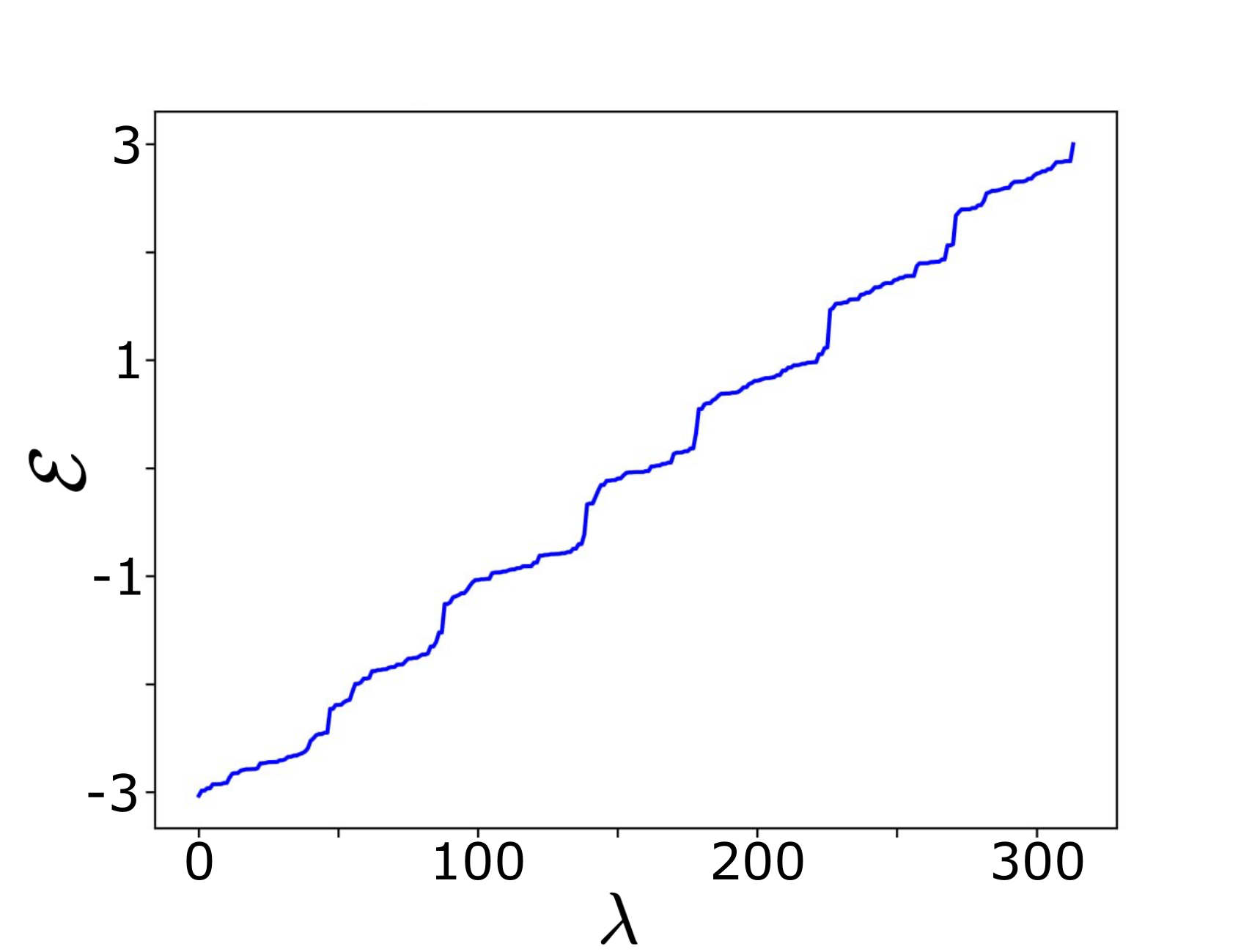}
    \caption{Plot for the {logarithm of} eigenvalues of
    $U_F$ (cf. Eq.~\eqref{eq:Floquet}) as {defined by} $U_F \ket{\psi_\lambda} = e^{-i \mathcal{E}_\lambda} \ket{\psi_\lambda}$ where $\lambda$ is the index of the eigenvalue for $\theta = \pi/3$ and $\theta_k = \pi/4$. The degenerate eigenvalues which were even below machine precision resolution of $10^{-16}$ are attributed the same index $\lambda$ (i.e. are plotted on top of each other).}
    \label{fig:dos-plot}
\end{figure}

Here we include complementary numerical data 
displaying additional information on top of those
highlighted in the main text. \figref{fig:ent-dist2} shows the von Neumann entanglement as a function of site position. We can easily see that the Kondo cloud in this case extends up to $n = 2$. {However, due to small system size, it is very hard to make any qualitative statements}. 


We also present the plots of oscillation range (which is taken by subtracting the minimum value from the maximum value in a time slice of 100 to 1000 Floquet steps) and {frequencies} with highest/second highest FFT peaks for the isotropic and Toulouse point in \figref{fig:thetak-1}. \figref{fig:theta-1} and \figref{fig:theta-2} highlights the variation of decay {times}, {frequencies}, oscillation range and magnetization asymptote with respect to $\theta$ keeping $\theta_K$ fixed at $\pi/6$. {Notice that from Eq.~\eqref{eq:AnalyticsDisc}, we see that even after small number of Floquet steps, we should observe no oscillations and thus there should not really be any effective oscillation range, which is a finite size effect in the numerics and the data for it seems to fluctuate in a random manner, cf. \figref{fig:thetak-1} and \figref{fig:theta-1}. In the upper panel of \figref{fig:theta-2}, we also highlight the analytical frequencies obtained (in black dashed curve), which follow a monotonically increasing trend, also observed in the numerical data. The trend observed in the lower panel of \figref{fig:theta-2} is also what the analytics (cf. Eq.~\eqref{eqn:nafull}) predicts}.

{\figref{fig:dos-plot} plots the natural logarithm of the eigenvalues of the Floquet unitary $U_F$ (cf. Eq.~\eqref{eq:Floquet}) in an increasing order. We can observe that while most of the eigenvalues are closely space, there appear jumps in values at certain indices.}

\bibliography{BibKondoQiskit.bib}

\begin{thebibliography}{67}%
\makeatletter
\providecommand \@ifxundefined [1]{%
 \@ifx{#1\undefined}
}%
\providecommand \@ifnum [1]{%
 \ifnum #1\expandafter \@firstoftwo
 \else \expandafter \@secondoftwo
 \fi
}%
\providecommand \@ifx [1]{%
 \ifx #1\expandafter \@firstoftwo
 \else \expandafter \@secondoftwo
 \fi
}%
\providecommand \natexlab [1]{#1}%
\providecommand \enquote  [1]{``#1''}%
\providecommand \bibnamefont  [1]{#1}%
\providecommand \bibfnamefont [1]{#1}%
\providecommand \citenamefont [1]{#1}%
\providecommand \href@noop [0]{\@secondoftwo}%
\providecommand \href [0]{\begingroup \@sanitize@url \@href}%
\providecommand \@href[1]{\@@startlink{#1}\@@href}%
\providecommand \@@href[1]{\endgroup#1\@@endlink}%
\providecommand \@sanitize@url [0]{\catcode `\\12\catcode `\$12\catcode `\&12\catcode `\#12\catcode `\^12\catcode `\_12\catcode `\%12\relax}%
\providecommand \@@startlink[1]{}%
\providecommand \@@endlink[0]{}%
\providecommand \url  [0]{\begingroup\@sanitize@url \@url }%
\providecommand \@url [1]{\endgroup\@href {#1}{\urlprefix }}%
\providecommand \urlprefix  [0]{URL }%
\providecommand \Eprint [0]{\href }%
\providecommand \doibase [0]{https://doi.org/}%
\providecommand \selectlanguage [0]{\@gobble}%
\providecommand \bibinfo  [0]{\@secondoftwo}%
\providecommand \bibfield  [0]{\@secondoftwo}%
\providecommand \translation [1]{[#1]}%
\providecommand \BibitemOpen [0]{}%
\providecommand \bibitemStop [0]{}%
\providecommand \bibitemNoStop [0]{.\EOS\space}%
\providecommand \EOS [0]{\spacefactor3000\relax}%
\providecommand \BibitemShut  [1]{\csname bibitem#1\endcsname}%
\let\auto@bib@innerbib\@empty
\bibitem [{\citenamefont {Preskill}(2018)}]{Preskill2018}%
  \BibitemOpen
  \bibfield  {author} {\bibinfo {author} {\bibfnamefont {J.}~\bibnamefont {Preskill}},\ }\bibfield  {title} {\bibinfo {title} {Quantum {C}omputing in the {NISQ} era and beyond},\ }\href {https://doi.org/10.22331/q-2018-08-06-79} {\bibfield  {journal} {\bibinfo  {journal} {{Quantum}}\ }\textbf {\bibinfo {volume} {2}},\ \bibinfo {pages} {79} (\bibinfo {year} {2018})}\BibitemShut {NoStop}%
\bibitem [{\citenamefont {Mori}(2023)}]{Mori2023}%
  \BibitemOpen
  \bibfield  {author} {\bibinfo {author} {\bibfnamefont {T.}~\bibnamefont {Mori}},\ }\bibfield  {title} {\bibinfo {title} {Floquet states in open quantum systems},\ }\href {https://doi.org/https://doi.org/10.1146/annurev-conmatphys-040721-015537} {\bibfield  {journal} {\bibinfo  {journal} {Annual Review of Condensed Matter Physics}\ }\textbf {\bibinfo {volume} {14}},\ \bibinfo {pages} {35} (\bibinfo {year} {2023})}\BibitemShut {NoStop}%
\bibitem [{\citenamefont {Marin~Bukov}\ and\ \citenamefont {Polkovnikov}(2015)}]{BukovPolkovnikov2015}%
  \BibitemOpen
  \bibfield  {author} {\bibinfo {author} {\bibfnamefont {L.~D.}\ \bibnamefont {Marin~Bukov}}\ and\ \bibinfo {author} {\bibfnamefont {A.}~\bibnamefont {Polkovnikov}},\ }\bibfield  {title} {\bibinfo {title} {Universal high-frequency behavior of periodically driven systems: from dynamical stabilization to {F}loquet engineering},\ }\href {https://doi.org/10.1080/00018732.2015.1055918} {\bibfield  {journal} {\bibinfo  {journal} {Advances in Physics}\ }\textbf {\bibinfo {volume} {64}},\ \bibinfo {pages} {139} (\bibinfo {year} {2015})}\BibitemShut {NoStop}%
\bibitem [{\citenamefont {Morvan}\ \emph {et~al.}(2022)\citenamefont {Morvan}, \citenamefont {Andersen}, \citenamefont {Mi}, \citenamefont {Neill}, \citenamefont {Petukhov}, \citenamefont {Kechedzhi}, \citenamefont {Abanin}, \citenamefont {Michailidis}, \citenamefont {Acharya}, \citenamefont {Arute},\ and\ \citenamefont {et~al.}}]{MorvanRoushan2022}%
  \BibitemOpen
  \bibfield  {author} {\bibinfo {author} {\bibfnamefont {A.}~\bibnamefont {Morvan}}, \bibinfo {author} {\bibfnamefont {T.~I.}\ \bibnamefont {Andersen}}, \bibinfo {author} {\bibfnamefont {X.}~\bibnamefont {Mi}}, \bibinfo {author} {\bibfnamefont {C.}~\bibnamefont {Neill}}, \bibinfo {author} {\bibfnamefont {A.}~\bibnamefont {Petukhov}}, \bibinfo {author} {\bibfnamefont {K.}~\bibnamefont {Kechedzhi}}, \bibinfo {author} {\bibfnamefont {D.~A.}\ \bibnamefont {Abanin}}, \bibinfo {author} {\bibfnamefont {A.}~\bibnamefont {Michailidis}}, \bibinfo {author} {\bibfnamefont {R.}~\bibnamefont {Acharya}}, \bibinfo {author} {\bibfnamefont {F.}~\bibnamefont {Arute}},\ and\ \bibinfo {author} {\bibnamefont {et~al.}},\ }\bibfield  {title} {\bibinfo {title} {Formation of robust bound states of interacting microwave photons},\ }\href {https://doi.org/10.1038/s41586-022-05348-y} {\bibfield  {journal} {\bibinfo  {journal} {Nature}\ }\textbf {\bibinfo {volume} {612}},\ \bibinfo {pages} {240–245} (\bibinfo {year} {2022})}\BibitemShut
  {NoStop}%
\bibitem [{\citenamefont {Rosenberg}\ \emph {et~al.}(2024)\citenamefont {Rosenberg}, \citenamefont {Andersen}, \citenamefont {Samajdar}, \citenamefont {Petukhov}, \citenamefont {Hoke}, \citenamefont {Abanin}, \citenamefont {Bengtsson}, \citenamefont {Drozdov}, \citenamefont {Erickson}, \citenamefont {Klimov} \emph {et~al.}}]{RosenbergRoushan2024}%
  \BibitemOpen
  \bibfield  {author} {\bibinfo {author} {\bibfnamefont {E.}~\bibnamefont {Rosenberg}}, \bibinfo {author} {\bibfnamefont {T.}~\bibnamefont {Andersen}}, \bibinfo {author} {\bibfnamefont {R.}~\bibnamefont {Samajdar}}, \bibinfo {author} {\bibfnamefont {A.}~\bibnamefont {Petukhov}}, \bibinfo {author} {\bibfnamefont {J.}~\bibnamefont {Hoke}}, \bibinfo {author} {\bibfnamefont {D.}~\bibnamefont {Abanin}}, \bibinfo {author} {\bibfnamefont {A.}~\bibnamefont {Bengtsson}}, \bibinfo {author} {\bibfnamefont {I.}~\bibnamefont {Drozdov}}, \bibinfo {author} {\bibfnamefont {C.}~\bibnamefont {Erickson}}, \bibinfo {author} {\bibfnamefont {P.}~\bibnamefont {Klimov}}, \emph {et~al.},\ }\bibfield  {title} {\bibinfo {title} {Dynamics of magnetization at infinite temperature in a {H}eisenberg spin chain},\ }\href {https://doi.org/10.1126/science.adi7877} {\bibfield  {journal} {\bibinfo  {journal} {Science}\ }\textbf {\bibinfo {volume} {384}},\ \bibinfo {pages} {48} (\bibinfo {year} {2024})}\BibitemShut {NoStop}%
\bibitem [{\citenamefont {Mi}\ \emph {et~al.}(2022)\citenamefont {Mi}, \citenamefont {Sonner}, \citenamefont {Niu}, \citenamefont {Lee}, \citenamefont {Foxen}, \citenamefont {Acharya}, \citenamefont {Aleiner}, \citenamefont {Andersen}, \citenamefont {Arute}, \citenamefont {Arya} \emph {et~al.}}]{MiRoushan2022}%
  \BibitemOpen
  \bibfield  {author} {\bibinfo {author} {\bibfnamefont {X.}~\bibnamefont {Mi}}, \bibinfo {author} {\bibfnamefont {M.}~\bibnamefont {Sonner}}, \bibinfo {author} {\bibfnamefont {M.~Y.}\ \bibnamefont {Niu}}, \bibinfo {author} {\bibfnamefont {K.~W.}\ \bibnamefont {Lee}}, \bibinfo {author} {\bibfnamefont {B.}~\bibnamefont {Foxen}}, \bibinfo {author} {\bibfnamefont {R.}~\bibnamefont {Acharya}}, \bibinfo {author} {\bibfnamefont {I.}~\bibnamefont {Aleiner}}, \bibinfo {author} {\bibfnamefont {T.~I.}\ \bibnamefont {Andersen}}, \bibinfo {author} {\bibfnamefont {F.}~\bibnamefont {Arute}}, \bibinfo {author} {\bibfnamefont {K.}~\bibnamefont {Arya}}, \emph {et~al.},\ }\bibfield  {title} {\bibinfo {title} {Noise-resilient edge modes on a chain of superconducting qubits},\ }\href {https://doi.org/10.1126/science.abq5769} {\bibfield  {journal} {\bibinfo  {journal} {Science}\ }\textbf {\bibinfo {volume} {378}},\ \bibinfo {pages} {785} (\bibinfo {year} {2022})}\BibitemShut {NoStop}%
\bibitem [{\citenamefont {Zhang}\ \emph {et~al.}(2022)\citenamefont {Zhang}, \citenamefont {Jiang}, \citenamefont {Deng}, \citenamefont {Wang}, \citenamefont {Chen}, \citenamefont {Zhang}, \citenamefont {Ren}, \citenamefont {Dong}, \citenamefont {Xu}, \citenamefont {Gao} \emph {et~al.}}]{ZhangWang2022}%
  \BibitemOpen
  \bibfield  {author} {\bibinfo {author} {\bibfnamefont {X.}~\bibnamefont {Zhang}}, \bibinfo {author} {\bibfnamefont {W.}~\bibnamefont {Jiang}}, \bibinfo {author} {\bibfnamefont {J.}~\bibnamefont {Deng}}, \bibinfo {author} {\bibfnamefont {K.}~\bibnamefont {Wang}}, \bibinfo {author} {\bibfnamefont {J.}~\bibnamefont {Chen}}, \bibinfo {author} {\bibfnamefont {P.}~\bibnamefont {Zhang}}, \bibinfo {author} {\bibfnamefont {W.}~\bibnamefont {Ren}}, \bibinfo {author} {\bibfnamefont {H.}~\bibnamefont {Dong}}, \bibinfo {author} {\bibfnamefont {S.}~\bibnamefont {Xu}}, \bibinfo {author} {\bibfnamefont {Y.}~\bibnamefont {Gao}}, \emph {et~al.},\ }\bibfield  {title} {\bibinfo {title} {Digital quantum simulation of {F}loquet symmetry-protected topological phases},\ }\href {https://doi.org/10.1038/s41586-022-04854-3} {\bibfield  {journal} {\bibinfo  {journal} {Nature}\ }\textbf {\bibinfo {volume} {607}},\ \bibinfo {pages} {468} (\bibinfo {year} {2022})}\BibitemShut {NoStop}%
\bibitem [{\citenamefont {Gritsev}\ and\ \citenamefont {Polkovnikov}(2017)}]{GritsevPolkovnikov2017}%
  \BibitemOpen
  \bibfield  {author} {\bibinfo {author} {\bibfnamefont {V.}~\bibnamefont {Gritsev}}\ and\ \bibinfo {author} {\bibfnamefont {A.}~\bibnamefont {Polkovnikov}},\ }\bibfield  {title} {\bibinfo {title} {Integrable {F}loquet dynamics},\ }\href {https://doi.org/10.21468/SciPostPhys.2.3.021} {\bibfield  {journal} {\bibinfo  {journal} {SciPost Physics}\ }\textbf {\bibinfo {volume} {2}},\ \bibinfo {pages} {021} (\bibinfo {year} {2017})}\BibitemShut {NoStop}%
\bibitem [{\citenamefont {Vanicat}\ \emph {et~al.}(2018)\citenamefont {Vanicat}, \citenamefont {Zadnik},\ and\ \citenamefont {Prosen}}]{VanicatProsen2018}%
  \BibitemOpen
  \bibfield  {author} {\bibinfo {author} {\bibfnamefont {M.}~\bibnamefont {Vanicat}}, \bibinfo {author} {\bibfnamefont {L.}~\bibnamefont {Zadnik}},\ and\ \bibinfo {author} {\bibfnamefont {T.}~\bibnamefont {Prosen}},\ }\bibfield  {title} {\bibinfo {title} {Integrable {T}rotterization: Local conservation laws and boundary driving},\ }\href {https://doi.org/10.1103/PhysRevLett.121.030606} {\bibfield  {journal} {\bibinfo  {journal} {Phys. Rev. Lett.}\ }\textbf {\bibinfo {volume} {121}},\ \bibinfo {pages} {030606} (\bibinfo {year} {2018})}\BibitemShut {NoStop}%
\bibitem [{\citenamefont {Ljubotina}\ \emph {et~al.}()\citenamefont {Ljubotina}, \citenamefont {Zadnik},\ and\ \citenamefont {Prosen}}]{LjubotinaProsen2019}%
  \BibitemOpen
  \bibfield  {author} {\bibinfo {author} {\bibfnamefont {M.}~\bibnamefont {Ljubotina}}, \bibinfo {author} {\bibfnamefont {L.}~\bibnamefont {Zadnik}},\ and\ \bibinfo {author} {\bibfnamefont {T.}~\bibnamefont {Prosen}},\ }\bibfield  {title} {\bibinfo {title} {Ballistic spin transport in a periodically driven integrable quantum system},\ }\href {https://doi.org/10.1103/PhysRevLett.122.150605} {\bibfield  {journal} {\bibinfo  {journal} {Phys. Rev. Lett.}\ }\textbf {\bibinfo {volume} {122}},\ \bibinfo {pages} {150605}}\BibitemShut {NoStop}%
\bibitem [{\citenamefont {Dupont}\ and\ \citenamefont {Moore}(2020)}]{Dupont2020}%
  \BibitemOpen
  \bibfield  {author} {\bibinfo {author} {\bibfnamefont {M.}~\bibnamefont {Dupont}}\ and\ \bibinfo {author} {\bibfnamefont {J.~E.}\ \bibnamefont {Moore}},\ }\bibfield  {title} {\bibinfo {title} {Universal spin dynamics in infinite-temperature one-dimensional quantum magnets},\ }\href {https://doi.org/10.1103/PhysRevB.101.121106} {\bibfield  {journal} {\bibinfo  {journal} {Phys. Rev. B}\ }\textbf {\bibinfo {volume} {101}},\ \bibinfo {pages} {121106} (\bibinfo {year} {2020})}\BibitemShut {NoStop}%
\bibitem [{\citenamefont {Aleiner}(2021)}]{Aleiner2021}%
  \BibitemOpen
  \bibfield  {author} {\bibinfo {author} {\bibfnamefont {I.~L.}\ \bibnamefont {Aleiner}},\ }\bibfield  {title} {\bibinfo {title} {{B}ethe ansatz solutions for certain periodic quantum circuits},\ }\href {https://doi.org/https://doi.org/10.1016/j.aop.2021.168593} {\bibfield  {journal} {\bibinfo  {journal} {Annals of Physics}\ }\textbf {\bibinfo {volume} {433}},\ \bibinfo {pages} {168593} (\bibinfo {year} {2021})}\BibitemShut {NoStop}%
\bibitem [{\citenamefont {Vernier}\ \emph {et~al.}(2024)\citenamefont {Vernier}, \citenamefont {Yeh}, \citenamefont {Piroli},\ and\ \citenamefont {Mitra}}]{VernierMitra2024}%
  \BibitemOpen
  \bibfield  {author} {\bibinfo {author} {\bibfnamefont {E.}~\bibnamefont {Vernier}}, \bibinfo {author} {\bibfnamefont {H.-C.}\ \bibnamefont {Yeh}}, \bibinfo {author} {\bibfnamefont {L.}~\bibnamefont {Piroli}},\ and\ \bibinfo {author} {\bibfnamefont {A.}~\bibnamefont {Mitra}},\ }\bibfield  {title} {\bibinfo {title} {Strong zero modes in integrable quantum circuits},\ }\href {https://doi.org/10.1103/PhysRevLett.133.050606} {\bibfield  {journal} {\bibinfo  {journal} {Phys. Rev. Lett.}\ }\textbf {\bibinfo {volume} {133}},\ \bibinfo {pages} {050606} (\bibinfo {year} {2024})}\BibitemShut {NoStop}%
\bibitem [{\citenamefont {Vernier}\ \emph {et~al.}(2023)\citenamefont {Vernier}, \citenamefont {Bertini}, \citenamefont {Giudici},\ and\ \citenamefont {Piroli}}]{VernierPiroli2023}%
  \BibitemOpen
  \bibfield  {author} {\bibinfo {author} {\bibfnamefont {E.}~\bibnamefont {Vernier}}, \bibinfo {author} {\bibfnamefont {B.}~\bibnamefont {Bertini}}, \bibinfo {author} {\bibfnamefont {G.}~\bibnamefont {Giudici}},\ and\ \bibinfo {author} {\bibfnamefont {L.}~\bibnamefont {Piroli}},\ }\bibfield  {title} {\bibinfo {title} {Integrable digital quantum simulation: Generalized {G}ebbs ensembles and {T}rotter transitions},\ }\href {https://doi.org/10.1103/PhysRevLett.130.260401} {\bibfield  {journal} {\bibinfo  {journal} {Phys. Rev. Lett.}\ }\textbf {\bibinfo {volume} {130}},\ \bibinfo {pages} {260401} (\bibinfo {year} {2023})}\BibitemShut {NoStop}%
\bibitem [{\citenamefont {Mitra}(2018)}]{Mitra2018}%
  \BibitemOpen
  \bibfield  {author} {\bibinfo {author} {\bibfnamefont {A.}~\bibnamefont {Mitra}},\ }\bibfield  {title} {\bibinfo {title} {Quantum quench dynamics},\ }\href {https://doi.org/https://doi.org/10.1146/annurev-conmatphys-031016-025451} {\bibfield  {journal} {\bibinfo  {journal} {Annual Review of Condensed Matter Physics}\ }\textbf {\bibinfo {volume} {9}},\ \bibinfo {pages} {245} (\bibinfo {year} {2018})}\BibitemShut {NoStop}%
\bibitem [{\citenamefont {McArdle}\ \emph {et~al.}(2020)\citenamefont {McArdle}, \citenamefont {Endo}, \citenamefont {Aspuru-Guzik}, \citenamefont {Benjamin},\ and\ \citenamefont {Yuan}}]{McArdleYuan2020}%
  \BibitemOpen
  \bibfield  {author} {\bibinfo {author} {\bibfnamefont {S.}~\bibnamefont {McArdle}}, \bibinfo {author} {\bibfnamefont {S.}~\bibnamefont {Endo}}, \bibinfo {author} {\bibfnamefont {A.}~\bibnamefont {Aspuru-Guzik}}, \bibinfo {author} {\bibfnamefont {S.~C.}\ \bibnamefont {Benjamin}},\ and\ \bibinfo {author} {\bibfnamefont {X.}~\bibnamefont {Yuan}},\ }\bibfield  {title} {\bibinfo {title} {Quantum computational chemistry},\ }\href {https://doi.org/10.1103/RevModPhys.92.015003} {\bibfield  {journal} {\bibinfo  {journal} {Rev. Mod. Phys.}\ }\textbf {\bibinfo {volume} {92}},\ \bibinfo {pages} {015003} (\bibinfo {year} {2020})}\BibitemShut {NoStop}%
\bibitem [{\citenamefont {Oftelie}\ \emph {et~al.}(2021)\citenamefont {Oftelie}, \citenamefont {Urbanek}, \citenamefont {Metcalf}, \citenamefont {Carter}, \citenamefont {Kemper},\ and\ \citenamefont {de~Jong}}]{OfteliedeJong2021}%
  \BibitemOpen
  \bibfield  {author} {\bibinfo {author} {\bibfnamefont {L.~B.}\ \bibnamefont {Oftelie}}, \bibinfo {author} {\bibfnamefont {M.}~\bibnamefont {Urbanek}}, \bibinfo {author} {\bibfnamefont {M.}~\bibnamefont {Metcalf}}, \bibinfo {author} {\bibfnamefont {J.}~\bibnamefont {Carter}}, \bibinfo {author} {\bibfnamefont {A.~F.}\ \bibnamefont {Kemper}},\ and\ \bibinfo {author} {\bibfnamefont {W.~A.}\ \bibnamefont {de~Jong}},\ }\bibfield  {title} {\bibinfo {title} {Simulating quantum materials with digital quantum computers},\ }\href {https://doi.org/10.1088/2058-9565/ac1ca6} {\bibfield  {journal} {\bibinfo  {journal} {Quantum Science and Technology}\ }\textbf {\bibinfo {volume} {6}},\ \bibinfo {pages} {043002} (\bibinfo {year} {2021})}\BibitemShut {NoStop}%
\bibitem [{\citenamefont {Ayral}\ \emph {et~al.}(2023)\citenamefont {Ayral}, \citenamefont {Besserve}, \citenamefont {Lacroix},\ and\ \citenamefont {Ruiz~Guzman}}]{AyralGuzman2023}%
  \BibitemOpen
  \bibfield  {author} {\bibinfo {author} {\bibfnamefont {T.}~\bibnamefont {Ayral}}, \bibinfo {author} {\bibfnamefont {P.}~\bibnamefont {Besserve}}, \bibinfo {author} {\bibfnamefont {D.}~\bibnamefont {Lacroix}},\ and\ \bibinfo {author} {\bibfnamefont {E.~A.}\ \bibnamefont {Ruiz~Guzman}},\ }\bibfield  {title} {\bibinfo {title} {Quantum computing with and for many-body physics},\ }\href {https://doi.org/10.1140/epja/s10050-023-01141-1} {\bibfield  {journal} {\bibinfo  {journal} {The European Physical Journal A}\ }\textbf {\bibinfo {volume} {59}},\ \bibinfo {pages} {227} (\bibinfo {year} {2023})}\BibitemShut {NoStop}%
\bibitem [{\citenamefont {Bravyi}\ and\ \citenamefont {Gosset}(2017)}]{Bravyi2017}%
  \BibitemOpen
  \bibfield  {author} {\bibinfo {author} {\bibfnamefont {S.}~\bibnamefont {Bravyi}}\ and\ \bibinfo {author} {\bibfnamefont {D.}~\bibnamefont {Gosset}},\ }\bibfield  {title} {\bibinfo {title} {Complexity of quantum impurity problems},\ }\href {https://doi.org/10.1007/s00220-017-2976-9} {\bibfield  {journal} {\bibinfo  {journal} {Communications in Mathematical Physics}\ }\textbf {\bibinfo {volume} {356}},\ \bibinfo {pages} {451} (\bibinfo {year} {2017})}\BibitemShut {NoStop}%
\bibitem [{\citenamefont {Hewson}(1997)}]{HewsonBook}%
  \BibitemOpen
  \bibfield  {author} {\bibinfo {author} {\bibfnamefont {A.~C.}\ \bibnamefont {Hewson}},\ }\href {https://doi.org/10.1017/CBO9780511470752} {\emph {\bibinfo {title} {The {K}ondo problem to heavy fermions}}},\ \bibinfo {number} {2}\ (\bibinfo  {publisher} {Cambridge university press},\ \bibinfo {year} {1997})\BibitemShut {NoStop}%
\bibitem [{\citenamefont {Wilson}(1975)}]{Wilson1975}%
  \BibitemOpen
  \bibfield  {author} {\bibinfo {author} {\bibfnamefont {K.~G.}\ \bibnamefont {Wilson}},\ }\bibfield  {title} {\bibinfo {title} {The renormalization group: Critical phenomena and the {K}ondo problem},\ }\href {https://doi.org/10.1103/RevModPhys.47.773} {\bibfield  {journal} {\bibinfo  {journal} {Rev. Mod. Phys.}\ }\textbf {\bibinfo {volume} {47}},\ \bibinfo {pages} {773} (\bibinfo {year} {1975})}\BibitemShut {NoStop}%
\bibitem [{\citenamefont {Andrei}(1980)}]{Andrei1980}%
  \BibitemOpen
  \bibfield  {author} {\bibinfo {author} {\bibfnamefont {N.}~\bibnamefont {Andrei}},\ }\bibfield  {title} {\bibinfo {title} {Diagonalization of the {K}ondo {H}amiltonian},\ }\href {https://doi.org/10.1103/PhysRevLett.45.379} {\bibfield  {journal} {\bibinfo  {journal} {Physical Review Letters}\ }\textbf {\bibinfo {volume} {45}},\ \bibinfo {pages} {379} (\bibinfo {year} {1980})}\BibitemShut {NoStop}%
\bibitem [{\citenamefont {Weigmann}(1980)}]{Wiegman1980}%
  \BibitemOpen
  \bibfield  {author} {\bibinfo {author} {\bibfnamefont {P.}~\bibnamefont {Weigmann}},\ }\bibfield  {title} {\bibinfo {title} {Exact solution of sd exchange model at {T}= 0},\ }\href {https://doi.org/10.1088/0022-3719/14/10/014} {\bibfield  {journal} {\bibinfo  {journal} {Soviet Journal of Experimental and Theoretical Physics Letters}\ }\textbf {\bibinfo {volume} {31}},\ \bibinfo {pages} {364} (\bibinfo {year} {1980})}\BibitemShut {NoStop}%
\bibitem [{\citenamefont {Coleman}(1983)}]{Coleman1983}%
  \BibitemOpen
  \bibfield  {author} {\bibinfo {author} {\bibfnamefont {P.}~\bibnamefont {Coleman}},\ }\bibfield  {title} {\bibinfo {title} {$\frac{1}{N}$ expansion for the kondo lattice},\ }\href {https://doi.org/10.1103/PhysRevB.28.5255} {\bibfield  {journal} {\bibinfo  {journal} {Phys. Rev. B}\ }\textbf {\bibinfo {volume} {28}},\ \bibinfo {pages} {5255} (\bibinfo {year} {1983})}\BibitemShut {NoStop}%
\bibitem [{\citenamefont {Coleman}(2015)}]{ColemanBook}%
  \BibitemOpen
  \bibfield  {author} {\bibinfo {author} {\bibfnamefont {P.}~\bibnamefont {Coleman}},\ }\href@noop {} {\emph {\bibinfo {title} {Introduction to many-body physics}}}\ (\bibinfo  {publisher} {Cambridge University Press},\ \bibinfo {year} {2015})\BibitemShut {NoStop}%
\bibitem [{\citenamefont {Emery}\ and\ \citenamefont {Kivelson}(1992)}]{EmeryKivelson1992}%
  \BibitemOpen
  \bibfield  {author} {\bibinfo {author} {\bibfnamefont {V.~J.}\ \bibnamefont {Emery}}\ and\ \bibinfo {author} {\bibfnamefont {S.}~\bibnamefont {Kivelson}},\ }\bibfield  {title} {\bibinfo {title} {Mapping of the two-channel {K}ondo problem to a resonant-level model},\ }\href {https://doi.org/10.1103/PhysRevB.46.10812} {\bibfield  {journal} {\bibinfo  {journal} {Phys. Rev. B}\ }\textbf {\bibinfo {volume} {46}},\ \bibinfo {pages} {10812} (\bibinfo {year} {1992})}\BibitemShut {NoStop}%
\bibitem [{\citenamefont {Affleck}\ and\ \citenamefont {Ludwig}(1991{\natexlab{a}})}]{AffleckLudwig1991}%
  \BibitemOpen
  \bibfield  {author} {\bibinfo {author} {\bibfnamefont {I.}~\bibnamefont {Affleck}}\ and\ \bibinfo {author} {\bibfnamefont {A.~W.}\ \bibnamefont {Ludwig}},\ }\bibfield  {title} {\bibinfo {title} {The {K}ondo effect, conformal field theory and fusion rules},\ }\href {https://doi.org/https://doi.org/10.1016/0550-3213(91)90109-B} {\bibfield  {journal} {\bibinfo  {journal} {Nuclear Physics B}\ }\textbf {\bibinfo {volume} {352}},\ \bibinfo {pages} {849} (\bibinfo {year} {1991}{\natexlab{a}})}\BibitemShut {NoStop}%
\bibitem [{\citenamefont {Pustilnik}\ and\ \citenamefont {Glazman}(2004)}]{PustilnikGlazman2004}%
  \BibitemOpen
  \bibfield  {author} {\bibinfo {author} {\bibfnamefont {M.}~\bibnamefont {Pustilnik}}\ and\ \bibinfo {author} {\bibfnamefont {L.}~\bibnamefont {Glazman}},\ }\bibfield  {title} {\bibinfo {title} {{K}ondo effect in quantum dots},\ }\href {https://doi.org/10.1088/0953-8984/16/16/R01} {\bibfield  {journal} {\bibinfo  {journal} {Journal of Physics: Condensed Matter}\ }\textbf {\bibinfo {volume} {16}},\ \bibinfo {pages} {R513} (\bibinfo {year} {2004})}\BibitemShut {NoStop}%
\bibitem [{\citenamefont {{Iftikhar}}\ \emph {et~al.}(2018)\citenamefont {{Iftikhar}}, \citenamefont {{Anthore}}, \citenamefont {{Mitchell}}, \citenamefont {{Parmentier}}, \citenamefont {{Gennser}}, \citenamefont {{Ouerghi}}, \citenamefont {{Cavanna}}, \citenamefont {{Mora}}, \citenamefont {{Simon}},\ and\ \citenamefont {{Pierre}}}]{IftikharPierre2018}%
  \BibitemOpen
  \bibfield  {author} {\bibinfo {author} {\bibfnamefont {Z.}~\bibnamefont {{Iftikhar}}}, \bibinfo {author} {\bibfnamefont {A.}~\bibnamefont {{Anthore}}}, \bibinfo {author} {\bibfnamefont {A.~K.}\ \bibnamefont {{Mitchell}}}, \bibinfo {author} {\bibfnamefont {F.~D.}\ \bibnamefont {{Parmentier}}}, \bibinfo {author} {\bibfnamefont {U.}~\bibnamefont {{Gennser}}}, \bibinfo {author} {\bibfnamefont {A.}~\bibnamefont {{Ouerghi}}}, \bibinfo {author} {\bibfnamefont {A.}~\bibnamefont {{Cavanna}}}, \bibinfo {author} {\bibfnamefont {C.}~\bibnamefont {{Mora}}}, \bibinfo {author} {\bibfnamefont {P.}~\bibnamefont {{Simon}}},\ and\ \bibinfo {author} {\bibfnamefont {F.}~\bibnamefont {{Pierre}}},\ }\bibfield  {title} {\bibinfo {title} {{Tunable quantum criticality and super-ballistic transport in a {\textquotedblleft}charge{\textquotedblright} {K}ondo circuit}},\ }\href {https://doi.org/10.1126/science.aan5592} {\bibfield  {journal} {\bibinfo  {journal} {Science}\ }\textbf {\bibinfo {volume} {360}},\ \bibinfo {pages} {1315}
  (\bibinfo {year} {2018})}\BibitemShut {NoStop}%
\bibitem [{\citenamefont {B\'eri}\ and\ \citenamefont {Cooper}(2012)}]{BeriCooper2012}%
  \BibitemOpen
  \bibfield  {author} {\bibinfo {author} {\bibfnamefont {B.}~\bibnamefont {B\'eri}}\ and\ \bibinfo {author} {\bibfnamefont {N.~R.}\ \bibnamefont {Cooper}},\ }\bibfield  {title} {\bibinfo {title} {Topological {K}ondo effect with {M}ajorana fermions},\ }\href {https://doi.org/10.1103/PhysRevLett.109.156803} {\bibfield  {journal} {\bibinfo  {journal} {Phys. Rev. Lett.}\ }\textbf {\bibinfo {volume} {109}},\ \bibinfo {pages} {156803} (\bibinfo {year} {2012})}\BibitemShut {NoStop}%
\bibitem [{\citenamefont {Mitchell}\ \emph {et~al.}(2021)\citenamefont {Mitchell}, \citenamefont {Liberman}, \citenamefont {Sela},\ and\ \citenamefont {Affleck}}]{MitchellAffleck2021}%
  \BibitemOpen
  \bibfield  {author} {\bibinfo {author} {\bibfnamefont {A.~K.}\ \bibnamefont {Mitchell}}, \bibinfo {author} {\bibfnamefont {A.}~\bibnamefont {Liberman}}, \bibinfo {author} {\bibfnamefont {E.}~\bibnamefont {Sela}},\ and\ \bibinfo {author} {\bibfnamefont {I.}~\bibnamefont {Affleck}},\ }\bibfield  {title} {\bibinfo {title} {{SO(5) Non-Fermi Liquid in a Coulomb Box Device}},\ }\href {https://doi.org/10.1103/PhysRevLett.126.147702} {\bibfield  {journal} {\bibinfo  {journal} {Phys. Rev. Lett.}\ }\textbf {\bibinfo {volume} {126}},\ \bibinfo {pages} {147702} (\bibinfo {year} {2021})}\BibitemShut {NoStop}%
\bibitem [{\citenamefont {Li}\ \emph {et~al.}(2023)\citenamefont {Li}, \citenamefont {K\"onig},\ and\ \citenamefont {V\"ayrynen}}]{LiVayrynen2023}%
  \BibitemOpen
  \bibfield  {author} {\bibinfo {author} {\bibfnamefont {G.}~\bibnamefont {Li}}, \bibinfo {author} {\bibfnamefont {E.~J.}\ \bibnamefont {K\"onig}},\ and\ \bibinfo {author} {\bibfnamefont {J.~I.}\ \bibnamefont {V\"ayrynen}},\ }\bibfield  {title} {\bibinfo {title} {Topological symplectic {K}ondo effect},\ }\href {https://doi.org/10.1103/PhysRevB.107.L201401} {\bibfield  {journal} {\bibinfo  {journal} {Phys. Rev. B}\ }\textbf {\bibinfo {volume} {107}},\ \bibinfo {pages} {L201401} (\bibinfo {year} {2023})}\BibitemShut {NoStop}%
\bibitem [{\citenamefont {K{\"o}nig}\ and\ \citenamefont {Tsvelik}(2023)}]{KoenigTsvelik2023}%
  \BibitemOpen
  \bibfield  {author} {\bibinfo {author} {\bibfnamefont {E.~J.}\ \bibnamefont {K{\"o}nig}}\ and\ \bibinfo {author} {\bibfnamefont {A.~M.}\ \bibnamefont {Tsvelik}},\ }\bibfield  {title} {\bibinfo {title} {Exact solution of the topological symplectic {K}ondo problem},\ }\href {https://doi.org/https://doi.org/10.1016/j.aop.2023.169231} {\bibfield  {journal} {\bibinfo  {journal} {Annals of Physics}\ }\textbf {\bibinfo {volume} {456}},\ \bibinfo {pages} {169231} (\bibinfo {year} {2023})}\BibitemShut {NoStop}%
\bibitem [{\citenamefont {Bollmann}\ \emph {et~al.}(2024)\citenamefont {Bollmann}, \citenamefont {V\"ayrynen},\ and\ \citenamefont {K\"onig}}]{BollmannKoenig2024}%
  \BibitemOpen
  \bibfield  {author} {\bibinfo {author} {\bibfnamefont {S.}~\bibnamefont {Bollmann}}, \bibinfo {author} {\bibfnamefont {J.~I.}\ \bibnamefont {V\"ayrynen}},\ and\ \bibinfo {author} {\bibfnamefont {E.~J.}\ \bibnamefont {K\"onig}},\ }\bibfield  {title} {\bibinfo {title} {Topological {K}ondo effect with spinful {M}ajorana fermions},\ }\href {https://doi.org/10.1103/PhysRevB.110.035136} {\bibfield  {journal} {\bibinfo  {journal} {Phys. Rev. B}\ }\textbf {\bibinfo {volume} {110}},\ \bibinfo {pages} {035136} (\bibinfo {year} {2024})}\BibitemShut {NoStop}%
\bibitem [{\citenamefont {Nordlander}\ \emph {et~al.}(1999)\citenamefont {Nordlander}, \citenamefont {Pustilnik}, \citenamefont {Meir}, \citenamefont {Wingreen},\ and\ \citenamefont {Langreth}}]{NordlanderLangreth1999}%
  \BibitemOpen
  \bibfield  {author} {\bibinfo {author} {\bibfnamefont {P.}~\bibnamefont {Nordlander}}, \bibinfo {author} {\bibfnamefont {M.}~\bibnamefont {Pustilnik}}, \bibinfo {author} {\bibfnamefont {Y.}~\bibnamefont {Meir}}, \bibinfo {author} {\bibfnamefont {N.~S.}\ \bibnamefont {Wingreen}},\ and\ \bibinfo {author} {\bibfnamefont {D.~C.}\ \bibnamefont {Langreth}},\ }\bibfield  {title} {\bibinfo {title} {How long does it take for the {K}ondo effect to develop?},\ }\href {https://doi.org/10.1103/PhysRevLett.83.808} {\bibfield  {journal} {\bibinfo  {journal} {Phys. Rev. Lett.}\ }\textbf {\bibinfo {volume} {83}},\ \bibinfo {pages} {808} (\bibinfo {year} {1999})}\BibitemShut {NoStop}%
\bibitem [{\citenamefont {Lobaskin}\ and\ \citenamefont {Kehrein}(2005)}]{LobaskinKehrein2005}%
  \BibitemOpen
  \bibfield  {author} {\bibinfo {author} {\bibfnamefont {D.}~\bibnamefont {Lobaskin}}\ and\ \bibinfo {author} {\bibfnamefont {S.}~\bibnamefont {Kehrein}},\ }\bibfield  {title} {\bibinfo {title} {Crossover from nonequilibrium to equilibrium behavior in the time-dependent {K}ondo model},\ }\href {https://doi.org/10.1103/PhysRevB.71.193303} {\bibfield  {journal} {\bibinfo  {journal} {Phys. Rev. B}\ }\textbf {\bibinfo {volume} {71}},\ \bibinfo {pages} {193303} (\bibinfo {year} {2005})}\BibitemShut {NoStop}%
\bibitem [{\citenamefont {Heyl}\ and\ \citenamefont {Kehrein}(2010{\natexlab{a}})}]{HeylKehrein2010}%
  \BibitemOpen
  \bibfield  {author} {\bibinfo {author} {\bibfnamefont {M.}~\bibnamefont {Heyl}}\ and\ \bibinfo {author} {\bibfnamefont {S.}~\bibnamefont {Kehrein}},\ }\bibfield  {title} {\bibinfo {title} {Interaction quench dynamics in the {K}ondo model in the presence of a local magnetic field},\ }\href {https://doi.org/10.1088/0953-8984/22/34/345604} {\bibfield  {journal} {\bibinfo  {journal} {Journal of Physics: Condensed Matter}\ }\textbf {\bibinfo {volume} {22}},\ \bibinfo {pages} {345604} (\bibinfo {year} {2010}{\natexlab{a}})}\BibitemShut {NoStop}%
\bibitem [{\citenamefont {Vasseur}\ \emph {et~al.}(2013)\citenamefont {Vasseur}, \citenamefont {Trinh}, \citenamefont {Haas},\ and\ \citenamefont {Saleur}}]{VasseurSaleur2013}%
  \BibitemOpen
  \bibfield  {author} {\bibinfo {author} {\bibfnamefont {R.}~\bibnamefont {Vasseur}}, \bibinfo {author} {\bibfnamefont {K.}~\bibnamefont {Trinh}}, \bibinfo {author} {\bibfnamefont {S.}~\bibnamefont {Haas}},\ and\ \bibinfo {author} {\bibfnamefont {H.}~\bibnamefont {Saleur}},\ }\bibfield  {title} {\bibinfo {title} {Crossover physics in the nonequilibrium dynamics of quenched quantum impurity systems},\ }\href {https://doi.org/10.1103/PhysRevLett.110.240601} {\bibfield  {journal} {\bibinfo  {journal} {Phys. Rev. Lett.}\ }\textbf {\bibinfo {volume} {110}},\ \bibinfo {pages} {240601} (\bibinfo {year} {2013})}\BibitemShut {NoStop}%
\bibitem [{\citenamefont {Averin}\ and\ \citenamefont {Pekola}(2010)}]{LobaskinKehrein2006}%
  \BibitemOpen
  \bibfield  {author} {\bibinfo {author} {\bibfnamefont {D.~V.}\ \bibnamefont {Averin}}\ and\ \bibinfo {author} {\bibfnamefont {J.~P.}\ \bibnamefont {Pekola}},\ }\bibfield  {title} {\bibinfo {title} {Violation of the fluctuation-dissipation theorem in time-dependent mesoscopic heat transport},\ }\href {https://doi.org/10.1103/PhysRevLett.104.220601} {\bibfield  {journal} {\bibinfo  {journal} {Phys. Rev. Lett.}\ }\textbf {\bibinfo {volume} {104}},\ \bibinfo {pages} {220601} (\bibinfo {year} {2010})}\BibitemShut {NoStop}%
\bibitem [{\citenamefont {Heyl}\ and\ \citenamefont {Kehrein}(2010{\natexlab{b}})}]{HeylKehrein2010b}%
  \BibitemOpen
  \bibfield  {author} {\bibinfo {author} {\bibfnamefont {M.}~\bibnamefont {Heyl}}\ and\ \bibinfo {author} {\bibfnamefont {S.}~\bibnamefont {Kehrein}},\ }\bibfield  {title} {\bibinfo {title} {Nonequilibrium steady state in a periodically driven {K}ondo model},\ }\href {https://doi.org/10.1103/PhysRevB.81.144301} {\bibfield  {journal} {\bibinfo  {journal} {Phys. Rev. B}\ }\textbf {\bibinfo {volume} {81}},\ \bibinfo {pages} {144301} (\bibinfo {year} {2010}{\natexlab{b}})}\BibitemShut {NoStop}%
\bibitem [{\citenamefont {Yao}\ \emph {et~al.}(2021)\citenamefont {Yao}, \citenamefont {Zhang}, \citenamefont {Wang}, \citenamefont {Ho},\ and\ \citenamefont {Orth}}]{YaoOrth2021}%
  \BibitemOpen
  \bibfield  {author} {\bibinfo {author} {\bibfnamefont {Y.}~\bibnamefont {Yao}}, \bibinfo {author} {\bibfnamefont {F.}~\bibnamefont {Zhang}}, \bibinfo {author} {\bibfnamefont {C.-Z.}\ \bibnamefont {Wang}}, \bibinfo {author} {\bibfnamefont {K.-M.}\ \bibnamefont {Ho}},\ and\ \bibinfo {author} {\bibfnamefont {P.~P.}\ \bibnamefont {Orth}},\ }\bibfield  {title} {\bibinfo {title} {{G}utzwiller hybrid quantum-classical computing approach for correlated materials},\ }\href {https://doi.org/10.1103/PhysRevResearch.3.013184} {\bibfield  {journal} {\bibinfo  {journal} {Phys. Rev. Res.}\ }\textbf {\bibinfo {volume} {3}},\ \bibinfo {pages} {013184} (\bibinfo {year} {2021})}\BibitemShut {NoStop}%
\bibitem [{\citenamefont {He}\ and\ \citenamefont {Lu}(2014)}]{He2014}%
  \BibitemOpen
  \bibfield  {author} {\bibinfo {author} {\bibfnamefont {R.-Q.}\ \bibnamefont {He}}\ and\ \bibinfo {author} {\bibfnamefont {Z.-Y.}\ \bibnamefont {Lu}},\ }\bibfield  {title} {\bibinfo {title} {Quantum renormalization groups based on natural orbitals},\ }\href {https://doi.org/10.1103/PhysRevB.89.085108} {\bibfield  {journal} {\bibinfo  {journal} {Phys. Rev. B}\ }\textbf {\bibinfo {volume} {89}},\ \bibinfo {pages} {085108} (\bibinfo {year} {2014})}\BibitemShut {NoStop}%
\bibitem [{\citenamefont {Yang}\ and\ \citenamefont {Feiguin}(2017)}]{Yang2017}%
  \BibitemOpen
  \bibfield  {author} {\bibinfo {author} {\bibfnamefont {C.}~\bibnamefont {Yang}}\ and\ \bibinfo {author} {\bibfnamefont {A.~E.}\ \bibnamefont {Feiguin}},\ }\bibfield  {title} {\bibinfo {title} {Unveiling the internal entanglement structure of the kondo singlet},\ }\href {https://doi.org/10.1103/PhysRevB.95.115106} {\bibfield  {journal} {\bibinfo  {journal} {Phys. Rev. B}\ }\textbf {\bibinfo {volume} {95}},\ \bibinfo {pages} {115106} (\bibinfo {year} {2017})}\BibitemShut {NoStop}%
\bibitem [{\citenamefont {Besserve}\ and\ \citenamefont {Ayral}(2022)}]{BesserveAyral2022}%
  \BibitemOpen
  \bibfield  {author} {\bibinfo {author} {\bibfnamefont {P.}~\bibnamefont {Besserve}}\ and\ \bibinfo {author} {\bibfnamefont {T.}~\bibnamefont {Ayral}},\ }\bibfield  {title} {\bibinfo {title} {Unraveling correlated material properties with noisy quantum computers: Natural orbitalized variational quantum eigensolving of extended impurity models within a slave-boson approach},\ }\href {https://doi.org/10.1103/PhysRevB.105.115108} {\bibfield  {journal} {\bibinfo  {journal} {Phys. Rev. B}\ }\textbf {\bibinfo {volume} {105}},\ \bibinfo {pages} {115108} (\bibinfo {year} {2022})}\BibitemShut {NoStop}%
\bibitem [{\citenamefont {Wu}\ \emph {et~al.}(2022)\citenamefont {Wu}, \citenamefont {Fishman}, \citenamefont {Pixley},\ and\ \citenamefont {Stoudenmire}}]{WuStoudenmire2022}%
  \BibitemOpen
  \bibfield  {author} {\bibinfo {author} {\bibfnamefont {A.-K.}\ \bibnamefont {Wu}}, \bibinfo {author} {\bibfnamefont {M.~T.}\ \bibnamefont {Fishman}}, \bibinfo {author} {\bibfnamefont {J.}~\bibnamefont {Pixley}},\ and\ \bibinfo {author} {\bibfnamefont {E.}~\bibnamefont {Stoudenmire}},\ }\bibfield  {title} {\bibinfo {title} {Disentangling interacting systems with fermionic {G}aussian circuits: Application to the single impurity {A}nderson model},\ }\href {https://arxiv.org/abs/2212.09798} {\bibfield  {journal} {\bibinfo  {journal} {arXiv preprint arXiv:2212.09798}\ } (\bibinfo {year} {2022})}\BibitemShut {NoStop}%
\bibitem [{\citenamefont {Albash}\ and\ \citenamefont {Lidar}(2018)}]{AlbashLidar2018}%
  \BibitemOpen
  \bibfield  {author} {\bibinfo {author} {\bibfnamefont {T.}~\bibnamefont {Albash}}\ and\ \bibinfo {author} {\bibfnamefont {D.~A.}\ \bibnamefont {Lidar}},\ }\bibfield  {title} {\bibinfo {title} {Adiabatic quantum computation},\ }\href {https://doi.org/10.1103/RevModPhys.90.015002} {\bibfield  {journal} {\bibinfo  {journal} {Rev. Mod. Phys.}\ }\textbf {\bibinfo {volume} {90}},\ \bibinfo {pages} {015002} (\bibinfo {year} {2018})}\BibitemShut {NoStop}%
\bibitem [{\citenamefont {Agarwal}\ \emph {et~al.}(2018)\citenamefont {Agarwal}, \citenamefont {Bhatt},\ and\ \citenamefont {Sondhi}}]{AgarwalSondhi2018}%
  \BibitemOpen
  \bibfield  {author} {\bibinfo {author} {\bibfnamefont {K.}~\bibnamefont {Agarwal}}, \bibinfo {author} {\bibfnamefont {R.~N.}\ \bibnamefont {Bhatt}},\ and\ \bibinfo {author} {\bibfnamefont {S.~L.}\ \bibnamefont {Sondhi}},\ }\bibfield  {title} {\bibinfo {title} {Fast preparation of critical ground states using superluminal fronts},\ }\href {https://doi.org/10.1103/PhysRevLett.120.210604} {\bibfield  {journal} {\bibinfo  {journal} {Phys. Rev. Lett.}\ }\textbf {\bibinfo {volume} {120}},\ \bibinfo {pages} {210604} (\bibinfo {year} {2018})}\BibitemShut {NoStop}%
\bibitem [{\citenamefont {Nielsen}\ and\ \citenamefont {Chuang}(2010)}]{NielsenChuang2010}%
  \BibitemOpen
  \bibfield  {author} {\bibinfo {author} {\bibfnamefont {M.~A.}\ \bibnamefont {Nielsen}}\ and\ \bibinfo {author} {\bibfnamefont {I.~L.}\ \bibnamefont {Chuang}},\ }\href {https://doi.org/https://doi.org/10.1017/CBO9780511976667} {\emph {\bibinfo {title} {Quantum Computation and Quantum Information: 10th Anniversary Edition}}}\ (\bibinfo  {publisher} {Cambridge University Press},\ \bibinfo {year} {2010})\BibitemShut {NoStop}%
\bibitem [{\citenamefont {Zhu}(2009)}]{Zhu2009}%
  \BibitemOpen
  \bibfield  {author} {\bibinfo {author} {\bibfnamefont {G.-Q.}\ \bibnamefont {Zhu}},\ }\bibfield  {title} {\bibinfo {title} {Average entanglement of spin 1 and 1/2 pair},\ }\href {https://doi.org/10.2478/s11534-008-0129-7} {\bibfield  {journal} {\bibinfo  {journal} {Open Physics}\ }\textbf {\bibinfo {volume} {7}},\ \bibinfo {pages} {135–140} (\bibinfo {year} {2009})}\BibitemShut {NoStop}%
\bibitem [{\citenamefont {Hill}\ and\ \citenamefont {Wootters}(1997)}]{HillWooters1997}%
  \BibitemOpen
  \bibfield  {author} {\bibinfo {author} {\bibfnamefont {S.~A.}\ \bibnamefont {Hill}}\ and\ \bibinfo {author} {\bibfnamefont {W.~K.}\ \bibnamefont {Wootters}},\ }\bibfield  {title} {\bibinfo {title} {Entanglement of a pair of quantum bits},\ }\href {https://doi.org/10.1103/PhysRevLett.78.5022} {\bibfield  {journal} {\bibinfo  {journal} {Phys. Rev. Lett.}\ }\textbf {\bibinfo {volume} {78}},\ \bibinfo {pages} {5022} (\bibinfo {year} {1997})}\BibitemShut {NoStop}%
\bibitem [{\citenamefont {Klich}\ and\ \citenamefont {Levitov}(2009)}]{KlichLevitov2009}%
  \BibitemOpen
  \bibfield  {author} {\bibinfo {author} {\bibfnamefont {I.}~\bibnamefont {Klich}}\ and\ \bibinfo {author} {\bibfnamefont {L.}~\bibnamefont {Levitov}},\ }\bibfield  {title} {\bibinfo {title} {Quantum noise as an entanglement meter},\ }\href {https://doi.org/10.1103/PhysRevLett.102.100502} {\bibfield  {journal} {\bibinfo  {journal} {Phys. Rev. Lett.}\ }\textbf {\bibinfo {volume} {102}},\ \bibinfo {pages} {100502} (\bibinfo {year} {2009})}\BibitemShut {NoStop}%
\bibitem [{\citenamefont {Affleck}(2008)}]{Affleck2008}%
  \BibitemOpen
  \bibfield  {author} {\bibinfo {author} {\bibfnamefont {I.}~\bibnamefont {Affleck}},\ }\bibfield  {title} {\bibinfo {title} {Quantum impurity problems in condensed matter physics},\ }\href@noop {} {\bibfield  {journal} {\bibinfo  {journal} {Exact Methods in Low-Dimensional Statistical Physics and Quantum Computing, Lecture Notes of the Les Houches Summer School}\ }\textbf {\bibinfo {volume} {89}},\ \bibinfo {pages} {3} (\bibinfo {year} {2008})}\BibitemShut {NoStop}%
\bibitem [{\citenamefont {Affleck}\ and\ \citenamefont {Ludwig}(1991{\natexlab{b}})}]{AffleckLudwig1991b}%
  \BibitemOpen
  \bibfield  {author} {\bibinfo {author} {\bibfnamefont {I.}~\bibnamefont {Affleck}}\ and\ \bibinfo {author} {\bibfnamefont {A.~W.~W.}\ \bibnamefont {Ludwig}},\ }\bibfield  {title} {\bibinfo {title} {Universal noninteger ``ground-state degeneracy'' in critical quantum systems},\ }\href {https://doi.org/10.1103/PhysRevLett.67.161} {\bibfield  {journal} {\bibinfo  {journal} {Phys. Rev. Lett.}\ }\textbf {\bibinfo {volume} {67}},\ \bibinfo {pages} {161} (\bibinfo {year} {1991}{\natexlab{b}})}\BibitemShut {NoStop}%
\bibitem [{\citenamefont {Zhou}\ \emph {et~al.}(2006)\citenamefont {Zhou}, \citenamefont {Barthel}, \citenamefont {Fj\ae{}restad},\ and\ \citenamefont {Schollw\"ock}}]{ZhouSchollwock2006}%
  \BibitemOpen
  \bibfield  {author} {\bibinfo {author} {\bibfnamefont {H.-Q.}\ \bibnamefont {Zhou}}, \bibinfo {author} {\bibfnamefont {T.}~\bibnamefont {Barthel}}, \bibinfo {author} {\bibfnamefont {J.~O.}\ \bibnamefont {Fj\ae{}restad}},\ and\ \bibinfo {author} {\bibfnamefont {U.}~\bibnamefont {Schollw\"ock}},\ }\bibfield  {title} {\bibinfo {title} {Entanglement and boundary critical phenomena},\ }\href {https://doi.org/10.1103/PhysRevA.74.050305} {\bibfield  {journal} {\bibinfo  {journal} {Phys. Rev. A}\ }\textbf {\bibinfo {volume} {74}},\ \bibinfo {pages} {050305} (\bibinfo {year} {2006})}\BibitemShut {NoStop}%
\bibitem [{\citenamefont {Calabrese}\ and\ \citenamefont {Cardy}(2004)}]{CalabreseCardy2004}%
  \BibitemOpen
  \bibfield  {author} {\bibinfo {author} {\bibfnamefont {P.}~\bibnamefont {Calabrese}}\ and\ \bibinfo {author} {\bibfnamefont {J.}~\bibnamefont {Cardy}},\ }\bibfield  {title} {\bibinfo {title} {Entanglement entropy and quantum field theory},\ }\href {https://doi.org/10.1088/1742-5468/2004/06/P06002} {\bibfield  {journal} {\bibinfo  {journal} {Journal of statistical mechanics: theory and experiment}\ }\textbf {\bibinfo {volume} {2004}},\ \bibinfo {pages} {P06002} (\bibinfo {year} {2004})}\BibitemShut {NoStop}%
\bibitem [{\citenamefont {Laflorencie}\ \emph {et~al.}(2008)\citenamefont {Laflorencie}, \citenamefont {S{\o}rensen},\ and\ \citenamefont {Affleck}}]{LaflorencieAffleck2008}%
  \BibitemOpen
  \bibfield  {author} {\bibinfo {author} {\bibfnamefont {N.}~\bibnamefont {Laflorencie}}, \bibinfo {author} {\bibfnamefont {E.~S.}\ \bibnamefont {S{\o}rensen}},\ and\ \bibinfo {author} {\bibfnamefont {I.}~\bibnamefont {Affleck}},\ }\bibfield  {title} {\bibinfo {title} {The {K}ondo effect in spin chains},\ }\href {https://doi.org/10.1088/1742-5468/2008/02/P02007} {\bibfield  {journal} {\bibinfo  {journal} {Journal of Statistical Mechanics: Theory and Experiment}\ }\textbf {\bibinfo {volume} {2008}},\ \bibinfo {pages} {P02007} (\bibinfo {year} {2008})}\BibitemShut {NoStop}%
\bibitem [{\citenamefont {Deschner}\ and\ \citenamefont {S{\o}rensen}(2011)}]{DeschnerSorensen2011}%
  \BibitemOpen
  \bibfield  {author} {\bibinfo {author} {\bibfnamefont {A.}~\bibnamefont {Deschner}}\ and\ \bibinfo {author} {\bibfnamefont {E.~S.}\ \bibnamefont {S{\o}rensen}},\ }\bibfield  {title} {\bibinfo {title} {Impurity entanglement in the ${J}--{J}2--\delta$ quantum spin chain},\ }\href {https://doi.org/https://doi.org/10.1088/1742-5468/2011/10/P10023} {\bibfield  {journal} {\bibinfo  {journal} {Journal of Statistical Mechanics: Theory and Experiment}\ }\textbf {\bibinfo {volume} {2011}},\ \bibinfo {pages} {P10023} (\bibinfo {year} {2011})}\BibitemShut {NoStop}%
\bibitem [{\citenamefont {Laflorencie}\ \emph {et~al.}(2006)\citenamefont {Laflorencie}, \citenamefont {S\o{}rensen}, \citenamefont {Chang},\ and\ \citenamefont {Affleck}}]{LaflorencieAffleck2006}%
  \BibitemOpen
  \bibfield  {author} {\bibinfo {author} {\bibfnamefont {N.}~\bibnamefont {Laflorencie}}, \bibinfo {author} {\bibfnamefont {E.~S.}\ \bibnamefont {S\o{}rensen}}, \bibinfo {author} {\bibfnamefont {M.-S.}\ \bibnamefont {Chang}},\ and\ \bibinfo {author} {\bibfnamefont {I.}~\bibnamefont {Affleck}},\ }\bibfield  {title} {\bibinfo {title} {Boundary effects in the critical scaling of entanglement entropy in 1d systems},\ }\href {https://doi.org/10.1103/PhysRevLett.96.100603} {\bibfield  {journal} {\bibinfo  {journal} {Phys. Rev. Lett.}\ }\textbf {\bibinfo {volume} {96}},\ \bibinfo {pages} {100603} (\bibinfo {year} {2006})}\BibitemShut {NoStop}%
\bibitem [{\citenamefont {Luca~D'Alessio}\ and\ \citenamefont {Rigol}(2016)}]{AlessioRigol2016}%
  \BibitemOpen
  \bibfield  {author} {\bibinfo {author} {\bibfnamefont {A.~P.}\ \bibnamefont {Luca~D'Alessio}, \bibfnamefont {Yariv~Kafri}}\ and\ \bibinfo {author} {\bibfnamefont {M.}~\bibnamefont {Rigol}},\ }\bibfield  {title} {\bibinfo {title} {From quantum chaos and eigenstate thermalization to statistical mechanics and thermodynamics},\ }\href {https://doi.org/10.1080/00018732.2016.1198134} {\bibfield  {journal} {\bibinfo  {journal} {Advances in Physics}\ }\textbf {\bibinfo {volume} {65}},\ \bibinfo {pages} {239} (\bibinfo {year} {2016})}\BibitemShut {NoStop}%
\bibitem [{\citenamefont {Deutsch}(2018)}]{Deutsch2018}%
  \BibitemOpen
  \bibfield  {author} {\bibinfo {author} {\bibfnamefont {J.~M.}\ \bibnamefont {Deutsch}},\ }\bibfield  {title} {\bibinfo {title} {Eigenstate thermalization hypothesis},\ }\href {https://doi.org/https://doi.org/10.1088/1361-6633/aac9f1} {\bibfield  {journal} {\bibinfo  {journal} {Reports on Progress in Physics}\ }\textbf {\bibinfo {volume} {81}},\ \bibinfo {pages} {082001} (\bibinfo {year} {2018})}\BibitemShut {NoStop}%
\bibitem [{\citenamefont {Affleck}()}]{Affleck2009}%
  \BibitemOpen
  \bibfield  {author} {\bibinfo {author} {\bibfnamefont {I.}~\bibnamefont {Affleck}},\ }\bibinfo {title} {The {K}ondo screening cloud: What it is and how to observe it},\ in\ \href {https://doi.org/10.1142/9789814299442_0001} {\emph {\bibinfo {booktitle} {Perspectives of Mesoscopic Physics}}},\ pp.\ \bibinfo {pages} {1--44}\BibitemShut {NoStop}%
\bibitem [{\citenamefont {Pixley}\ \emph {et~al.}(2015)\citenamefont {Pixley}, \citenamefont {Chowdhury}, \citenamefont {Miecnikowski}, \citenamefont {Stephens}, \citenamefont {Wagner},\ and\ \citenamefont {Ingersent}}]{PixleyIngersent2015}%
  \BibitemOpen
  \bibfield  {author} {\bibinfo {author} {\bibfnamefont {J.~H.}\ \bibnamefont {Pixley}}, \bibinfo {author} {\bibfnamefont {T.}~\bibnamefont {Chowdhury}}, \bibinfo {author} {\bibfnamefont {M.~T.}\ \bibnamefont {Miecnikowski}}, \bibinfo {author} {\bibfnamefont {J.}~\bibnamefont {Stephens}}, \bibinfo {author} {\bibfnamefont {C.}~\bibnamefont {Wagner}},\ and\ \bibinfo {author} {\bibfnamefont {K.}~\bibnamefont {Ingersent}},\ }\bibfield  {title} {\bibinfo {title} {Entanglement entropy near {K}ondo-destruction quantum critical points},\ }\href {https://doi.org/10.1103/PhysRevB.91.245122} {\bibfield  {journal} {\bibinfo  {journal} {Phys. Rev. B}\ }\textbf {\bibinfo {volume} {91}},\ \bibinfo {pages} {245122} (\bibinfo {year} {2015})}\BibitemShut {NoStop}%
\bibitem [{\citenamefont {Wille}\ \emph {et~al.}(2019)\citenamefont {Wille}, \citenamefont {Van~Meter},\ and\ \citenamefont {Naveh}}]{WilleNaveh2019}%
  \BibitemOpen
  \bibfield  {author} {\bibinfo {author} {\bibfnamefont {R.}~\bibnamefont {Wille}}, \bibinfo {author} {\bibfnamefont {R.}~\bibnamefont {Van~Meter}},\ and\ \bibinfo {author} {\bibfnamefont {Y.}~\bibnamefont {Naveh}},\ }\bibfield  {title} {\bibinfo {title} {{I}{B}{M}’s qiskit tool chain: Working with and developing for real quantum computers},\ }in\ \href {https://doi.org/10.23919/DATE.2019.8715261} {\emph {\bibinfo {booktitle} {2019 Design, Automation I\& Test in Europe Conference I\& Exhibition (DATE)}}}\ (\bibinfo {year} {2019})\ pp.\ \bibinfo {pages} {1234--1240}\BibitemShut {NoStop}%
\bibitem [{\citenamefont {Zar\'and}\ and\ \citenamefont {von Delft}(2000)}]{ZarandVonDelft2000}%
  \BibitemOpen
  \bibfield  {author} {\bibinfo {author} {\bibfnamefont {G.}~\bibnamefont {Zar\'and}}\ and\ \bibinfo {author} {\bibfnamefont {J.}~\bibnamefont {von Delft}},\ }\bibfield  {title} {\bibinfo {title} {Analytical calculation of the finite-size crossover spectrum of the anisotropic two-channel {K}ondo model},\ }\href {https://doi.org/10.1103/PhysRevB.61.6918} {\bibfield  {journal} {\bibinfo  {journal} {Phys. Rev. B}\ }\textbf {\bibinfo {volume} {61}},\ \bibinfo {pages} {6918} (\bibinfo {year} {2000})}\BibitemShut {NoStop}%
\bibitem [{\citenamefont {Affleck}(1995)}]{Affleck1995}%
  \BibitemOpen
  \bibfield  {author} {\bibinfo {author} {\bibfnamefont {I.}~\bibnamefont {Affleck}},\ }\bibfield  {title} {\bibinfo {title} {Conformal field theory approach to the {K}ondo effect},\ }\href {https://doi.org/https://doi.org/10.48550/arXiv.cond-mat/9512099} {\bibfield  {journal} {\bibinfo  {journal} {Acta Phys. Polon. B}\ }\textbf {\bibinfo {volume} {26}},\ \bibinfo {pages} {1869} (\bibinfo {year} {1995})}\BibitemShut {NoStop}%
\bibitem [{\citenamefont {Heyl}(2012)}]{Heyl2012}%
  \BibitemOpen
  \bibfield  {author} {\bibinfo {author} {\bibfnamefont {M.~P.~L.}\ \bibnamefont {Heyl}},\ }\href {http://nbn-resolving.de/urn:nbn:de:bvb:19-145838} {\bibinfo {title} {Nonequilibrium phenomena in many-body quantum systems}} (\bibinfo {year} {2012})\BibitemShut {NoStop}%
\bibitem [{\citenamefont {Bertrand}\ \emph {et~al.}(2024)\citenamefont {Bertrand}, \citenamefont {Besserve}, \citenamefont {Ferrero},\ and\ \citenamefont {Ayral}}]{BertrandAyral2024}%
  \BibitemOpen
  \bibfield  {author} {\bibinfo {author} {\bibfnamefont {C.}~\bibnamefont {Bertrand}}, \bibinfo {author} {\bibfnamefont {P.}~\bibnamefont {Besserve}}, \bibinfo {author} {\bibfnamefont {M.}~\bibnamefont {Ferrero}},\ and\ \bibinfo {author} {\bibfnamefont {T.}~\bibnamefont {Ayral}},\ }\bibfield  {title} {\bibinfo {title} {Turning qubit noise into a blessing: Automatic state preparation and long-time dynamics for impurity models on quantum computers},\ }\href {https://arxiv.org/abs/2412.13711} {\bibfield  {journal} {\bibinfo  {journal} {arXiv preprint arXiv:2412.13711}\ } (\bibinfo {year} {2024})}\BibitemShut {NoStop}%
\end{thebibliography}%

\end{document}